\documentclass[12pt]{article}
\usepackage{amsmath,amssymb}
\usepackage{bm}
\usepackage{color}
\usepackage[dvipdfmx]{graphicx}
\usepackage{cite}
\usepackage{mathtools}

\usepackage{multirow}
\usepackage {diagbox}
\usepackage{dcolumn}
\usepackage{mathdots}
\usepackage{here}

\def\slash#1{\not\!\!#1}

\begin{document}

\title{
\begin{flushright}
\ \\*[-80pt]
\begin{minipage}{0.2\linewidth}
\normalsize
%arXiv:YYMM.NNNN \\
EPHOU-22-018\\*[50pt]
\end{minipage}
\end{flushright}
% Title
{\Large \bf
Quark and lepton flavor structure in magnetized orbifold models at residual modular symmetric points
\\*[20pt]}}
% /Title

\author{
Kouki Hoshiya,
%\footnote{A's mail}
Shota Kikuchi,
%\footnote{B's mail}
~Tatsuo Kobayashi, and
%\footnote{C's mail}
~Hikaru Uchida
%\footnote{D's mail}
\\*[20pt]
\centerline{
\begin{minipage}{\linewidth}
\begin{center}
{\it \normalsize
Department of Physics, Hokkaido University, Sapporo 060-0810, Japan} \\*[5pt]
\end{center}
\end{minipage}}
\\*[50pt]}

\date{
\centerline{\small \bf Abstract}
\begin{minipage}{0.9\linewidth}
\medskip
\medskip
\small
We study quark and lepton mass matrices derived from magnetized $T^2/\mathbb{Z}_2$ orbifold models.
Quark and lepton masses have a large hierarchy.
In addition, mixing angles are large in the lepton sector, while 
those are small in the quark sector.
We find that this structure can be realized in certain flavor models, which are identified by the zero points of the zero-mode wave functions of fermions and Higgs modes.
We classify such realistic flavor models.
Fixed points $\tau=i$, $e^{2\pi i/3}$ and $i\infty$ of the modulus $\tau$ play a role 
in realizing a large mass hierarchy
through our scenario, 
where residual $S$, $ST$, and $T$ symmetries remain and the lightest Higgs modes 
can correspond to eigenstates of residual symmetries at the leading order.
As a result, we find that there are 24  flavor models in total which can be realistic in a vicinity of $S$-symmetric vacuum but no flavor models for $ST$ and $T$-symmetric vacua.
%As an illustrative example, we show the realization of the quark and lepton mass ratios and mixings.
\end{minipage}
}

\begin{titlepage}
\maketitle
\thispagestyle{empty}
\end{titlepage}

\newpage

% ------------------------------------------------------ %
% ------------------------------------------------------ %
% ------------------------------------------------------ %
% ------------------------------------------------------ %

\section{Introduction}
\label{Intro}

The standard model (SM) established by the discovery of Higgs paricle is a successful theory describing almost all observations from current experiments.
However it still contains several unsolved issues.
The origin of the quark and lepton flavors is one of such issues.
Quark and  lepton masses are required to be hierarchical, and  neutrino masses 
must be extremely light compared with other fermions by observational results.
In addition, quark mixings are required to be small but lepton mixings are large.
Also we have CP-violating phases.
To describe these observables in the quark sector, the SM needs ten real parameters: six quark masses, three mixing angles and one CP-violating phase.
For the lepton sector, it needs twelve real parameters: six lepton masses, three mixing angles, and three Dirac and Majorana CP-violating phases.
Understanding the origin of these parameters is one of the most fundamental and challenging issues in the current particle physics.

Superstring theory is a promising candidate for the unified theory.
This theory predicts ten-dimensional (10D) space-time.
The extra six-dimensional (6D) space must be compactified to be unobserved.
The low-energy effective theory of superstring theory leads 10D super-Yang-Mills theory and its compactification can lead to solutions to the issues in particle physics.
Hence, we can expect that the quark and lepton flavor structures are originated from this 6D space.
For example, torus and orbifold compactifications with magnetic flux background give the four-dimensional chiral theory where wave functions have a generation structure determined by the size of the magnetic flux and Yukawa couplings depend on moduli \cite{Cremades:2004wa,Abe:2008fi,Abe:2013bca,Abe:2014noa}.
In this sense, superstring theory on torus compactification and its orbifoldings 
with magnetic fluxes are attractive.
Indeed, several numerical studies have shown that realistic flavor structures can be realized \cite{Abe:2012fj,Abe:2014vza,Fujimoto:2016zjs,Kobayashi:2016qag,Kikuchi:2021yog,Kikuchi:2022geu}.

Moreover, torus compactification has another important aspect.
The geometrical symmetry of torus is the modular symmetry $\Gamma \equiv SL(2,\mathbb{Z})$ as well as $\bar{\Gamma} \equiv SL(2,\mathbb{Z})/\mathbb{Z}_2$, which has recently drawn  attention from the bottom-up approach.
It is well known that the finite modular subgroups $\Gamma_N$ for $N=2,3,4$ and $5$ are isomorphic to $S_3$, $A_4$, $S_4$ and $A_5$, respectively \cite{deAdelhartToorop:2011re}.
Inspired by these aspects, flavor models with $\Gamma_N$ have been studied intensively in the bottom-up approach.
(See e.g. Refs.~\cite{Feruglio:2017spp,Kobayashi:2018vbk,Penedo:2018nmg,Novichkov:2018nkm,Criado:2018thu,
Kobayashi:2018scp,
Ding:2019zxk,Novichkov:2018ovf,
Kobayashi:2019mna,Wang:2019ovr,Ding:2019xna,
Liu:2019khw,Chen:2020udk,Novichkov:2020eep,Liu:2020akv,
deMedeirosVarzielas:2019cyj,
  	Asaka:2019vev,Ding:2020msi,Asaka:2020tmo,deAnda:2018ecu,Kobayashi:2019rzp,Novichkov:2018yse,Kobayashi:2018wkl,Okada:2018yrn,Okada:2019uoy,Nomura:2019jxj, Okada:2019xqk,
  	Nomura:2019yft,Nomura:2019lnr,Criado:2019tzk,
  	King:2019vhv,Gui-JunDing:2019wap,deMedeirosVarzielas:2020kji,Zhang:2019ngf,Nomura:2019xsb,Kobayashi:2019gtp,Lu:2019vgm,Wang:2019xbo,King:2020qaj,Abbas:2020qzc,Okada:2020oxh,Okada:2020dmb,Ding:2020yen,Okada:2020rjb,Okada:2020ukr,Nagao:2020azf,Wang:2020lxk,
  	Okada:2020brs,Yao:2020qyy}.)
In those models, matter fields are assumed to transform non-trivially under the modular symmetry.
In addition, Yukawa couplings are also the modular forms and depend on the modulus $\tau$.
Especially, the modulus at the modular fixed points, $\tau=i$, $e^{2\pi i/3}$ and $i\infty$, are important.
Yukawa couplings as well as mass matrices have $S(\mathbb{Z}_2)$, $ST(\mathbb{Z}_3)$ and $T(\mathbb{Z}_N)$-symmetries at $\tau=i$, $e^{2\pi i/3}$ and $i\infty$, respectively, and these residual symmetries make the structure of Yukawa couplings specific patterns.
In Refs.~\cite{Novichkov:2018yse,Okada:2019uoy,Gui-JunDing:2019wap,Okada:2020rjb,Okada:2020ukr}, realistic results were obtained at the vicinity of the modualr fixed points.
(See also Ref.~\cite{Kikuchi:2022svo}.)

Here, we study the quark and lepton mass matrices derived from magnetized $T^2/\mathbb{Z}_2$ orbifold models.
Zero-modes in magnetized $T^2/\mathbb{Z}_2$ orbifold models transform 
non-trivially under the modular symmetry 
\cite{Kobayashi:2018rad,Kobayashi:2018bff,Kariyazono:2019ehj,Ohki:2020bpo,Kikuchi:2020frp,Kikuchi:2020nxn,
Kikuchi:2021ogn,Almumin:2021fbk}.
In general,  compactifications of superstring theory lead to more than one candidates for  Higgs modes, which have the same quantum numbers under $SU(3)\times SU(2)\times U(1)$ SM gauge group and can couple with quarks and leptons.
It is true for torus and orbifold compactification with magnetic flux.
Note that again the generation number of wave functions are determined by the size of magnetic flux.
Then, the number of Higgs pairs is also determined by the magnetic fluxes in the quark and lepton sectors and is larger than one, in general.
They can couple to both quarks and leptons and give multi-pair Yukawa couplings among them.
Thus mass matrices of quarks and leptons are given by the liner combination of these Yukawa couplings and vacuum expectation values (VEVs) of multi-pair Higgs fields.
We expect that those Higgs modes may have mass terms, i.e. the so-called $\mu$-terms, 
and the lightest linear combination develops its VEV.
Thus, the Higgs VEV direction is determined by the lightest direction.
The mass terms  of Higgs fields are forbidden at the perturbative level in superstring theory.
They can be induced by non-perturbative effects such as D-brane instanton effects.
However, such analyses are not straightforward in explicit models, and the lightest direction is not clear.
In our analysis, as in Refs.~\cite{Abe:2012fj,Abe:2014vza,Fujimoto:2016zjs,Kobayashi:2016qag}, we use the direction of Higgs VEVs as parameters and find the model conditions to lead to hierarchical masses of quarks and charged lepton, small mixing of quarks and large mixing of leptons.
In addition, we assume the vicinity  at either of three modular fixed points, $\tau=i$, $e^{2\pi i/3}$ and $i\infty$, and Higgs VEVs are aligned in the eigenbasis of $S$, $ST$ and $T$-transformations, respectively.
As we will show, such vacuum can be led by the leading order Higgs $\mu$-term due to D-brane instanton effects.

Also we need to solve the smallness of neutrino masses.
It can be realized by see-saw mechanism.
%In this paper, we consider the see-saw mechanism.
Majorana mass terms of right-handed neutrino can be induced by 
D-brane instanton effects \cite{Blumenhagen:2006xt,Ibanez:2006da,Ibanez:2007rs,Antusch:2007jd,Kobayashi:2015siy}.
In particular, in Ref.~\cite{Hoshiya:2021nux}, possible forms of right-handed neutrino mass terms induced by D-brane instanton effects on magnetized $T^2/\mathbb{Z}_2$ orbifold were studied.
Assuming same D-brane instanton effects, we obtain light neutrino masses through the see-saw mechanism.

The paper is organized as follows.
In section \ref{sec:Magnetized_orbifold_model}, we review the zero-mode wave functions and flavor models on the torus and orbifold with magnetic fluxes.
In section \ref{sec:conditions}, we study the conditions to realize quark and lepton flavor structure  and classify the flavor models consistent with them.
In section \ref{sec:Modular_symmetric_models}, we also classify the flavor models consistent with the modular symmetric vacuum.
In section \ref{sec:Numerical_example}, we give the numerical studies for the quark and lepton mass matrices in our models. 
Section \ref{sec:Conclusion} concludes this study.
In Appendix \ref{App:Neutrino_Majorana_mass_term} and \ref{App:Higgs_mu_term}, we review the Majorana mass terms of right-handed neutrinos and Higgs $\mu$-terms induced by D-brane instanton effects, respectively.
In Appendix \ref{App:The_model_in_the_numerical_example}, we summarize the Yukawa couplings and Majorana mass terms on the model in the numerical studies in section \ref{sec:Numerical_example}.
In Appendix \ref{apx:CLASSMODELS}, we show the list of flavor models satisfying the conditions shown in section \ref{sec:conditions}.

%-----------------------------------------------------
%-----------------------------------------------------
%-----------------------------------------------------

\section{Magnetized orbifold model}
\label{sec:Magnetized_orbifold_model}

%-----------------------------------------------------
%-----------------------------------------------------

\subsection{Torus and orbifold compactifications}

First of all, we briefly review the zero-mode wave functions on the magnetized $T^2$.
We assume non-trivial background magnetic flux on $T^2$.
To make our analysis simple, we concentrate on $U(1)$ background magnetic flux:
\begin{align}
  F = dA = \frac{\pi iM}{{\rm Im}\tau} dz \wedge d\bar{z},
\end{align}
where $z$ denotes the complex coordinate on $T^2$, $\tau$ denotes the complex structure modulus and $M$ is a quantity quantized by the torus boundary condition.
This flux is led by the following vector potential one-form,
\begin{align}
  A = \frac{\pi M}{{\rm Im}\tau} {\rm Im} (\bar{z}dz).
\end{align}
Then the torus identification $z\sim z+m+n\tau,~m,n\in\mathbb{Z}$, makes the value of flux $M$ be integer, which is called Dirac quantization condition.

The two-dimensional spinor with $U(1)$ unit charge $q=1$, $\psi=(\psi_+,\psi_-)^T$, must satisfy the following boundary conditions,
\begin{align}
  \psi (z+1) = e^{2\pi i\alpha_1} e^{\pi iM\frac{{\rm Im}z}{{\rm Im}\tau}} \psi (z), \quad \psi (z+\tau) = e^{2\pi i\alpha_2} e^{\pi iM\frac{{\rm Im}\bar{\tau}z}{{\rm Im}\tau}} \psi (z), \label{eq:B.C.PSI}
\end{align}
where $\alpha_i$, $i=1,2$, denote Scherk-Schwarz (SS) phases, which cannot be removed by the gauge transformation.
Imposing these conditions on the massless Dirac equation, $i\slash{D}\psi = 0$, it is found that 
when $M>0$ ($M<0$), $\psi_+$ ($\psi_-$) has the $|M|$ number of degenerate solutions but $\psi_-$ ($\psi_+$) has no solution.
Then the $j$-th zero-modes of $\psi_+$ for $M>0$  and $\psi_-$ for $M<0$ are  expressed as
\begin{align}
  &\psi_+^{(j+\alpha_1,\alpha_2),|M|}(z,\tau)
  = \left(\frac{|M|}{{\cal A}^2}\right)^{1/4} e^{2\pi i\frac{(j+\alpha_1)\alpha_2}{|M|}} e^{\pi i|M|z\frac{{\rm Im}z}{{\rm Im}\tau}}
  \vartheta \begin{bmatrix}
    \frac{j+\alpha_1}{|M|} \\
    -\alpha_2 \\
  \end{bmatrix}
  (|M|z,|M|\tau), \label{eq:psi_formula+} \\
  &\psi_-^{(j+\alpha_1,\alpha_2),|M|}(z,\tau)
  = \left(\psi_+^{(-j-\alpha_1,\alpha_2),|M|}(z,\tau)\right)^*, \quad j = 0,1,...,|M|-1,
\end{align}
where ${\cal A}$ denotes the area of $T^2$ and $\vartheta$ denotes the Jacobi theta function defined by
\begin{align}
  \vartheta
  \begin{bmatrix}
    a \\ b \\
  \end{bmatrix}
  (\nu,\tau)
  = \sum_{\ell\in\mathbb{Z}} e^{\pi i(a+\ell)^2\tau} e^{2\pi i(a+\ell)(\nu+b)}.
\end{align}
This function has the following property:
\begin{align}
  \vartheta
  \begin{bmatrix}
    \frac{j}{M_1} \\ 0 \\
  \end{bmatrix}
  (\nu_1,M_1\tau)
  \times
  \vartheta
  \begin{bmatrix}
    \frac{k}{M_2} \\ 0 \\
  \end{bmatrix}
  (\nu_2,M_2\tau)
  &= \sum_{m\in\mathbb{Z}_{M_1+M_2}}
  \vartheta
  \begin{bmatrix}
    \frac{j+k+M_1m}{M_1+M_2} \\ 0 \\
  \end{bmatrix}
  (\nu_1+\nu_2,(M_1+M_2)\tau) \notag \\
  \times
  \vartheta &
  \begin{bmatrix}
    \frac{M_2j-M_1k+M_1M_2m}{M_1M_2(M_1+M_2)} \\ 0 \\
  \end{bmatrix}
  (\nu_1M_2-\nu_2M_1,M_1M_2(M_1+M_2)\tau),
\end{align}
and it follows that the zero-modes satisfy the normalization and product expansions,
\begin{align}
  &\int d^2z \psi^{(i+\alpha_1,\alpha_2),|M|}_{\pm}(z,\tau) \left(\psi^{(j+\alpha_1,\alpha_2),|M|}_{\pm} (z,\tau)\right)^* = (2{\rm Im}\tau)^{-1/2}\delta_{i,j}, \label{eq:NormalizationPSI}\\
  &\psi^{(i+\alpha_1,\alpha_2),|M_1|}_\pm (z,\tau) \cdot \psi^{(j+\beta_1,\beta_2),|M_2|}_\pm (z,\tau)
  = \sum_{k\in\mathbb{Z}_{|M_3|}} y^{ijk}_{T^2} \psi^{(k+\gamma_1,\gamma_2),|M_3|}_\pm (z,\tau), \label{eq:ProductPSI}
\end{align}
where $|M_3|=|M_1|+|M_2|$, $(\gamma_1,\gamma_2)=(\alpha_1,\alpha_2)+(\beta_1,\beta_2)$ and
\begin{align}
  y^{ijk}_{T^2} = (2{\rm Im}\tau)^{1/2}\int_{T^2} d^2z \psi^{(i+\alpha_{1L},\alpha_{2L}),M_L}_{T^2}(z) \cdot \psi^{(j+\alpha_{1R},\alpha_{2R}),M_R}_{T^2}(z) \cdot \left(\psi^{(k+\alpha_{1H},\alpha_{2H}),M_H}_{T^2}(z)\right)^*.
\end{align}
Hereafter we assume $M>0$ and denote $\psi_+^{(j+\alpha_1,\alpha_2),|M|}$ as $\psi_{T^2}^{(j+\alpha_1,\alpha_2),M}$ although one can study $M<0$ case in the same way.

Second, we review the zero-mode wave functions on the $T^2/\mathbb{Z}_2$ twisted orbifold.
The $T^2/\mathbb{Z}_2$ twisted orbifold is obtained by the $\mathbb{Z}_2$ twist identification, $z\sim -z$, in addition to torus identification.
Then the zero-modes on $T^2/\mathbb{Z}_2$ are given by
\begin{align}
  \psi_{T^2/\mathbb{Z}_2^m}^{(j+\alpha_1,\alpha_2),M} (z)
  &= {\cal N}^{(j+\alpha_1)} \left(
  \psi_{T^2}^{(j+\alpha_1,\alpha_2),M}(z) + (-1)^m \psi_{T^2}^{(j+\alpha_1,\alpha_2),M}(-z)
  \right) \label{eq:z-z} \\
  &= {\cal N}^{(j+\alpha_1)} \left(
  \psi_{T^2}^{(j+\alpha_1,\alpha_2),M}(z) + (-1)^{m-2\alpha_2} \psi_{T^2}^{(M-(j+\alpha_1),\alpha_2),M}(z)
  \right) \notag \\
  &\equiv O^{jk,\alpha_1,\alpha_2,M}_m \psi_{T^2}^{(k+\alpha_1,\alpha_2),M}(z), \notag
\end{align}
where $m\in\{0,1\}$ denotes parities under $\mathbb{Z}_2$ twist, ${\cal N}^{(j+\alpha_1)}$ is a normalization factor defined by
\begin{align}
  {\cal N}^{(j+\alpha_1)} = \left\{
  \begin{array}{l}
    1/2 \quad (j+\alpha_1=0,M/2), \\
    1/\sqrt{2} \quad ({\rm otherwise}), \\
  \end{array} \right.
\end{align}
and
\begin{align}
  O^{jk,\alpha_1,\alpha_2,M}_m = {\cal N}^{(j+\alpha_1)} (\delta_{j,k}+(-1)^{m-2\alpha_2}\delta_{M-j-2\alpha_1,k}). \label{eq:OforT2/Z2}
\end{align}
Note that SS phases on $T^2/\mathbb{Z}_2$ are restricted to
\begin{align}
  (\alpha_1,\alpha_2) = (0,0),~(1/2,0),~(0,1/2),~(1/2,1/2),
\end{align}
due to the $\mathbb{Z}_2$ twist identification.
The number of zero-modes is summarized in Table \ref{tab:NumZero-modesT2/Z2}.
\begin{table}[h]
  \begin{center}
    \begin{tabular}{|c||c|c|} \hline
      ($\mathbb{Z}_2$ parity, $\alpha_1$, $\alpha_2$) & $M={\rm even}$ & $M={\rm odd}$ \\ \hline
      (0,0,0) & $\frac{M}{2}+1$ & $\frac{M+1}{2}$ \\ \hline
      (1,0,0) & $\frac{M}{2}-1$ & $\frac{M-1}{2}$ \\ \hline
      (0,1/2,0) & $\frac{M}{2}$ & $\frac{M+1}{2}$ \\ \hline
      (1,1/2,0) & $\frac{M}{2}$ & $\frac{M-1}{2}$ \\ \hline
      (0,0,1/2) & $\frac{M}{2}$ & $\frac{M+1}{2}$ \\ \hline
      (1,0,1/2) & $\frac{M}{2}$ & $\frac{M-1}{2}$ \\ \hline
      (0,1/2,1/2) & $\frac{M}{2}$ & $\frac{M-1}{2}$ \\ \hline
      (1,1/2,1/2) & $\frac{M}{2}$ & $\frac{M+1}{2}$ \\ \hline
    \end{tabular}
  \end{center}
  \caption{The number of zero-modes on $T^2/\mathbb{Z}_2$.
  $\mathbb{Z}_2$ parities 0 and 1 denote even and odd modes, respectively.}
  \label{tab:NumZero-modesT2/Z2}
\end{table}

%-----------------------------------------------------
%-----------------------------------------------------

\subsection{Zero points of zero-modes on $T^2/\mathbb{Z}_2$}

Here we study the zero points of zero-mode wave functions on $T^2/\mathbb{Z}_2$.
Zero points of zero-mode wave functions are the coordinates on $T^2/\mathbb{Z}_2$ where all zero-mode wave functions vanish, $\psi_{T^2/\mathbb{Z}_2^m}^{(j+\alpha_1,\alpha_2),M} (z) =0$ 
for all of $j$.
We focus on zero points at the fixed points which are invariant points under $\mathbb{Z}_2$ twist up to lattice translations of torus.
As we will see soon, all zero-mode wave functions become zero or non-zero at each fixed point.
The fixed points on $T^2/\mathbb{Z}_2$ are obtained as follows:
\begin{align}
  P_F = \left\{0,~\frac{1}{2},~\frac{\tau}{2},~\frac{1+\tau}{2}\right\}.
\end{align}
First we consider the boundary conditions at the fixed points in order to find the zero points.
From Eq.~(\ref{eq:psi_formula+}), we can check that the zero-modes on $T^2$ satisfy the following conditions,
\begin{align}
  \psi_{T^2}^{(j+\alpha_1,\alpha_2),M} (z+\frac{n_1+n_2\tau}{2}) 
  =e^{\pi i(j+\alpha_1)n_1}e^{\frac{\pi iM}{4}n_1n_2} e^{\frac{\pi iM}{2}\frac{{\rm Im}(n_1+n_2\bar{\tau})z}{{\rm Im}\tau}} \cdot \psi_{T^2}^{(j+\alpha_1+\frac{M}{2}n_2,\alpha_2+\frac{M}{2}n_1),M} (z),
\end{align}
where $n_1,n_2\in\mathbb{Z}.$
They lead to
\begin{align}
  &(-1)^{m-2\alpha_2}\psi_{T^2}^{(M-(j+\alpha_1),\alpha_2),M} (z+\frac{n_1+n_2\tau}{2}) 
  = 
    e^{\pi i(j+\alpha_1)n_1}e^{\frac{\pi iM}{4}n_1n_2} e^{\frac{\pi iM}{2}\frac{{\rm Im}(n_1+n_2\bar{\tau})z}{{\rm Im}\tau}} \notag \\
    &\qquad \qquad \qquad \times (-1)^{(m-Mn_1n_2-2(\alpha_1n_1+\alpha_2n_2))-2(\alpha_2+\frac{M}{2}n_1)}\psi_{T^2}^{(M-(j+\alpha_1+\frac{M}{2}n_2),\alpha_2+\frac{M}{2}n_1),M} (z).
\end{align}
Thus, the zero-modes on $T^2/\mathbb{Z}_2$ also satisfy the following conditions,
\begin{align}
  &\psi_{T^2/\mathbb{Z}_2^m}^{(j+\alpha_1,\alpha_2),M} (z+\frac{n_1+n_2\tau}{2}) =e^{\pi i(j+\alpha_1)n_1}e^{\frac{\pi iM}{4}n_1n_2} e^{\frac{\pi iM}{2}\frac{{\rm Im}(n_1+n_2\bar{\tau})z}{{\rm Im}\tau}} \psi_{T^2/\mathbb{Z}_2^{m'}}^{(j+\alpha_1+\frac{M}{2}n_2,\alpha_2+\frac{M}{2}n_1),M} (z),
  \label{eq:BCxxx}
\end{align}
where 
\begin{align}
m' = m-Mn_1n_2-2(\alpha_1n_1+\alpha_2n_2)~~~~~({\rm mod}~2).
\end{align}
Note that $m,m'=0$ and $1$ denote $\mathbb{Z}_2$ even and odd modes, respectively.
At $z=0$, we obtain
\begin{align}
  &\psi_{T^2/\mathbb{Z}_2^m}^{(j+\alpha_1,\alpha_2),M} (\frac{n_1+n_2\tau}{2}) 
  = 
  e^{\pi i(j+\alpha_1)n_1}e^{\frac{\pi iM}{4}n_1n_2} \psi_{T^2/\mathbb{Z}_2^{m'}}^{(j+\alpha_1+\frac{M}{2}n_2,\alpha_2+\frac{M}{2}n_1),M} (0). \label{eq:BC+alphaT2/Z2}
\end{align}
This means that the zero-mode wave function values on $T^2/\mathbb{Z}_2$ at the fixed points are given in terms  of ones at $z=0$; therefore we can find whether zero-modes at the fixed points vanish or not by the values of zero-modes at $z=0$.

It follows from zero-modes formula in Eq.~(\ref{eq:z-z}) that $\mathbb{Z}_2$ even and odd modes satisfy
\begin{align}
  &\psi^{(j+\alpha_1,\alpha_2),M}_{T^2/\mathbb{Z}_2^0}(0) \neq 0, \quad \psi^{(j+\alpha_1,\alpha_2),M}_{T^2/\mathbb{Z}_2^1}(0) = 0, \label{eq:T2/Z2at0}
\end{align}
at $z=0$ for all $j,\alpha_1,\alpha_2$ and $M$.
This means that if the $\mathbb{Z}_2$ parity of the zero-mode on the right-hand-side in Eq.~(\ref{eq:BC+alphaT2/Z2}) is odd, the zero-mode on the left-hand-side vanishes.
Thus the zero-modes with $(m,\alpha_1,\alpha_2)$ at the fixed point $z=\frac{n_1+n_2\tau}{2}$ become zero if
\begin{align}
  m'=m-Mn_1n_2-2(\alpha_1n_1+\alpha_2n_2) = 1~~~~~({\rm mod}~2), 
\end{align}
are satisfied.
In Table \ref{tab:ZeroPoints} we have summarized the zero points of each zero-mode at the fixed points.
\begin{table}[h]
  \begin{center}
    \begin{tabular}{|c||c|c|} \hline
      ($\mathbb{Z}_2$ parity,$\alpha_1$,$\alpha_2$) & $M={\rm even}$ & $M={\rm odd}$ \\ \hline
      (0,0,0) & None & $\frac{1+\tau}{2}$ \\ \hline
      (1,0,0) & $0,~\frac{1}{2},~\frac{\tau}{2},~\frac{1+\tau}{2}$ & $0,~\frac{1}{2},~\frac{\tau}{2}$ \\ \hline
      (0,1/2,0) & $\frac{1}{2},~\frac{1+\tau}{2}$ & $\frac{1}{2}$ \\ \hline
      (1,1/2,0) & $0,~\frac{\tau}{2}$ & $0,~\frac{\tau}{2},~\frac{1+\tau}{2}$ \\ \hline
      (0,0,1/2) & $\frac{\tau}{2},~\frac{1+\tau}{2}$ & $\frac{\tau}{2}$ \\ \hline
      (1,0,1/2) & $0,~\frac{1}{2}$ & $0,~\frac{1}{2},~\frac{1+\tau}{2}$ \\ \hline
      (0,1/2,1/2) & $\frac{1}{2},~\frac{\tau}{2}$ & $\frac{1}{2},~\frac{\tau}{2},~\frac{1+\tau}{2}$ \\ \hline
      (1,1/2,1/2) & $0,~\frac{1+\tau}{2}$ & $0$ \\ \hline
    \end{tabular}
  \end{center}
  \caption{Zero points of each zero-mode at the fixed points.
  $\mathbb{Z}_2$ parities 0 and 1 denote even and odd modes, respectively.}
  \label{tab:ZeroPoints}
\end{table}

Also we find the zero points of the derivative of zero-modes wave functions.
The boundary condition in Eq.~(\ref{eq:BCxxx}) leads to the following relation:
\begin{align}
  &\frac{\partial}{\partial z} \psi^{(j+\alpha_1,\alpha_2),M}_{T^2/\mathbb{Z}_2^m}(z+\frac{n_1+n_2\tau}{2}) \notag \\
  &= 
    e^{\pi i(j+\alpha_1)n_1}e^{\frac{\pi iM}{4}n_1n_2}e^{\frac{\pi iM}{2}\frac{{\rm Im}(n_1+n_2\bar{\tau})z}{{\rm Im}\tau}}
  \left(\frac{\partial}{\partial z}+\frac{\pi M}{4{\rm Im}\tau}(n_1+n_2\bar{\tau})\right)\psi^{(j+\alpha_1+\frac{M}{2}n_2,\alpha_2+\frac{M}{2}n_1),M}_{T^2/\mathbb{Z}_2^{m'}}(z). \label{eq:B.C.Derivative}
\end{align}
At $z=0$, we obtain
\begin{align}
  &\frac{\partial}{\partial z} \psi^{(j+\alpha_1,\alpha_2),M}_{T^2/\mathbb{Z}_2^m}(\frac{n_1+n_2\tau}{2}) \notag \\
  &= 
    e^{\pi i(j+\alpha_1)n_1}e^{\frac{\pi iM}{4}n_1n_2}\left(\frac{\partial}{\partial z}+\frac{\pi M}{4{\rm Im}\tau}(n_1+n_2\bar{\tau})\right)\psi^{(j+\alpha_1+\frac{M}{2}n_2,\alpha_2+\frac{M}{2}n_1),M}_{T^2/\mathbb{Z}_2^{m'}}(0). 
  \label{eq:BCderivative}
\end{align}
Since the derivatives of $\mathbb{Z}_2$ even and odd modes are $\mathbb{Z}_2$ odd and even modes respectively, they satisfy
\begin{align}
  &\frac{\partial}{\partial z} \psi^{(j+\alpha_1,\alpha_2),M}_{T^2/\mathbb{Z}_2^0}(0) = 0, \quad \frac{\partial}{\partial z} \psi^{(j+\alpha_1,\alpha_2),M}_{T^2/\mathbb{Z}_2^1}(0) \neq 0,
\end{align}
at $z=0$ for all $j,\alpha_1,\alpha_2$ and $M$.
This and Eq.~(\ref{eq:T2/Z2at0}) mean that either the first or second terms on the right-hand-side in Eq.~(\ref{eq:BCderivative}) vanishes but remaining term does not vanish.
Thus only the derivatives of $\mathbb{Z}_2$ even modes vanish at $z=0$ and do not vanish at $z=\frac{1}{2},\frac{\tau}{2},\frac{1+\tau}{2}$, and other derivatives do not vanish at all fixed points.
In Table \ref{tab:ZeroPointsD} we have summarized the results.
\begin{table}[h]
  \begin{center}
    \begin{tabular}{|c||c|c|} \hline
      ($\mathbb{Z}_2$ parity;SS phases) & $M={\rm even}$ & $M={\rm odd}$ \\ \hline
      (0,0,0) & $0$ & $0$ \\ \hline
      (1,0,0) & None & None \\ \hline
      (0,1/2,0) & $0$ & $0$ \\ \hline
      (1,1/2,0) & None & None \\ \hline
      (0,0,1/2) & $0$ & $0$ \\ \hline
      (1,0,1/2) & None & None \\ \hline
      (0,1/2,1/2) & $0$ & $0$ \\ \hline
      (1,1/2,1/2) & None & None \\ \hline
    \end{tabular}
  \end{center}
  \caption{Zero points of derivatives of each zero-mode at the fixed points.}
  \label{tab:ZeroPointsD}
\end{table}

%-----------------------------------------------------
%-----------------------------------------------------

\subsection{Yukawa couplings}

Here we study Yukawa couplings which are obtained by the overlap integrals of the wave functions of left-handed fermion, right-handed fermion and Higgs fields.
First we study Yukawa couplings on $T^2$ instead of ones on $T^2/\mathbb{Z}_2$.
Yukawa couplings on $T^2$ are given by the overlap integral of zero-modes on $T^2$:
\begin{align}
  Y^{ijk}_{T^2} = g(2{\rm Im}\tau)^{1/2} \int_{T^2} d^2z \psi^{(i+\alpha_{1L},\alpha_{2L}),M_L}_{T^2}(z) \cdot \psi^{(j+\alpha_{1R},\alpha_{2R}),M_R}_{T^2}(z) \cdot \left(\psi^{(k+\alpha_{1H},\alpha_{2H}),M_H}_{T^2}(z)\right)^*,
\end{align}
where $(M_f,\alpha_{1f},\alpha_{2f})$ with $f\in\{L,R,H\}$ are (flux, SS phases) of left-handed fermion $(L)$, right-handed fermion $(R)$ and Higgs fields $(H)$, and $g$ denotes the 3-point 
coupling in higher dimensional theory.
Using the normalization in Eq.~(\ref{eq:NormalizationPSI}) and the product expansion in Eq.~(\ref{eq:ProductPSI}), we find
\begin{align}
  &Y^{ijk}_{T^2} = g{\cal A}^{-1/2} \left|\frac{M_LM_R}{M_H}\right|^{1/4} e^{2\pi i\left(\frac{(i+\alpha_{1L})\alpha_{2L}}{M_L}+\frac{(j+\alpha_{1R})\alpha_{2R}}{M_R}-\frac{(k+\alpha_{1H})\alpha_{2H}}{M_H}\right)} \notag \\
  &\times \sum^{M_H-1}_{m=0} \vartheta
  \begin{bmatrix}
    \frac{M_R(i+\alpha_{1L})-M_L(j+\alpha_{1R})+M_LM_Rm}{M_LM_RM_H} \\ 0 \\
  \end{bmatrix}
  (M_L\alpha_{2R}-M_R\alpha_{2L},M_LM_RM_H\tau) \cdot \delta_{i+j-k,M_H\ell-M_Lm},
\end{align}
where $M_L+M_R=M_H$, $(\alpha_{1L},\alpha_{2L})+(\alpha_{1R},\alpha_{2LR})=(\alpha_{1H},\alpha_{2H})$ and $\ell\in\mathbb{Z}$.

Similarly Yukawa couplings on $T^2/\mathbb{Z}_2$ are given by the overlap integral of zero-modes on $T^2/\mathbb{Z}_2$:
\begin{align}
  Y^{ijk}_{T^2/\mathbb{Z}_2} &= g(2{\rm Im}\tau)^{1/2}\int d^2z \psi_{T^2/\mathbb{Z}_2^{m_L}}^{(i+\alpha_{1L},\alpha_{2L}),M_L}(z) \cdot \psi_{T^2/\mathbb{Z}_2^{m_R}}^{(j+\alpha_{1R},\alpha_{2R}),M_R}(z) \cdot \left(\psi_{T^2/\mathbb{Z}_2^{m_H}}^{(k+\alpha_{1H},\alpha_{2H}),M_H}(z) \right)^*,
\end{align}
where $m_f$ with $f\in\{L,R,H\}$ is $\mathbb{Z}_2$ parity of left-handed fermion $(L)$, right-handed fermion $(R)$ and Higgs fields $(H)$ and we have $m_f=$0 for $\mathbb{Z}_2$ even and 
$m_f= 1$ for $\mathbb{Z}_2$ odd.
From Eq.~(\ref{eq:OforT2/Z2}), this can be rewritten by Yukawa couplings on $T^2$ as
\begin{align}
  Y^{ijk}_{T^2/\mathbb{Z}_2} = O_{m_L}^{ii',\alpha_{1L},\alpha_{2L},M_L} O_{m_R}^{jj',\alpha_{1R},\alpha_{2R},M_R} \left(O_{m_H}^{kk',\alpha_{1H},\alpha_{2H},M_H}\right)^* Y^{i'j'k'}_{T^2}.
\end{align}
Hereafter we denote Yukawa couplings on $T^2/\mathbb{Z}_2$ as $Y^{ijk}$ instead of $Y^{ijk}_{T^2/\mathbb{Z}_2}$.

%-----------------------------------------------------
%-----------------------------------------------------

\subsection{Quark and lepton flavor models}

Here we study quark and lepton flavor models on the magnetized orbifold model.
We start with higher dimensional theory with larger gauge group, e.g. 
$SU(3) \times SU(2) \times U(1)_Y \times U(1)^n$.
We introduce magnetic flux background along $U(1)$ directions so as to 
obtain three generations of quarks and leptons.
$U(1)$ gauge bosons may become massive except $U(1)_Y$.
See for details of model building Refs.~ \cite{Abe:2014vza,Abe:2017gye}.
We consider all possible zero-mode assignments into left-handed quark doublets $Q=(u_L,d_L)^T$, right-handed up-sector (down-sector) quark singlets $u_R$ $(d_R)$, left-handed lepton doublets $L=(\nu_L,e_L)^T$, right-handed neutrino (charged lepton) singlets $\nu_R$ $(e_R)$ and up and down type Higgs fields $H_{u,d}$.
Here and in what follows we denote (Flux, $\mathbb{Z}_2$ parity, SS phases) of zero-modes assigned into each field of $f \in \{Q=(u_L,d_L)^T,u_R,d_R|L=(\nu_L,e_L)^T,\nu_R,e_R|H_u,H_d\}$ 
by $B_f$.
In addition, we denote the $j$-th zero-mode wave function of each field as $\psi_{f^j}$.
Then mass matrices for up-sector quarks, down-sector quarks and charged leptons, $M_{u}$, $M_{d}$ and $M_e$, are given by
\begin{align}
  M_u = Y^{ijk}_u\langle H_u^k\rangle, \quad M_{d} = Y^{ijk}_{d}\langle H_d^k\rangle, \quad
  M_{e} = Y^{ijk}_{e}\langle H_d^k\rangle,
\end{align}
where
\begin{align}
  &Y^{ijk}_u = g(2{\rm Im}\tau)^{1/2}\int d^2z \psi_{u_L^i} \cdot \psi_{u_R^j} \cdot \left(\psi_{H_u^k}\right)^* =g (2{\rm Im}\tau)^{1/2}\int d^2z \psi_{Q^i} \cdot \psi_{u_R^j} \cdot \left(\psi_{H_u^k}\right)^*, \\
  &Y^{ijk}_{d} =g (2{\rm Im}\tau)^{1/2}\int d^2z \psi_{d_L^i} \cdot \psi_{d_R^j} \cdot \left(\psi_{H_d^k}\right)^* = g(2{\rm Im}\tau)^{1/2}\int d^2z \psi_{Q^i} \cdot \psi_{d_R^j} \cdot \left(\psi_{H_d^k}\right)^*, \\
  &Y^{ijk}_{e} = g(2{\rm Im}\tau)^{1/2}\int d^2z \psi_{e_L^i} \cdot \psi_{e_R^j} \cdot \left(\psi_{H_d^k}\right)^* =g (2{\rm Im}\tau)^{1/2}\int d^2z \psi_{L^i} \cdot \psi_{e_R^j} \cdot \left( \psi_{H_d^k}\right)^*,
\end{align}
and $\langle H_{u,d}^k\rangle$ denote Higgs VEVs.
On the other hand the light neutrino mass, $M_\nu$, can be induced through the seesaw mechanism as follows,
\begin{align}
  M_\nu = M_D M_{RR}^{-1} M_D^T, \label{eq;MDNMDT}
\end{align}
where $M_{RR}$ is Majorana mass matrix of right-handed neutrinos, $M_D^{ij}=Y^{ijk}_\nu\langle H^k_u\rangle$ is Dirac mass and
\begin{align}
  Y^{ijk}_\nu = g(2{\rm Im}\tau)^{1/2}\int d^2z \psi_{\nu_L^i} \cdot \psi_{\nu_R^j} \cdot \left(\psi_{H_u^k}\right)^*.
\end{align}
In Appendix \ref{App:Neutrino_Majorana_mass_term}, we briefly review Majorana mass terms of right-handed neutrinos induced by the D-brane instanton effects on the magnetized $T^2/\mathbb{Z}_2$ model.
To obtain non-vanishing Yukawa couplings for quarks and leptons, (Flux, $\mathbb{Z}_2$ parity, SS phases) of  each field must satisfy
\begin{align}
  &B_Q + B_{u_R} = B_L + B_{\nu_R} = B_{H_u}, \label{eq:FluxParitySS1} \\
  &B_Q + B_{d_R} = B_L + B_{e_R} = B_{H_d}. \label{eq:FluxParitySS2}
\end{align}
Because of these conditions, the number of Higgs modes is larger than one, in general.
Furthermore, to cancel the chiral anomaly the number of up and down type Higgs fields must be same in four dimensional supersymmetric models.
Under these conditions, we have obtained 6,460 flavor models in total on the magnetized orbifold.
However realistic quark and lepton flavor structure cannot be realized in most of models.
In the following section, we will see the difficulties realizing realistic quark and lepton flavor structure, and find the conditions to avoid them.
We will also classify the flavor models satisfying such conditions.

%-----------------------------------------------------
%-----------------------------------------------------
%-----------------------------------------------------

\section{Conditions to realize quark and lepton flavors}
\label{sec:conditions}

As we have seen in the previous section, we have obtained 6,460 candidate models for quark and lepton flavors on the magnetized orbifold models.
However it is not easy to realize realistic flavors due to mass hierarchies and the differences of mixing angles between quarks and leptons.
Here we study the conditions to realize realistic quark and lepton flavor observables.

%-----------------------------------------------------
%-----------------------------------------------------

\subsection{Conditions for flavors}

First we show four conditions to realize realistic quark and lepton flavor structure.
Here, we assume a linear combination $H^k_{u,d}$ corresponds to the lightest mode, 
and it develops its VEV.
\begin{enumerate}
  % Condition I
  {\bf \item{Condition for up quark masses:}} \\
  Since the up-sector quarks have large mass hierarchy, its mass matrix can be regarded 
approximately as a rank one matrix.
  \begin{align}
    M_u &= Y^{ijk}_u \langle H^k_u\rangle =
    (U_L^u)^\dagger
    \begin{pmatrix}
      m_u & & \\
      & m_c & \\
      & & m_t \\
    \end{pmatrix}
    U_R^u \propto
    (U_L^u)^\dagger
    \begin{pmatrix}
      {\cal O}(10^{-6}) & & \\
      & {\cal O}(10^{-3}) & \\
      & & 1 \notag \\
    \end{pmatrix}
    U_R^u \\
    &\sim
    (U_L^u)^\dagger
    \begin{pmatrix}
      0 & & \\
      & 0 & \\
      & & 1 \\
    \end{pmatrix}
    U_R^u,
  \end{align}
  where $U_L^u$ and $U_R^u$ are unitary matrices to diagonalize $M_u$.
  That is,
  \begin{align}
    M_u^{ij} = Y^{ijk}_u\langle H_u^k\rangle \sim M_{\rm rank-1}, \label{eq:rank_one_sim_u}
  \end{align}
  are required, 
where $M_{\rm rank-1}$ denotes a rank one matrix.
  To realize such a mass matrix , the following direction $h^k_{u}$ must exist 
  \begin{align}
    ^\exists h^k_u~{\rm s.t.}~Y^{ijk}_{u}h^k_u = M_{\rm rank-1} \quad  ({\bf Condition~I}). \label{eq:rank_one_conditionUP2}
  \end{align}
  Then it is possible to realize the mass hierarchy in the up-sector by taking $\langle H^k_{u}\rangle = h^k_{u}+\varepsilon^k_{u}$ such that $\varepsilon_u/h_u\sim {\cal O}(\frac{m_c}{m_t}) \sim {\cal O}(10^{-3})$.
  % Condition II
  {\bf \item{Condition for down quark and charged lepton masses:}} \\
  The down-sector quarks and charged leptons also have large hierarchies, and their mass matrices can be regarded approximately as rank one matrices.
  \begin{align}
    M_{d} &= Y^{ijk}_{d} \langle H^k_d\rangle =
    (U_L^{d})^\dagger
    \begin{pmatrix}
      m_d & & \\
      & m_s & \\
      & & m_b \\
    \end{pmatrix}
    U_R^{d} \propto
    (U_L^{d})^\dagger
    \begin{pmatrix}
      {\cal O}(10^{-4}) & & \\
      & {\cal O}(10^{-2}) & \\
      & & 1 \\
    \end{pmatrix}
    U_R^{d} \notag \\
    &\sim
    (U_L^{d})^\dagger
    \begin{pmatrix}
      0 & & \\
      & 0 & \\
      & & 1 \\
    \end{pmatrix}
    U_R^{d}, \\
    M_{e} &= Y^{ijk}_{e} \langle H^k_d\rangle =
    (U_L^{e})^\dagger
    \begin{pmatrix}
      m_e & & \\
      & m_\mu & \\
      & & m_\tau \\
    \end{pmatrix}
    U_R^e \propto
    (U_L^{e})^\dagger
    \begin{pmatrix}
      {\cal O}(10^{-4}) & & \\
      & {\cal O}(10^{-2}) & \\
      & & 1 \\
    \end{pmatrix}
    U_R^e \notag \\
    &\sim
    (U_L^{e})^\dagger
    \begin{pmatrix}
      0 & & \\
      & 0 & \\
      & & 1 \\
    \end{pmatrix}
    U_R^e,
  \end{align}
  where $U_L^{d}$ and $U_R^{d}$ are unitary matrices to diagonalize $M_{d}$, and $U_L^e$ and $U_R^e$ are ones for $M_e$.
  That is,
  \begin{align}
    M_{d}^{ij} = Y^{ijk}_{d} \langle H_d^k\rangle \sim M_{\rm Rank-1}, \quad
    M_e^{ij} = Y^{ijk}_e \langle H_d^k \rangle \sim M_{\rm rank-1}, \label{eq:rank_one_sim_d}
  \end{align}
  are required.
  To realize such mass matrices, the following direction $h_d^k$ must exist 
  \begin{align}
    ^\exists h_d^k~{\rm s.t.}~Y_{d}^{ijk} h_d^k =M_ {\rm rank-1}, \quad Y_{e}^{ijk} h_d^k =M_ {\rm rank-1} \quad
    ({\bf Condition~II}). \label{eq:Condition_II}
  \end{align}
  Then it is possible to realize the down-sector quark and charged lepton mass hierarchies by taking $\langle H_d^k\rangle = h_d^k + \varepsilon_d^k$ such that $\varepsilon_d/h_d\sim{\cal O}(\frac{m_s}{m_b})\sim{\cal O}(\frac{m_\mu}{m_\tau})\sim{\cal O}(10^{-2})$.
  % Condition III
  {\bf \item{Condition for quark mixing:}} \\
  The absolute values of the Cabibbo-Kobayashi-Maskawa (CKM) matrix elements are observed as
  \begin{align}
    |V_{\rm CKM}| \equiv |(U_L^u)^\dagger U_L^{d}| =
    \begin{pmatrix}
      0.974 & 0.227 & 0.00361 \\
      0.226 & 0.973 & 0.0405 \\
      0.00854 & 0.0398 & 0.999 \\
    \end{pmatrix}, \label{eq:CKMob}
  \end{align}
  which is approximately a unit matrix.
  This implies that the following relation,
  \begin{align}
    U_L^u \sim U_L^{d},
  \end{align}
  are required.
  To realize this relation, here we introduce the unitary matrices $u_{L,R}^{{u,d}}$ which diagonalize rank one matrices $Y^{ijk}_{{u,d}}h_{u,d}^k$:
  \begin{align}
    [(u_L^{u,d})^\dagger]^{ii'}Y^{i'j'k}_{{u,d}}h_{u,d}^k [u_R^{u,d}]^{j'j} \propto
    \begin{pmatrix}
      0&&\\
      &0&\\
      &&1\\
    \end{pmatrix}^{ij},
  \end{align}
  and impose the following condition,
  \begin{align}
    u_L^u = u_L^{d} \quad ({\bf Condition~III}). \label{eq:Condition_III}
  \end{align}
  Under this condition, in a basis where $u_L^u$, $u_L^{d}$, $u_R^u$ and $u_R^{d}$ are unit matrices, the quark mass matrices should have the following forms,
  \begin{align}
    M_u &= Y^{ijk}_u\langle H_u^k\rangle = Y^{ijk}_u (h_u^k+\varepsilon_u^k) \propto 
    \begin{pmatrix}
      0&&\\&0&\\&&1\\
    \end{pmatrix}
    +
    {\cal O}(\tfrac{m_c}{m_t}),\\
    M_{d} &= Y^{ijk}_{d}\langle H_d^k\rangle = Y^{ijk}_{d} (h_d^k+\varepsilon_d^k) \propto 
    \begin{pmatrix}
      0&&\\&0&\\&&1\\
    \end{pmatrix}
    +
    {\cal O}(\tfrac{m_s}{m_b}),
  \end{align}
  because of the mass hierarchies.
  From the above, the unitary matrices $U_{L,R}^{{u,d}}$ which diagonalize $M_{{u,d}}$ can be estimated to be
  \begin{align}
    &U_{L,R}^u \sim
    \begin{pmatrix}
      1&0&0\\
      0&1&{\cal O}(\tfrac{m_c}{m_t})\\
      0&{\cal O}(\tfrac{m_c}{m_t})&1\\
    \end{pmatrix}
    \begin{pmatrix}
      1&0&{\cal O}(\tfrac{m_c}{m_t})\\
      0&1&0\\
      {\cal O}(\tfrac{m_c}{m_t})&0&1\\
    \end{pmatrix}
    \begin{pmatrix}
      \ast&\ast&0\\
      \ast&\ast&0\\
      0&0&1\\
    \end{pmatrix}
    \sim
    \begin{pmatrix}
      \ast&\ast&{\cal O}(\tfrac{m_c}{m_t})\\
      \ast&\ast&{\cal O}(\tfrac{m_c}{m_t})\\
      {\cal O}(\tfrac{m_c}{m_t})&{\cal O}(\tfrac{m_c}{m_t})&1\\
    \end{pmatrix}, \\
    &U_{L,R}^{d} \sim
    \begin{pmatrix}
      1&0&0\\
      0&1&{\cal O}(\tfrac{m_s}{m_b})\\
      0&{\cal O}(\tfrac{m_s}{m_b})&1\\
    \end{pmatrix}
    \begin{pmatrix}
      1&0&{\cal O}(\tfrac{m_s}{m_b})\\
      0&1&0\\
      {\cal O}(\tfrac{m_s}{m_b})&0&1\\
    \end{pmatrix}
    \begin{pmatrix}
      \ast&\ast&0\\
      \ast&\ast&0\\
      0&0&1\\
    \end{pmatrix}
    \sim
    \begin{pmatrix}
      \ast&\ast&{\cal O}(\tfrac{m_s}{m_b})\\
      \ast&\ast&{\cal O}(\tfrac{m_s}{m_b})\\
      {\cal O}(\tfrac{m_s}{m_b})&{\cal O}(\tfrac{m_s}{m_b})&1\\
    \end{pmatrix},
  \end{align}
  where $\ast$ denotes unestimated values.
  Then the CKM matrix can be estimated to be
  \begin{align}
    V_{\rm CKM} \sim &
    \begin{pmatrix}
      \ast&\ast&{\cal O}(\tfrac{m_s}{m_b}) \\
      \ast&\ast&{\cal O}(\tfrac{m_s}{m_b}) \\
      {\cal O}(\tfrac{m_s}{m_b})&{\cal O}(\tfrac{m_s}{m_b})&1 \\
    \end{pmatrix}
    \sim
    \begin{pmatrix}
      \ast&\ast&{\cal O}(10^{-2}) \\
      \ast&\ast&{\cal O}(10^{-2}) \\
      {\cal O}(10^{-2})&{\cal O}(10^{-2})&1 \\
    \end{pmatrix},
  \end{align}
  and this is consistent with the observations in Eq.~(\ref{eq:CKMob}).
  Thus it is possible to realize realistic quark mixing under the condition III in Eq.~(\ref{eq:Condition_III}).
  % Condition IV
  {\bf \item{Condition for lepton mixing:}} \\
  If we take the direction of up type Higgs VEVs, $\langle H^k_{u}\rangle = h^k_{u}+\varepsilon_{u}^k$, satisfying Eq.~(\ref{eq:rank_one_sim_d}), the light neutrino mass matrix becomes
  \begin{align}
    M_\nu^{ij} &= Y^{ii'm}_\nu \langle H^m_u\rangle (M_{RR}^{i'j'})^{-1} (Y^{j'jn}_\nu)^T\langle H^n_u \rangle \notag \\
    &= Y^{ii'm}_\nu h^m_u (M_{RR}^{i'j'})^{-1} (Y^{j'jn}_\nu)^T h^n_u + {\cal O}(\varepsilon_u) + {\cal O}(\varepsilon_u^2) .
%\\
 %   &\sim  Y^{ii'm}_\nu h^m_u (M_{RR}^{i'j'})^{-1} (Y^{j'jn}_\nu)^T h^n_u. \notag
  \end{align}
If the first term is non-vanishing and $\varepsilon_u$ is small enough, 
the first term is dominant.
%  This means that the structure of light neutrino mass matrix is almost independent on $\varepsilon_u$.
 In this case, the direction of $h_u$ is determined to satisfy Eq.~(\ref{eq:rank_one_sim_u}) and there are no parameters to be used for realizing lepton mixing.
  Thus it is difficult to realize realistic lepton mixing unless $M_{RR}$ and $Y^{ijk}_\nu h^k_u$ have ideal structures.
  To avoid this difficulty, here we impose the following condition,
  \begin{align}
    Y^{ijk}_u h^k_u = M_{\rm rank-1} \Rightarrow Y^{ijk}_\nu h^k_u = 0 \quad ({\bf Condition~IV}). \label{eq:Condition_IV}
  \end{align}
  In such case the mass matrix of light neutrino is given by
  \begin{align}
    M_\nu^{ij} = Y^{ii'm}_\nu \varepsilon_{u}^m (M_{RR}^{i'j'})^{-1} (Y^{j'jn}_\nu)^T \varepsilon_u^n,
  \end{align}
  and we have the possibility to realize realistic lepton mixing by taking appropriate directions of $\varepsilon_u$.
\end{enumerate}
In Table \ref{tab:Conditions}, we summarize all conditions.
\begin{table}[h]
  \begin{center}
  \renewcommand{\arraystretch}{1.2}
    \begin{tabular}{|c|c|} \hline
      & Conditions \\ \hline
      I & $^\exists h^k_{u}$ s.t. $Y^{ijk}_{u}h^k_u = {\rm rank~one~matrix}$ \\
      II & $^\exists h_d^k~{\rm s.t.}~Y_{d}^{ijk} h_d^k = {\rm rank~one~matrix}, \quad Y_{e}^{ijk} h_d^k = {\rm rank~one~matirx}$ \\
      III & $u_L^u = u_L^{d}$ \\
      IV & $Y^{ijk}_u h^k_u = {\rm rank~one~matrix }$ $\Rightarrow$ $Y^{ijk}_\nu h^k_u = 0$ \\ \hline
    \end{tabular}
  \end{center}
  \caption{The conditions I, II, III and IV.}
  \label{tab:Conditions}
\end{table}

%-----------------------------------------------------
%-----------------------------------------------------

\subsection{Zero point analysis}
\label{subsec:3.2}

In the previous subsection, we saw four conditions, I, II, III and IV, to realize  quark and lepton  masses and their mixing angles.
Then Higgs VEV directions $h_{u,d}^k$ leading to rank one or vanishing mass matrices have been required.
In this subsection, we show such directions can be realized in several cases by checking the zero points of zero-modes.
The procedure is as follows.
First we start from Yukawa couplings between left-handed fermion zero-modes $\psi^i_L$, right-handed fermion zero-modes $\psi^j_R$ and Higgs field zero-modes $\psi^k_H$.
We consider zero points of zero-modes $\psi^i_L$, $\psi^j_R$ and $\psi^k_H$.
As we will see soon, zero points patterns on Yukawa couplings have the information what liner combinations of Yukawa matrices lead to rank one or vanishing mass matrix.
Second, we will construct unitary matrices for Higgs field zero-modes which correspond to this liner combination.
Finally, we classify the structure of mass matrices in each pattern of zero points.

Yukawa couplings between $\psi_L^i$, $\psi_R^j$ and $\psi_H$ are given by
\begin{align}
  Y^{ijk} = gy^{ijk}=g(2{\rm Im}\tau)^{1/2}\int d^2z \psi^{i}_L(z) \cdot \psi^j_R(z) \cdot \left(\psi^k_H(z)\right)^*.
\end{align}
This leads to the product expansion,
\begin{align}
  \psi^i_L(z) \cdot \psi^j_R(z)
  = y^{ijk} \psi^k_H(z). \label{eq:psiLpsiRpsiH}
\end{align}
Here we denote sets of the zero points at the fixed points of $\psi_L^i$, $\psi_R^j$ and $\psi_H^k$ as $P_{\psi_L}$, $P_{\psi_R}$ and $P_{\psi_H}$, and ones of the derivatives of $\psi_L^i$, $\psi_R^j$ and $\psi_H^k$ as $P'_{\psi_L}$, $P'_{\psi_R}$ and $P'_{\psi_H}$.

Next, we choose one point $p$ on $T^2/\mathbb{Z}_2$ (not necessary to be fixed points) and consider a unitary transformation for $\psi_L^i$ such as
\begin{align}
  &\psi_L^i \rightarrow \hat{\psi}_L^i = U_{\psi_L}^{ij}(p) \psi_L^j, \\
  &U_{\psi_L}^{ij}(p) =
  \begin{pmatrix}
    \cos\theta_2 & 0 & -\sin\theta_2 \\
    0 & 1 & 0 \\
    \sin\theta_2 & 0 & \cos\theta_2 \\
  \end{pmatrix}
  \begin{pmatrix}
    1 & 0 & 0 \\
    0 & \cos \theta_1 & -\sin\theta_1 \\
    0 & \sin\theta_1 & \cos\theta_1 \\
  \end{pmatrix}
  \begin{pmatrix}
    e^{-i\alpha_0} & 0 & 0 \\
    0 & e^{-i\alpha_1} & 0 \\
    0 & 0 & e^{-i\alpha_2} \\
  \end{pmatrix}, \label{eq:U_psi_1(p)}
\end{align}
where
\begin{align}
  &\left\{
  \begin{array}{l}
    \alpha^i = {\rm Ang}(\psi_L^i(p)) \quad {\rm for}~p \notin P_{\psi_L}, \\
    \alpha^i = {\rm Ang}(\frac{\partial}{\partial z}\psi_L^i(p)) \quad {\rm for}~p \in P_{\psi_L}, \\
  \end{array}
  \right. \\
  &\left\{
  \begin{array}{l}
    \theta_1 = \tan^{-1}\dfrac{|\psi_L^1(p)|}{|\psi_L^2(p)|}, \quad \theta_2 = \tan^{-1}\dfrac{|\psi_L^0(p)|}{\sin\theta_1 |\psi_L^1(p)|+ \cos\theta_1 |\psi_L^2(p)|} \quad {\rm for}~p \notin P_{\psi_L}, \\
    \theta_1 = \tan^{-1}\dfrac{|\frac{\partial}{\partial z}\psi_L^1(p)|}{|\frac{\partial}{\partial z}\psi_L^2(p)|}, \quad \theta_2 = \tan^{-1}\dfrac{|\frac{\partial}{\partial z}\psi_L^0(p)|}{\sin\theta_1 |\frac{\partial}{\partial z}\psi_L^1(p)|+ \cos\theta_1 |\frac{\partial}{\partial z}\psi_L^2(p)|} \quad {\rm for}~p \in P_{\psi_L}. \\
  \end{array}
  \right.
\end{align}
For $p\notin P_{\psi_L}$ redefined zero-modes $\hat{\psi}_L^i(z)$ $(i\neq 2)$ become zero at $z=p$ while for $p\in P_{\psi_L}$ the derivative of redefined zero-modes $\frac{\partial}{\partial z}\hat{\psi}_L^i(z)$ $(i\neq 2)$ become zero at $z=p$.
In a similar way we can obtain redefined zero-modes, $\hat{\psi}_R$ and $\hat{\psi}_H$, for $\psi_R$ and $\psi_H$ by unitary transformations $U_{\psi_R}(p)$ and $U_{\psi_H}(p)$ such that only $\hat{\psi}_R^2$ and $\hat{\psi}_H^{(g_H-1)}$ are non-vanishing.
Then we consider the structures of redefined Yukawa couplings,
\begin{align}
  \hat{Y}^{ijk} = g(2{\rm Im}\tau)^{1/2}\int d^2z \hat{\psi}_L^i(z) \cdot \hat{\psi}_R^j(z) \cdot \left( \hat{\psi}_H^k(z)\right)^*.
\end{align}
From Table \ref{tab:ZeroPoints}, we can find that non-vanishing Yukawa couplings conditions in Eqs.~(\ref{eq:FluxParitySS1}) and (\ref{eq:FluxParitySS2}) mean that when one point $p$ is in $P_L$, it is also in either $P_R$ or $P_H$; when $p$ is in $P_R$, it is also in either $P_L$ or $P_H$; when $p$ is in $P_H$, it is also in either $P_L$ or $P_R$.
That is, there are four possible patterns of $p$:
\begin{align}
  &(1)~p\notin P_{\psi_L},~ p\notin P_{\psi_R},~ p\notin P_{\psi_H}, \notag\\
  &(2)~p\in P_{\psi_L},~ p\in P_{\psi_R},~ p\notin P_{\psi_H}, \notag\\
  &(3)~p\in P_{\psi_L},~ p\notin P_{\psi_R},~ p\in P_{\psi_H}~(p\notin P_{\psi_L}',~p\notin P_{\psi_H}'), \notag\\
  &(4)~p\notin P_{\psi_L},~ p\in P_{\psi_R},~ p\in P_{\psi_H}~(p\notin P_{\psi_R}',~p\notin P_{\psi_H}'). \notag
\end{align}
Note that the derivatives of zero-modes which vanish at $z=p\in P_F$ do not vanish at $z=p$ as can be read from Tables \ref{tab:ZeroPoints} and \ref{tab:ZeroPointsD}.
In each pattern, we focus on the structures of $\hat{Y}^{ij(g_H-1)}=U_{\psi_L}^{ii'}(p)U_{\psi_R}^{jj'}(p)Y^{i'j'k}(U_{\psi_H}^{(g_H-1)k}(p))^*$, where $g_H$ denotes the number of Higgs fields, 
because we use the Higgs mode basis such that the wave function for the $(g_H-1)$ th Higgs mode is non-vanishing at $p$ 
 and the others vanish.
\begin{enumerate}
  \item[(1)] $p\notin P_{\psi_L},~ p\notin P_{\psi_R},~ p\notin P_{\psi_H}$ \\
  Table \ref{tab:pattern1} shows the zero points of redefined zero-modes.
  \begin{table}[h]
    \begin{center}
    \renewcommand{\arraystretch}{1.2}
      \begin{tabular}{|c|c|c|c|c|c|c|c|} \hline
        & $j=0$ & 1 & 2 & 3 & $\cdots$ & $g_H-2$ & $g_H-1$ \\ \hline \hline
        $\hat{\psi}_L^j=U_{\psi_1}^{jk}\psi_L^k$ & $P_{\psi_1},~p$ & $P_{\psi_L},~p$ & $P_{\psi_L}$ & \diagbox[width=1.55 cm, height=1.2\line, dir=NE]{}{} & $\cdots$ & \diagbox[width=1.55cm, height=1.2\line, dir=NE]{}{} & \diagbox[width=1.55cm, height=1.2\line, dir=NE]{}{} \\ \hline
        $\hat{\psi}_R^j=U_{\psi_2}^{jk}\psi_R^k$ & $P_{\psi_R},~p$ & $P_{\psi_R},~p$ & $P_{\psi_R}$ & \diagbox[width=1.55cm, height=1.2\line, dir=NE]{}{} & $\cdots$ & \diagbox[width=1.55cm, height=1.2\line, dir=NE]{}{} & \diagbox[width=1.55cm, height=1.2\line, dir=NE]{}{} \\ \hline
        $\hat{\psi}_H^j=U_{\psi_H}^{jk}\psi_H^k$ & $P_{\psi_H},~p$ & $P_{\psi_H},~p$ & $P_{\psi_H},~p$ & $P_{\psi_H},~p$ & $\cdots$ & $P_{\psi_H},~p$ & $P_{\psi_H}$ \\ \hline
      \end{tabular}
    \end{center}
    \caption{Zero points of redefined zero-modes in pattern (1).}
    \label{tab:pattern1}
  \end{table}
  In this case, the product expansion in Eq.~(\ref{eq:psiLpsiRpsiH}) at $z=p$ leads to
  \begin{align}
    \underbrace{\hat{\psi}^i_L(p)}_{\propto \delta^{i,2}} \cdot \underbrace{\hat{\psi}^j_R(p)}_{\propto \delta^{j,2}}
    = \hat{y}^{ijk} \underbrace{\hat{\psi}^k_H(p)}_{\propto \delta^{k,(g_H-1)}} \quad
    &\Leftrightarrow \quad \hat{Y}^{ij(g_H-1)} \propto \delta^{i,2}\delta^{j,2} \quad ({\rm rank~one~matrix}) \\
    &\Leftrightarrow \quad Y^{ijk}(U_{\psi_H}^{(g_H-1)k}(p))^* = M_{\rm rank-1}. \label{eq:YU=RankOne1}
  \end{align}
  \item[(2)] $p\in P_{\psi_L},~ p\in P_{\psi_R},~ p\notin P_{\psi_H}$ \\
  Table \ref{tab:pattern2} shows the zero points of redefined zero-modes.
  \begin{table}[h]
    \begin{center}
    \renewcommand{\arraystretch}{1.2}
      \begin{tabular}{|c|c|c|c|c|c|c|c|} \hline
        & $j=0$ & 1 & 2 & 3 & $\cdots$ & $g_H-2$ & $g_H-1$ \\ \hline \hline
        $\hat{\psi}_L^j=U_{\psi_L}^{jk}\psi_L^k$ & $P_{\psi_L}$ & $P_{\psi_L}$ & $P_{\psi_L}$ & \diagbox[width=1.55 cm, height=1.2\line, dir=NE]{}{} & $\cdots$ & \diagbox[width=1.55 cm, height=1.2\line, dir=NE]{}{} & \diagbox[width=1.55cm, height=1.2\line, dir=NE]{}{} \\ \hline
        $\hat{\psi}_2^j=U_{\psi_R}^{jk}\psi_R^k$ & $P_{\psi_R}$ & $P_{\psi_R}$ & $P_{\psi_R}$ & \diagbox[width=1.55cm, height=1.2\line, dir=NE]{}{} & $\cdots$ & \diagbox[width=1.55cm, height=1.2\line, dir=NE]{}{} & \diagbox[width=1.55 cm, height=1.2\line, dir=NE]{}{} \\ \hline
        $\hat{\psi}_H^j=U_{\psi_H}^{jk}\psi_H^k$ & $P_{\psi_H},~p$ & $P_{\psi_H},~p$ & $P_{\psi_H},~p$ & $P_{\psi_H},~p$ & $\cdots$ & $P_{\psi_H},~p$ & $P_{\psi_H}$ \\ \hline
      \end{tabular}
    \end{center}
    \caption{Zero points of redefined zero-modes in pattern (2).}
    \label{tab:pattern2}
  \end{table}
  In this case, the product expansion in Eq.~(\ref{eq:psiLpsiRpsiH}) at $z=p$ leads to
  \begin{align}
    \underbrace{\hat{\psi}^i_L(p)}_{=0} \cdot \underbrace{\hat{\psi}^j_R(p)}_{=0}
    = \hat{y}^{ijk} \underbrace{\hat{\psi}^k_H(p)}_{\propto \delta^{k,(g_H-1)}} \quad
    &\Leftrightarrow \quad \hat{Y}^{ij(g_H-1)} =0 \\
    &\Leftrightarrow \quad Y^{ijk}(U_{\psi_H}^{(g_H-1)k}(p))^* = 0.
  \end{align}
  \item[(3)] $p\in P_{\psi_L},~ p\notin P_{\psi_R},~ p\in P_{\psi_H}$ $(p\notin P_{\psi_L}',~p\notin P_{\psi_H}')$ \\
  Table \ref{tab:pattern3} shows the zero points of redefined zero-mode wave functions and 
their derivatives.
  \begin{table}[H]
    \begin{center}
    \renewcommand{\arraystretch}{1.2}
      \begin{tabular}{|c|c|c|c|c|c|c|c|} \hline
        & $j=0$ & 1 & 2 & 3 & $\cdots$ & $g_H-2$ & $g_H-1$ \\ \hline \hline
        $\hat{\psi}_L^j=U_{\psi_L}^{jk}\psi_L^k$ & $P_{\psi_L}$ & $P_{\psi_L}$ & $P_{\psi_L}$ & \diagbox[width=1.55 cm, height=1.2\line, dir=NE]{}{} & $\cdots$ & \diagbox[width=1.55cm, height=1.2\line, dir=NE]{}{} & \diagbox[width=1.55cm, height=1.2\line, dir=NE]{}{} \\ \hline
        $\hat{\psi}_R^j=U_{\psi_R}^{jk}\psi_R^k$ & $P_{\psi_R},~p$ & $P_{\psi_R},~p$ & $P_{\psi_R}$ & \diagbox[width=1.55cm, height=1.2\line, dir=NE]{}{} & $\cdots$ & \diagbox[width=1.55cm, height=1.2\line, dir=NE]{}{} & \diagbox[width=1.55cm, height=1.2\line, dir=NE]{}{} \\ \hline
        $\hat{\psi}_H^j=U_{\psi_H}^{jk}\psi_H^k$ & $P_{\psi_H}$ & $P_{\psi_H}$ & $P_{\psi_H}$ & $P_{\psi_H}$ & $\cdots$ & $P_{\psi_H}$ & $P_{\psi_H}$ \\ \hline
        $\tfrac{\partial}{\partial z}\hat{\psi}_L^j=U_{\psi_L}^{jk}\tfrac{d}{dz}\psi_L^k$ & $P_{\psi_L}',~p$ & $P_{\psi_L}',~p$ & $P_{\psi_L}'$ & \diagbox[width=1.55 cm, height=1.2\line, dir=NE]{}{} & $\cdots$ & \diagbox[width=1.55cm, height=1.2\line, dir=NE]{}{} & \diagbox[width=1.55cm, height=1.2\line, dir=NE]{}{} \\ \hline
        $\tfrac{\partial}{\partial z}\hat{\psi}_H^j=U_{\psi_H}^{jk}\tfrac{d}{dz}\psi_H^k$ & $P_{\psi_H}',~p$ & $P_{\psi_H}'~p$ & $P_{\psi_H}',~p$ & $P_{\psi_H}',~p$ & $\cdots$ & $P_{\psi_H}',~p$ & $P_{\psi_H}'$ \\ \hline
      \end{tabular}
    \end{center}
    \caption{Zero points of redefined zero-modes and its derivatives in pattern (3).}
    \label{tab:pattern3}
  \end{table}
  In this case, the product expansion in Eq.~(\ref{eq:psiLpsiRpsiH}) at $z=p$ give no information for $\hat{Y}^{ij(g_H-1)}$.
  Instead of Eq.~(\ref{eq:psiLpsiRpsiH}) we consider the derivative of Eq.(\ref{eq:psiLpsiRpsiH}).
  At $z=p$, it leads to
  \begin{align}
    \underbrace{\hat{\psi}^i_L(p)}_{=0} \cdot \underbrace{\hat{\psi}^j_R(p)}_{\propto \delta^{j,2}}
    = \hat{y}^{ijk} \underbrace{\hat{\psi}^k_H(p)}_{=0} \quad
    &\Rightarrow \quad \underbrace{\textstyle\frac{\partial}{\partial z}\hat{\psi}^i_L(p)}_{\propto\delta^{i,2}} \cdot \underbrace{\hat{\psi}^j_R(p)}_{\propto\delta^{j,2}} + \underbrace{\hat{\psi}^i_L(p)}_{=0} \cdot \textstyle\frac{\partial}{\partial z}\hat{\psi}^j_R(p)
    = \hat{y}^{ijk} \underbrace{\textstyle\frac{\partial}{\partial z}\hat{\psi}^k_H(p)}_{\propto \delta^{k,(g_H-1)}} \\
    &\Leftrightarrow \quad \hat{Y}^{ij(g_H-1)} \propto \delta^{i,2}\delta^{j,2} \quad ({\rm rank~one~matrix}) \\
    &\Leftrightarrow \quad Y^{ijk}(U_{\psi_H}^{(g_H-1)k}(p))^* = M_{\rm rank-1}. \label{eq:YU=RankOne2}
  \end{align}
  \item[(4)] $p\notin P_{\psi_L},~ p\in P_{\psi_R},~ p\in P_{\psi_H}$ $(p\notin P_{\psi_R}',~p\notin P_{\psi_H}')$ \\
  This case is flipping between $\psi_L$ and $\psi_R$ in the pattern (3); therefore it gives same result as the pattern (3),
  \begin{align}
    Y^{ijk}(U_{\psi_H}^{(g_H-1)k}(p))^* = M_{\rm rank-1}.
  \end{align}
\end{enumerate}
In Eqs.~(\ref{eq:YU=RankOne1}) and (\ref{eq:YU=RankOne2}), note that unitary transformations $U_{\psi_L}$ and $U_{\psi_R}$ do not change the rank of the matrix.
Thus we can obtain Higgs VEV directions $h_{u,d}^k=v_{u,d}(U_{H_{u,d}}^{(g_H-1)k})^*$ leading to rank one fermion mass matrices in three patterns (1), (3) and (4).
On the other hand, the pattern (2) gives vanishing mass matrices.

%-----------------------------------------------------
%-----------------------------------------------------

\subsection{Classification of models}

In this subsection, we classify all of the quark and lepton flavor models on the magnetized orbifold model which satisfy the conditions I, II, III and IV.
In what follows we denote sets of the zero points at the fixed points of each field $f$ as $P_f$ for
\begin{align}
  f \in \{Q=(u_L,d_L),~u_R,~d_R|~L=(\nu_L,e_L),~\nu_R,~e_R|~H_u,~H_d\}.
\end{align}
First, we show the constraints of $P_f$ to satisfy each condition.
\begin{enumerate}
% Condition I
{\bf\item Condition I}
~\\
The condition I is that the up type Higgs VEV direction leading to the up-sector quark mass matrix with the rank one must exist.
Hence,
\begin{align}
  Y^{ijk}_{u}h_u^k = M_{\rm rank-1},
\end{align}
is required.
As shown in the previous subsection, this requirement means that the following point $p_u$ must exist:
\begin{align}
  ^\exists p_u=p ~{\rm s.t.~} \left\{
  \begin{array}{l}
    (1)~p\notin P_{\psi_L},~ p\notin P_{\psi_R},~ p\notin P_{\psi_H} \\
    (3)~p\in P_{\psi_L},~ p\notin P_{\psi_R},~ p\in P_{\psi_H}~(p\notin P_{\psi_L}',~p\notin P_{\psi_H}') \\
    (4)~p\notin P_{\psi_L},~ p\in P_{\psi_R},~ p\in P_{\psi_H}~(p\notin P_{\psi_R}',~p\notin P_{\psi_H}') \\
  \end{array}
  \right., \label{eq:Constraint_I}
\end{align}
for $(\psi_L,~\psi_R,~\psi_H) = (Q,~u_R,~H_u)$, i.e., ${\bf Constraint~I}$.
% Condition II
{\bf\item Condition II}
~\\
The condition II is that the down type Higgs VEV direction leading to the down-sector quark and charged lepton mass matrices with the rank one must exist.
Hence,
\begin{align}
  Y^{ijk}_{d}h_d^k =M_{\rm rank-1}, \quad Y^{ijk}_eh_d^k =M_{\rm rank-1},
\end{align}
are required.
As shown in the previous subsection, this requirement means that the following point $p_d$ must exist:
\begin{align}
  ^\exists p_d=p ~{\rm s.t.~} \left\{
  \begin{array}{l}
    (1)~p\notin P_{\psi_L},~ p\notin P_{\psi_R},~ p\notin P_{\psi_H} \\
    (3)~p\in P_{\psi_L},~ p\notin P_{\psi_R},~ p\in P_{\psi_H}~(p\notin P_{\psi_L}',~p\notin P_{\psi_H}') \\
    (4)~p\notin P_{\psi_L},~ p\in P_{\psi_R},~ p\in P_{\psi_H}~(p\notin P_{\psi_R}',~p\notin P_{\psi_H}') \\
  \end{array}
  \right., \label{eq:Constraint_II}
\end{align}
for $(\psi_L,~\psi_R,~\psi_H) = (Q,~d_R,~H_d)$, i.e., ${\bf Constraint~II_1}$ and for $(\psi_L,~\psi_R,~\psi_H) = (L,~e_R,~H_d)$, i.e., ${\bf Constraint~II_2}$.
% Condition III
{\bf\item Condition III}
~\\
The condition III is that unitary matrices $u_{L}^{u,d}$ which diagonalize rank one matrices $Y^{ijk}_{u,d}h_{u,d}^k$ must satisfy $u_L^u=u_L^{d}$.
When the condition I is satisfied, that is, the constraint I is satisfied, we can find the up-sector quark mass matrix with the rank one,
\begin{align}
  &Y^{ijk}_u h_u^k = v_{u}Y^{ijk}_u (U_{\psi_H}^{(g_H-1)k}(p_u))^*,
\end{align}
and they are diagonalized as
\begin{align}
  U_{\psi_L}^{ii'}(p_u)U_{\psi_R}^{jj'}(p_u)v_{u}Y^{i'j'k}_u (U_{\psi_H}^{(g_H-1)k}(p_u))^* \propto \delta^{i,2}\delta^{j,2},
\end{align}
for $(\psi_L,~\psi_R,~\psi_H)=(Q,~u_R,~H_u)$.
Note that $U_{\psi_L}$, $U_{\psi_R}$ and $U_{\psi_H}$ are defined by Eq.~(\ref{eq:U_psi_1(p)}) and the sentence below.
Similarly, when the condition II is satisfied, that is, the constraints II$_1$ and II$_2$ are satisfied, we can find the  down-sector quark mass matrix with the rank one,
\begin{align}
  &Y^{ijk}_{d} h_d^k = v_{d}Y^{ijk}_{d} (U_{\psi_H}^{(g_H-1)k}(p_d))^*,
\end{align}
and they are diagonalized as
\begin{align}
  U_{\psi_L}^{ii'}(p_d)U_{\psi_R}^{jj'}(p_d)v_dY^{i'j'k}_{d} (U_{\psi_H}^{(g_H-1)k}(p_d))^* \propto \delta^{i,2}\delta^{j,2},
\end{align}
for $(\psi_L,~\psi_R,~\psi_H)=(Q,~d_R,~H_d)$.
Then $u_L^u=u_L^{d}$ is equivalent to the equation,
\begin{align}
  U_{\psi_L}^{ij}(p_u) = U_{\psi_L}^{ij}(p_d) \quad {\rm for~} \psi_L = Q.
\end{align}
Obviously this can be satisfied by
\begin{align}
  p_u = p_d \quad ({\bf Constraint~III}). \label{eq:Constraint_III}
\end{align}
%i.e., ${\bf Constraint~III}$.
% Condition IV
{\bf\item Condition IV}
~\\
The condition IV is that when the up-sector quark mass matrix is a rank one matrix, the neutrino Dirac mass matrix must vanish.
Hence,
\begin{align}
  Y^{ijk}_{u}h_u^k = M_{\rm rank-1}~\Rightarrow ~Y^{ijk}_\nu h_u^k = 0,
\end{align}
is required.
As shown in the previous subsection, this requirement means that the following point $p_u$ must exist:
\begin{align}
  ^\exists p_u = p ~{\rm s.t.}~(2)~p\in P_{\psi_L},~ p\in P_{\psi_R},~ p\notin P_{\psi_H}, \label{eq:Constraint_IV}
\end{align}
for $(\psi_L,~\psi_R,~\psi_H)=(L,~\nu_R,~H_u)$, i.e., ${\bf Constraint~IV}$.
\end{enumerate}
In Table \ref{tab:Constraints}, we summarize all constraints.
\begin{table}[h]
  \begin{center}
  \renewcommand{\arraystretch}{1.2}
    \begin{tabular}{|c|c|c|c|c|c|c|c|c|} \hline
      & $P_Q$ & $P_{u_R}$ & $P_{d_R}$ & $P_{L}$ & $P_{\nu_R}$ & $P_{e_R}$ & $P_{H_u}$ & $P_{H_d}$ \\ \hline
      \multirow{3}{*}{I: $p_u$ is $\left\{ \begin{matrix} (1)\\(3)\\(4)\\\end{matrix}\right.$} & {\bf not in} & {\bf not in} & - & - & - & - & {\bf not in} & - \\ 
      & in & not in & - & - & - & - & in & - \\
      & not in & in & - & - & - & - & in & - \\ \hline
      \multirow{3}{*}{II$_1$: $p_d$ is $\left\{ \begin{matrix} (1)\\(3)\\(4)\\ \end{matrix} \right.$} & not in & - & not in & - & - & - & - & not in \\
      & in & - & not in & - & - & - & - & in \\
      & {\bf not in} & - & {\bf in} & - & - & - & - & {\bf in} \\ \hline
      \multirow{3}{*}{II$_2$: $p_d$ is $\left\{ \begin{matrix} (1)\\(3)\\(4)\\ \end{matrix} \right.$} & - & - & - & not in & - & not in & - & not in \\
      & - & - & - & {\bf in} & - & {\bf not in} & - & {\bf in} \\
      & - & - & - & not in & - & in & - & in \\ \hline
      III: $p_u=p_d$ & - & - & - & - & - & - & - & - \\ \hline
      IV: $p_u$ is & - & - & - & {\bf in} & {\bf in} & - & {\bf not in} & - \\ \hline
    \end{tabular}
  \end{center}
  \caption{The constraints I, II$_1$, II$_2$, III and IV.
For example, if $p_u$ corresponds to $(1)$, it is not included in $P_Q$.
If $p_u$ corresponds to $(3)$, it is included in $P_Q$.
  The bold texts denote the choices in Eq.~(\ref{eq:consistent_p}) which are consistent with all constraints.}
  \label{tab:Constraints}
\end{table}

Next, we classify all possible flavor models satisfying the above constraints.
See Table \ref{tab:Constraints}.
From the constraint III, $p_u=p_d\equiv p$ must consist.
Furthermore, from the constraint IV, $p$ must be in $P_F$ and satisfy
\begin{align}
  p \in P_L,~ p\in P_{\nu_R},~p\notin P_{H_u}.
\end{align}
From the constraint I, this makes $p$ be the pattern (1) for $(\psi_L,~\psi_R,~\psi_H)=(Q,~u_R,~H_u)$ in Eq.~(\ref{eq:Constraint_I}), 
i.e., 
\begin{align}
  p \notin P_Q,~p\notin P_{u_R},~p\notin P_{H_u}.
\end{align}
Similarly, from the constraint II$_2$, $p$ must be the pattern (3) for $(\psi_L,~\psi_R,~\psi_H)=(L,~e_R,~H_d)$ in Eq.~(\ref{eq:Constraint_II}),
i.e.,
\begin{align}
  p \in P_L,~p\notin P_{e_R},~p\in P_{H_d}.
\end{align}
Finally, from the constraint II$_1$, $p$ must be the pattern (4) for $(\psi_L,~\psi_R,~\psi_H)=(Q,~d_R,~H_d)$ in Eq.~(\ref{eq:Constraint_II}),
i.e.,
\begin{align}
  p \notin P_Q,~p\in P_{d_R},~p\in P_{H_d}.
\end{align}
Thus, the point $p$ consistent with all conditions must satisfy
\begin{align}
  p\in P_L\cup P_{d_R} \cup P_{\nu_R} \cup P_{H_d} \subset P_F, \quad
  p\notin P_Q \cup P_{u_R} \cup P_{e_R} \cup P_{H_u} \subset P_F. \label{eq:consistent_p}
\end{align}

Now we are ready to classify all possible flavor models satisfying the conditions I, II, III and IV.
We again note that $p_u=p_d=p\in P_F$ and therefore we can find flavor models with consistent $p$ by checking the zero points of zero-modes of each field from Table \ref{tab:ZeroPoints}.
Flavor models are picked up by Eq.~(\ref{eq:consistent_p}) in addition to the non-vanishing Yukawa coupling conditions, Eqs.~(\ref{eq:FluxParitySS1}) and (\ref{eq:FluxParitySS2}) and the anomaly cancellation condition which makes the number of up and down type Higgs fields same.
The results are shown in Appendix \ref{apx:CLASSMODELS}.
There are 408  flavor models in total.

%-----------------------------------------------------
%-----------------------------------------------------
%-----------------------------------------------------

\section{Modular symmetric models}
\label{sec:Modular_symmetric_models}

In this section, we classify the flavor models, which have a specific property under the $S$-transformation.
To calculate fermion flavors, we need to identify two types of VEVs; one is the VEV of modulus and another one is the VEVs of Higgs fields.
In the former, we consider the vacuum where the modulus lies on either of three modular fixed points; (1) $\tau= i$ is invariant under $S$-transformation; (2) $\tau = e^{2\pi i/3}\equiv \omega$ is invariant under $ST$-transformation; (3) $\tau = i\infty$ is invariant under $T$-transformation.
In the latter, we consider Higgs VEVs aligned in eigenbasis of the modular transformation corresponding to each fixed point.
%We call these vacua for both of modulus and Higgs fields as $S$, $ST$ and $T$-symmetric vacuum respectively, or modular symmetric vacuum.
We will show that some flavor models have the possibility to lead to realistic flavor observations in a vicinity of $S$-symmetric vacuum but there are no consistent flavor models for $ST$ and $T$-symmetric vacua.

%-----------------------------------------------------
%-----------------------------------------------------

\subsection{Higgs $\mu$-term}

First, we start from assuming that the value of modulus is fixed at either of $\tau=i$, $\omega$ and $i\infty$.
In this subsection, we study which direction Higgs VEVs are aligned at these three modular fixed points.

Higgs VEVs are aligned in the lightest mass direction.
Supersymmetric mass term ($\mu$-term) of Higgs fields can be generated by D-brane instanton effects \cite{Blumenhagen:2006xt,Ibanez:2006da,Ibanez:2007rs,Antusch:2007jd,Kobayashi:2015siy}.
As shown in Appendix \ref{App:Higgs_mu_term}, actually in a leading order, D-brane instanton effects give the following Higgs $\mu$-term,
\begin{align}
  \mu^{ij}\varepsilon_{nm}H_{um}^iH_{dn}^j
  &= \Lambda e^{-S_{\rm inst}}(2{\rm Im}\tau)^{-1}(Y_u^iY_d^j)\varepsilon_{nm}H_{um}^iH_{dn}^j,
\end{align}
where $\Lambda $ denotes a typical scale such as the compactification scale and 
$S_{\rm inst}$ denotes the instanton action.
Here, 
$Y_u^i$ ($Y_d^j$) are the 3-point couplings among instanton zero-modes $\alpha$, $\beta$ ($\gamma$) and Higgs fields $H_u^i$ ($H_d^j$) given by
\begin{align}
  Y_u^{i} = g({\rm Im}\tau)^{1/2} \int d^2z \psi_\alpha(z) \cdot \psi_\beta(z) \cdot (\psi_{H_u}^i(z))^*, \quad
  Y_d^{j} = g({\rm Im}\tau)^{1/2} \int d^2z \psi_\alpha(z) \cdot \psi_\gamma(z) \cdot (\psi_{H_d}^j(z))^*,
\end{align}
where $\psi$s are the zero-mode wave functions on $T^2/\mathbb{Z}_2$ corresponding to instanton zero-modes $\alpha$, $\beta$ $(\gamma)$ and Higgs fields $H_u$ ($H_d$).

Here, let us consider the modular transformation of this leading mass term.
Under modular transformation, the zero-modes of $\alpha,\beta,\gamma$ and Higgs fields behave as the modular forms of weight 1/2:
\begin{align}
  \psi_{\alpha,\beta,\gamma} \rightarrow \widetilde{J}_{1/2}(\widetilde{\gamma},\tau) \widetilde{\rho}_{\alpha,\beta,\gamma}(\widetilde{\gamma}) \psi_{\alpha,\beta,\gamma},
  \quad \psi_{H_{u,d}}^i \rightarrow \widetilde{J}_{1/2}(\widetilde{\gamma},\tau) \widetilde{\rho}_{H_{u,d}}(\widetilde{\gamma})_{ij} \psi_{H_{u,d}}^j,
\end{align}
where $\widetilde{J}_{1/2}(\widetilde{\gamma},\tau)$ is the automophy factor given by
\begin{align}
  \widetilde{J}_{1/2}(\widetilde{S},\tau) = (-\tau)^{1/2}, \quad \widetilde{J}_{1/2}(\widetilde{T},\tau) = 1, \quad \widetilde{J}_{1/2}(\widetilde{ST},\tau) = (-(\tau+1))^{1/2}, \label{eq;automophy}
\end{align}
and $\widetilde{\rho}_{\alpha,\beta,\gamma}$ and $\widetilde{\rho}_{H_{u,d}}$ denote $1\times 1$ and $g_H\times g_H$ unitary matrices for $\alpha$, $\beta$, $\gamma$ and $H_{u,d}$.
Then, the modular transformations of 3-points couplings $Y_u^i$ and $Y_d^j$ are obtained as
\begin{align}
  &Y_u^i \rightarrow \widetilde{J}_{1/2}(\widetilde{\gamma},\tau)\widetilde{\rho}_\alpha(\widetilde{\gamma}) \cdot \widetilde{\rho}_\beta(\widetilde{\gamma}) \cdot (\widetilde{\rho}_{H_u}(\widetilde{\gamma})_{ik})^*Y^k_u, \\
  &Y_d^j \rightarrow \widetilde{J}_{1/2}(\widetilde{\gamma},\tau)\widetilde{\rho}_\alpha(\widetilde{\gamma}) \cdot  \widetilde{\rho}_\gamma(\widetilde{\gamma}) \cdot (\widetilde{\rho}_{H_d}(\widetilde{\gamma})_{jk})^*Y^k_d,
\end{align}
and it follows from these that the $\mu$ matrix is transformed as
\begin{align}
  &\mu^{ij}(\tau) \rightarrow \widetilde{J}_{1/2}(\widetilde{\gamma},\tau)^4\widetilde{J}_{1/2}^*(\widetilde{\gamma},\tau)^2 [\widetilde{\rho}_\alpha(\widetilde{\gamma})]^2 \widetilde{\rho}_\beta(\widetilde{\gamma}) \widetilde{\rho}_\gamma(\widetilde{\gamma}) \cdot (\widetilde{\rho}_{H_u}(\widetilde{\gamma})_{ii'})^* (\widetilde{\rho}_{H_d}(\widetilde{\gamma})_{jj'})^* \mu^{i'j'}(\tau). \label{eq:modtransmu}
\end{align}
At the modular fixed points $\tau=i$, $\omega$ and $i\infty$, the leading mass term becomes invariant under $S$, $ST$ and $T$-transformations respectively since $S:\tau=-1/\tau$, $ST:\tau=-1/(\tau+1)$ and $T:\tau=\tau+1$.
That is, the $\mu^{ij}$ matrix obeys the following modular invariance relations,
\begin{align}
  &\mu^{ij}(i) = (\widetilde{J}_{1/2}(\widetilde{S},i)\widetilde{\rho}_{H_u}(\widetilde{S})_{ii'})^* \notag \\
  &\cdot [\widetilde{J}_{1/2}(\widetilde{S},i)\widetilde{\rho}_{\alpha}(\widetilde{S})]^2 \widetilde{J}_{1/2}(\widetilde{S},i)\widetilde{\rho}_{\beta}(\widetilde{S}) \widetilde{J}_{1/2}(\widetilde{S},i)\widetilde{\rho}_{\gamma}(\widetilde{S}) \mu^{ij}(i) \cdot (\widetilde{J}_{1/2}(\widetilde{S},i)\widetilde{\rho}_{H_d}(\widetilde{S})_{j'j})^\dagger, \label{eq:MuAti} \\
  &\mu^{ij}(\omega) = (\widetilde{J}_{1/2}(\widetilde{ST},\omega)\widetilde{\rho}_{H_u}(\widetilde{ST})_{ii'})^* \notag \\
  &\cdot [\widetilde{J}_{1/2}(\widetilde{ST},\omega)\widetilde{\rho}_{\alpha}(\widetilde{ST})]^2 \widetilde{J}_{1/2}(\widetilde{ST},\omega)\widetilde{\rho}_{\beta}(\widetilde{ST}) \widetilde{J}_{1/2}(\widetilde{ST},\omega)\widetilde{\rho}_{\gamma}(\widetilde{ST}) \mu^{ij}(\omega) \cdot (\widetilde{J}_{1/2}(\widetilde{ST},\omega)\widetilde{\rho}_{H_d}(\widetilde{ST})_{j'j})^\dagger, \\
  &\mu^{ij}(i\infty) = (\widetilde{J}_{1/2}(\widetilde{T},i\infty)\widetilde{\rho}_{H_u}(\widetilde{T})_{ii'})^* \notag \\
  &\cdot [\widetilde{J}_{1/2}(\widetilde{T},i\infty)\widetilde{\rho}_{\alpha}(\widetilde{T})]^2 \widetilde{J}_{1/2}(\widetilde{T},i\infty)\widetilde{\rho}_{\beta}(\widetilde{T}) \widetilde{J}_{1/2}(\widetilde{T},i\infty)\widetilde{\rho}_{\gamma}(\widetilde{T}) \mu^{ij}(i\infty) \cdot (\widetilde{J}_{1/2}(\widetilde{T},i\infty)\widetilde{\rho}_{H_d}(\widetilde{T})_{j'j})^\dagger, \label{eq:MuAtiinfty} 
\end{align}
where we have used Eq.~(\ref{eq;automophy}).
These relations mean that the mass eigenbasis of the leading mass term at each modular fixed point $\tau=i$, $\omega$ and $i\infty$ are also $S$, $ST$ and $T$-transformation eigenbasis, respectively.
Let us check this conclusion at $\tau=i$ as an example.
Let us consider the simple case that $\widetilde{J}_{1/2}(\widetilde{S},i)\widetilde{\rho}_{\alpha}(\widetilde{S})=\widetilde{J}_{1/2}(\widetilde{S},i)\widetilde{\rho}_{\beta}(\widetilde{S})=\widetilde{J}_{1/2}(\widetilde{S},i)\widetilde{\rho}_{\gamma}(\widetilde{S})=1$ and ${\rm diag}(\widetilde{J}_{1/2}(\widetilde{S},i)\widetilde{\rho}_{H_{u,d}}(\widetilde{S}))=(1,-1)$, hence two pair Higgs fields.
Then the relation Eq.~(\ref{eq:MuAti}) in $S$-eigenbasis is given by
\begin{align}
  \begin{pmatrix}
    \mu^{00}(i) & \mu^{01}(i) \\
    \mu^{10}(i) & \mu^{11}(i) \\
  \end{pmatrix}
  =
  \begin{pmatrix}
    1 & 0 \\
    0 & -1 \\
  \end{pmatrix}
  \begin{pmatrix}
    \mu^{00}(i) & \mu^{01}(i) \\
    \mu^{10}(i) & \mu^{11}(i) \\
  \end{pmatrix}
  \begin{pmatrix}
    1 & 0 \\
    0 & -1 \\
  \end{pmatrix},
  \label{eq:mu_term_2}
\end{align}
and obviously the leading mass matrix is diagonalized due to $\mu^{01}(i)=\mu^{10}(i)=0$.
In a similar way, we can show the leading mass eigenbasis at each modular fixed point are also eigenbasis of each residual symmetry of the modular transformation.

In general, there would exist some configurations giving a single instanton zero-mode; therefore the leading mass term should be rewritten by the liner combination of them as
\begin{align}
  \mu^{ij}(\tau) = \sum_a d_aY_u^{ia}Y_d^{ja} \equiv \sum_a c_a\mu^{ij}_a(\tau), \label{eq:mu_leading}
\end{align}
where $a$ runs all possible instanton zero-mode configurations and $Y_u^{ia}$ ($Y_d^{ja}$) denotes 3-point couplings among instanton zero-modes $\alpha_a$, $\beta_a$ ($\gamma_a$) and Higgs fields $H_u$ ($H_d$).
Under modular transformation, $\mu^{ij}_a$ is transformed as Eq.~(\ref{eq:modtransmu}) and obeys the same modular invariance relations in Eqs.~(\ref{eq:MuAti})-(\ref{eq:MuAtiinfty}).
Hence, the general leading mass eigenbasis at each modular fixed point are also eigenbasis of corresponding modular transformation.
Thus, at the leading order, Higgs VEVs which are aligned in the lightest mass direction at $\tau=i$, $\omega$ and $i\infty$ must be eigenbasis of $S$, $ST$ and $T$-transformations, respectively.
%These are just the modular symmetric vacuum.

Unfortunately, on the magnetized $T^2/\mathbb{Z}_2$ orbifold models, we cannot find the leading order Higgs $\mu$-term being able to determine the lightest mass direction uniquely because of the shortage of number of instanton zero-mode configurations which couple to Higgs fields.
In what follows, we assume that Higgs VEVs are aligned along eigenvectors of residual modular symmetry as the leading order although we do not know the full order $\mu$-term structure.
Such Higgs VEVs can be realized as long as $\mu$-term transforms under modular transformation as 
\begin{align}
  \sum_ac_a\mu_a^{ij} H_u^i H_d^j \xrightarrow{\gamma} \sum_aA(\widetilde{\gamma},\tau,a) \cdot c_a \mu^{ij}_a H_u^i H_d^j, \label{eq:muA}
\end{align}
where $A(\widetilde{\gamma},\tau,a)$ means a modular symmetry anomaly on $\mu$-term.
In fact, the leading mass term, Eq~(\ref{eq:mu_leading}), is transformed as
\begin{align}
  \sum_a c_a \mu_a^{ij}(\tau) H_u^iH_d^j \xrightarrow{\gamma} \widetilde{J}_{1/2}(\widetilde{\gamma},\tau)^4\sum_a [\widetilde{\rho}_{\alpha_a}(\widetilde{\gamma})]^2\widetilde{\rho}_{\beta_a}(\widetilde{\gamma})\widetilde{\rho}_{\gamma_a}(\widetilde{\gamma}) \cdot c_a\mu_a^{ij} H_u^iH_d^j,
\end{align}
under modular transformation since $H_{u,d}$ behaves as the modular form of weight -1/2:
\begin{align}
  H_{u,d}^i \xrightarrow{\gamma} \widetilde{J}_{-1/2}^*(\widetilde{\gamma},\tau) \widetilde{\rho}_{H_{u,d}}(\widetilde{\gamma})_{ij} H_{u,d}^j.
\end{align}
The modular transformation Eq.~(\ref{eq:muA}) ensures that the modular invariances of $\mu^{ij}_a$ at the fixed points:
\begin{align}
  &\mu^{ij}_a(i) = A(\widetilde{S},\tau,a) (\widetilde{J}_{1/2}(\widetilde{S},i) \widetilde{\rho}_{H_{u}}(\widetilde{S})_{ii'})^* (\widetilde{J}_{1/2}(\widetilde{S},i) \widetilde{\rho}_{H_{d}}(\widetilde{S})_{jj'})^* \mu^{i'j'}_a(i), \\
  &\mu^{ij}_a(\omega) = A(\widetilde{ST},\tau,a) (\widetilde{J}_{1/2}(\widetilde{ST},\omega) \widetilde{\rho}_{H_{u}}(\widetilde{ST})_{ii'})^* (\widetilde{J}_{1/2}(\widetilde{ST},\omega) \widetilde{\rho}_{H_{d}}(\widetilde{ST})_{jj'})^* \mu^{i'j'}_a(\omega), \\
  &\mu^{ij}_a(i\infty) = A(\widetilde{T},\tau,a) (\widetilde{J}_{1/2}(\widetilde{T},i\infty) \widetilde{\rho}_{H_{u}}(\widetilde{T})_{ii'})^* (\widetilde{J}_{1/2}(\widetilde{T},i\infty) \widetilde{\rho}_{H_{d}}(\widetilde{T})_{jj'})^* \mu^{i'j'}_a(i\infty).
\end{align}
Thus, there is no mixing bewteen Higgs modes with different eiganvalues of residual symmetry in the $\mu$-matrix as seen in Eq.~(\ref{eq:mu_term_2}).
Therefore, the Higgs VEVs are aligned along eigencectors of residual modular symmetry.

%-----------------------------------------------------
%-----------------------------------------------------

\subsection{Classification of the modular symmetric models}

In this subsection, we investigate the conditions to realize the Higgs VEVs which
correspond to 
% are the modular symmetric vacuum, that is, 
the eigenvectors of the residual modular transformation at the modular fixed points.
Note that we ignore $T$-symmetric vacuum because the values of elements of Yukawa matrices at $\tau=i\infty$ are strictly restricted by $T$-symmetry and it is difficult to realize realistic flavor observations.
In addition, the fixed point, $\tau=i\infty$, corresponds the decompactification limit, 
and it is not valid from the viewpoint of four-dimensional effective theory.

Under $S$ and $T$-transformations, the complex coordinate on $T^2/\mathbb{Z}_2$, $z$, and the modulus, $\tau$, are transformed as
\begin{align}
  &(z,\tau)\xrightarrow{S}(-\frac{z}{\tau},-\frac{1}{\tau}), \quad (z,\tau)\xrightarrow{T}(z,\tau+1).
\end{align}
This gives the following $S$ and $T$-transformations of zero-modes,
\begin{align}
  &\psi^{(j+\alpha_1,\alpha_2),M}_{T^2/\mathbb{Z}_2^m}(z,\tau) \xrightarrow{S}
  \psi^{(j+\alpha_1,\alpha_2),M}_{T^2/\mathbb{Z}_2^m}(S:(z,\tau))
  = (-\tau)^{1/2} \rho(S)^{jk\alpha_1\alpha_1'\alpha_2\alpha_2'} \psi^{(k+\alpha_1,\alpha_2),M}_{T^2/\mathbb{Z}_2^m}(z,\tau),
  \label{eq:StransPSI} \\
  &\psi^{(j+\alpha_1,\alpha_2),M}_{T^2/\mathbb{Z}_2^m}(z,\tau) \xrightarrow{T}
  \psi^{(j+\alpha_1,\alpha_2),M}_{T^2/\mathbb{Z}_2^m}(T:(z,\tau))
  = \rho(T)^{jk\alpha_1\alpha_1'\alpha_2\alpha_2'} \psi^{(k+\alpha_1,\alpha_2),M}_{T^2/\mathbb{Z}_2^m}(z,\tau),
  \label{eq:TtransPSI}
\end{align}
where
\begin{align}
  &\rho(S)^{jk\alpha_1\alpha_1'\alpha_2\alpha_2'} = \left\{
  \begin{array}{l}
   {\cal N}^{(j+\alpha_1)}{\cal N}^{(k+\alpha_1')} \frac{4e^{\pi i/4}}{\sqrt{M}} \cos \left(\frac{2\pi (j+\alpha_1)(k+\alpha_1')}{M}\right)\delta_{(\alpha_2,\alpha_1),(\alpha_1',\alpha_2')} \quad (m=0), \\
   {\cal N}^{(j+\alpha_1)}{\cal N}^{(k+\alpha_1')} \frac{4ie^{\pi i/4}}{\sqrt{M}} \sin \left(\frac{2\pi (j+\alpha_1)(k+\alpha_1')}{M}\right)\delta_{(\alpha_2,\alpha_1),(\alpha_1',\alpha_2')} \quad (m=1), \\
  \end{array} \right. \label{eq:StransRHO} \\
  &\rho(T)^{jk\alpha_1\alpha_1'\alpha_2\alpha_2'} = e^{\frac{\pi i(j+\alpha_1)^2}{M}} \delta_{j,k} \delta_{(\alpha_1,\alpha_2-\alpha_1+\frac{M}{2}),(\alpha_1',\alpha_2')}.
\end{align}
Obviously zero-modes are mapped into ones with same SS phases only if $\alpha_1=\alpha_2$ under $S$-transformation and only if $\alpha_1=\alpha_2=M/2$ (mod 1) under $ST$-transformation.
Therefore, modular symmetric Higgs VEVs at least have the following SS phases,
\begin{align}
  \left\{
  \begin{array}{l}
    (\alpha_1,\alpha_2) = (0,0)~{\rm or}~(1/2,1/2) \quad {\rm for}~S{\text -}{\rm symmetric~vacuum}, \\
    (\alpha_1,\alpha_2) = (M/2,M/2)~({\rm mod}~1) \quad {\rm for}~ST{\text -}{\rm symmetric~vacuum}.
  \end{array}\right.
  \label{eq:conditions_for_SSmodsym}
\end{align}
In each case, we can find Higgs fields which are eigenbasis of $S$ and $ST$-transformations, respectively.
On the other hand, it is not clear whether the realistic flavor structure is realized in these 
vacua or not.

Next we study the condition to realize the direction $h_{u,d}^k=v_{u,d}(U_{H_{u,d}}^{(g_H-1)k}(p))^*$ which are eigenvectors of residual symmetries at modular fixed points.
We have the possibilities of realizing realistic flavor structure by assuming the vicinity of such an eigenvector directions since fermion mass hierarchies can be realized near $h_{u,d}^k$ as described in the end of subsection \ref{subsec:3.2}.
The conditions for the modular eigenvectors $h_{u,d}^k$ are given by
\begin{align}
  \left\{
  \begin{array}{l}
    p = 0 ~{\rm or}~\frac{1+i}{2} \quad {\rm for}~S{\text -}{\rm symmetric~vacuum}, \\
    p = 0 \quad {\rm for}~ST{\text -}{\rm symmetric~vacuum}.
  \end{array}\right. \label{eq:conditions_for_modsym}
\end{align}
Let us  prove the condition for $S$-symmetric vacuum.
To make the direction $h_{u,d}^k=v_{u,d}(U_{H_{u,d}}^{(g_H-1)k}(p))^*$ $S$-eigenstate at $\tau=i$, the non-vanishing redefined zero-modes of Higgs fields defined in Eq.~(\ref{eq:U_psi_1(p)}), $\hat{\psi}^{(g_H-1)}_{H_{u,d}}(z,\tau)=U_{H_{u,d}}^{(g_H-1)k}(p)\psi_{H_{u,d}}^k(z,\tau)$, must be eigenbasis of $S$-transformation.
We will check this by calculating $S$-transformation of $\hat{\psi}^{(g_H-1)}_{H_{u,d}}(p,i)$ for $p\notin P_{H_{u,d}}$ and $\frac{\partial}{\partial z}\hat{\psi}^{(g_H-1)}_{H_{u,d}}(p,i)$ for $p\in P_{H_{u,d}}$.
Note that the redefined zero-modes satisfy
\begin{align}
  \left\{
  \begin{array}{l}
    \hat{\psi}^{j\neq (g_H-1)}(p,\tau) = 0, \quad \hat{\psi}^{(g_H-1)}(p,\tau) \neq 0 \quad {\rm for}~p\notin P_{H_{u,d}}, \\
    \frac{\partial}{\partial z}\hat{\psi}^{j\neq (g_H-1)}(p,\tau) = 0, \quad \frac{\partial}{\partial z}\hat{\psi}^{(g_H-1)}(p,\tau) \neq 0, \quad (\hat{\psi}^j(p,\tau)=0) \quad {\rm for}~p\in P_{H_{u,d}}, \\
  \end{array} \right.
\end{align}
as defined in Eq.~(\ref{eq:U_psi_1(p)}).
For $p=0$ or $\frac{1+i}{2}$ and $p\notin P_{H_{u,d}}$, $S$-transformation of non-vanishing mode $\hat{\psi}^{(g_H-1)}_{H_{u,d}}(p,i)$ is given by
\begin{align}
  \hat{\psi}^{(g_H-1)}_{H_{u,d}}(p,i) &\xrightarrow{S} \hat{\psi}^{(g_H-1)}_{H_{u,d}}(S:(p,i)) \notag \\
  &=
  \left\{
  \begin{array}{l}
    \hat{\psi}^{(g_H-1)}_{H_{u,d}}(0,i) \quad {\rm for}~p=0 \\
    \hat{\psi}^{(g_H-1)}_{H_{u,d}}(\frac{-1+i}{2},i) \quad {\rm for}~p=\frac{1+i}{2} \\
  \end{array}
  \right. \\
  &=
  \left\{
  \begin{array}{l}
    \hat{\psi}^{(g_H-1)}_{H_{u,d}}(0,i) \quad {\rm for}~p=0 \\
    e^{-2\pi i\alpha_{1H_{u,d}}}e^{-\pi i\frac{M_{H_{u,d}}}{2}}\hat{\psi}^{(g_H-1)}_{H_{u,d}}(\frac{1+i}{2},i) \quad {\rm for}~p=\frac{1+i}{2} \\
  \end{array}
  \right.,  \notag
\end{align}
from the boundary condition in Eq.~(\ref{eq:B.C.PSI}).
Similarly, for $p=0$ or $\frac{1+i}{2}$ and $p\in P_{H_{u,d}}$, $S$-transformation of $\frac{\partial}{\partial z}\hat{\psi}^{(g_H-1)}_{H_{u,d}}(p,i)$ is given by,
\begin{align}
  \textstyle\frac{\partial}{\partial z} \hat{\psi}^{(g_H-1)}_{H_{u,d}}(p,i) &\xrightarrow{S} (-i)\textstyle\frac{\partial}{\partial z} \hat{\psi}^{(g_H-1)}_{H_{u,d}}(S:(p,i)) \notag \\
  &=
  \left\{
  \begin{array}{l}
    (-i)\frac{\partial}{\partial z}\hat{\psi}^{(g_H-1)}_{H_{u,d}}(0,i) \quad {\rm for}~p=0 \\
    (-i)\frac{\partial}{\partial z}\hat{\psi}^{(g_H-1)}_{H_{u,d}}(\frac{-1+i}{2},i) \quad {\rm for}~p=\frac{1+i}{2} \\
  \end{array}
  \right. \\
  &=
  \left\{
  \begin{array}{l}
    (-i)\frac{\partial}{\partial z}\hat{\psi}^{(g_H-1)}_{H_{u,d}}(0,i) \quad {\rm for}~p=0 \\
    (-i)e^{-2\pi i\alpha_{1H_{u,d}}}e^{-\pi i\frac{M_{H_{u,d}}}{2}} \frac{\partial}{\partial z}\hat{\psi}^{(g_H-1)}_{H_{u,d}}(\frac{1+i}{2},i) \quad {\rm for}~p=\frac{1+i}{2} \\
  \end{array}
  \right., \notag
\end{align}
from the boundary condition in Eq.~(\ref{eq:B.C.Derivative}).
As shown in Eq.~(\ref{eq:StransPSI}), the transformation law is independent of $z$; therefore same relations consist on $z\neq p$,
\begin{align}
  &\hat{\psi}^{(g_H-1)}_{H_{u,d}}(z,i) \xrightarrow{S} 
  \left\{
  \begin{array}{l}
    \hat{\psi}^{(g_H-1)}_{H_{u,d}}(z,i) \quad {\rm for}~p=0\notin P_{H_{u,d}} \\
    e^{-2\pi i\alpha_{1H_{u,d}}}e^{-\pi i\frac{M_{H_{u,d}}}{2}}\hat{\psi}^{(g_H-1)}_{H_{u,d}}(z,i) \quad {\rm for}~p=\frac{1+i}{2} \notin P_{H_{u,d}} \\
  \end{array}
  \right.,
  \\
  &\textstyle\frac{\partial}{\partial z} \hat{\psi}^{(g_H-1)}_{H_{u,d}}(z,i) \xrightarrow{S} 
  \left\{
  \begin{array}{l}
    (-i)\frac{\partial}{\partial z}\hat{\psi}^{(g_H-1)}_{H_{u,d}}(z,i) \quad {\rm for}~p=0 \in P_{H_{u,d}} \\
    (-i)e^{-2\pi i\alpha_{1H_{u,d}}}e^{-\pi i\frac{M_{H_{u,d}}}{2}} \frac{\partial}{\partial z}\hat{\psi}^{(g_H-1)}_{H_{u,d}}(z,i) \quad {\rm for}~p=\frac{1+i}{2} \in P_{H_{u,d}} \\
  \end{array}
  \right..
\end{align}
Using $\frac{\partial}{\partial z}\xrightarrow{S}(-i)\frac{\partial}{\partial z}$, finally we obtain
\begin{align}
  \hat{\psi}^{(g_H-1)}_{H_{u,d}}(z,i) \xrightarrow{S} 
  \left\{
  \begin{array}{l}
    \hat{\psi}^{(g_H-1)}_{H_{u,d}}(z,i) \quad {\rm for}~p=0\notin P_{H_{u,d}} \\
    e^{-2\pi i\alpha_{1H_{u,d}}}e^{-\pi i\frac{M_{H_{u,d}}}{2}}\hat{\psi}^{(g_H-1)}_{H_{u,d}}(z,i) \quad {\rm for}~p=\frac{1+i}{2} \notin P_{H_{u,d}} \\
    \hat{\psi}^{(g_H-1)}_{H_{u,d}}(z,i) \quad {\rm for}~p=0\in P_{H_{u,d}} \\
    e^{-2\pi i\alpha_{1H_{u,d}}}e^{-\pi i\frac{M_{H_{u,d}}}{2}}\hat{\psi}^{(g_H-1)}_{H_{u,d}}(z,i) \quad {\rm for}~p=\frac{1+i}{2} \in P_{H_{u,d}} \\
  \end{array}
  \right..
\end{align}
This means that $\hat{\psi}^{(g_H-1)}_{H_{u,d}}(z,i)$ with $p=0$ or $\frac{1+i}{2}$ becomes eigenbasis of $S$-transformation.
This is because $z=0$ and $\frac{1+i}{2}$ are invariant under $S$-transformation up to lattice translations of torus.

On the other hand, the cases $p=\frac{1}{2}$ and $\frac{i}{2}$ are complicated.
The boundary conditions Eqs.~(\ref{eq:BCxxx}) and (\ref{eq:B.C.Derivative}) may give the relations
\begin{align}
  \hat{\psi}^{(g_H-1)}_{H_{u,d}}(\textstyle\frac{1}{2},i)\propto \hat{\psi}^{(g_H-1)}_{H_{u,d}}(\textstyle\frac{i}{2},i), \quad \frac{\partial}{\partial z}\hat{\psi}^{(g_H-1)}_{H_{u,d}}(\textstyle\frac{1}{2},i)\propto \frac{\partial}{\partial z}\hat{\psi}^{(g_H-1)}_{H_{u,d}}(\textstyle\frac{i}{2},i),
  \label{eq:1/2i/2}
\end{align}
in certain patterns of flux, SS phases and $\mathbb{Z}_2$ parity.
If these consist, $\hat{\psi}^{(g_H-1)}_{H_{u,d}}(z,i)$ with $p=\frac{1}{2}$ or $\frac{i}{2}$ can be eigenbasis of $S$-transformation since $S:z=S:\frac{1}{2}=\frac{i}{2}$ at $\tau=i$.
However it is unclear and difficult to show whether the relations in Eq.~(\ref{eq:1/2i/2}) consist or not.
Instead, we directly calculate whether the directions $\hat{\psi}^{(g_H-1)}_{H_{u,d}}(z,i)$ in each model is eigenbasis of $S$-transformation or not by using Eq.~(\ref{eq:StransPSI}).
As a result, there are no models where $\hat{\psi}^{(g_H-1)}_{H_{u,d}}(z,i)$ with $p=\frac{1}{2}$ or $\frac{i}{2}$ is $S$-eigenbasis.
Thus, the direction $h_{u,d}^k=v_{u,d}(U_{H_{u,d}}^{(g_H-1)k}(p))^*$ with $S$-invariant points $p=0$ and $\frac{1+i}{2}$ is $S$-symmetric vacuum.

In a similar way, we can check the condition for $ST$-symmetric vacuum in Eq.~(\ref{eq:conditions_for_modsym}).
The direction $h_{u,d}^k$ with $ST$-invariant point $p=0$ is $ST$-eigenstate but other points $p\in P_F$ which are not $ST$-invariant lead to not $ST$-symmetric vacuum.

Now, we are ready to classify the flavor models whose $h_{u,d}^k$ is modular symmetric.
The conditions are Eqs.~(\ref{eq:conditions_for_SSmodsym}) and (\ref{eq:conditions_for_modsym}).
As a result, we cannot find the flavor models satisfying conditions for $ST$-symmetric vacuum but can find models for $S$-symmetric vacuum.
The results for $S$-symmetric vacuum are shown in Table \ref{tab:classification_S_eig}.
There are 24  flavor models in total.
\begin{table}[H]
\begin{align}
  \begin{array}{c|c|c|c|c|c|c|c|c|c} \hline
    B_Q & B_{u_R} & B_{d_R} & B_L & B_{\nu_R} & B_{e_R} & B_{H_u} & B_{H_d} & g_H & p \\ 
\hline \hline
5,0,0,0 & 7,0,\frac{1}{2},\frac{1}{2} & 6,1,\frac{1}{2},\frac{1}{2} & 6,1,\frac{1}{2},0 & 6,1,0,\frac{1}{2} & 5,0,0,\frac{1}{2} & 12,0,\frac{1}{2},\frac{1}{2} & 11,1,\frac{1}{2},\frac{1}{2} & 6 & 0 \\
5,0,0,0 & 7,0,\frac{1}{2},\frac{1}{2} & 6,1,\frac{1}{2},\frac{1}{2} & 6,1,0,\frac{1}{2} & 6,1,\frac{1}{2},0 & 5,0,\frac{1}{2},0 & 12,0,\frac{1}{2},\frac{1}{2} & 11,1,\frac{1}{2},\frac{1}{2} & 6 & 0 \\
5,0,\frac{1}{2},0 & 6,0,\frac{1}{2},0 & 6,1,0,\frac{1}{2} & 6,1,\frac{1}{2},\frac{1}{2} & 5,1,\frac{1}{2},\frac{1}{2} & 5,0,0,0 & 11,0,0,0 & 11,1,\frac{1}{2},\frac{1}{2} & 6 & 0 \\
5,0,\frac{1}{2},0 & 6,1,0,\frac{1}{2} & 6,0,\frac{1}{2},0 & 6,1,\frac{1}{2},\frac{1}{2} & 5,0,0,0 & 5,1,\frac{1}{2},\frac{1}{2} & 11,1,\frac{1}{2},\frac{1}{2} & 11,0,0,0 & 6 & \frac{1+i}{2} \\
5,0,0,\frac{1}{2} & 6,0,0,\frac{1}{2} & 6,1,\frac{1}{2},0 & 6,1,\frac{1}{2},\frac{1}{2} & 5,1,\frac{1}{2},\frac{1}{2} & 5,0,0,0 & 11,0,0,0 & 11,1,\frac{1}{2},\frac{1}{2} & 6 & 0 \\
5,0,0,\frac{1}{2} & 6,1,\frac{1}{2},0 & 6,0,0,\frac{1}{2} & 6,1,\frac{1}{2},\frac{1}{2} & 5,0,0,0 & 5,1,\frac{1}{2},\frac{1}{2} & 11,1,\frac{1}{2},\frac{1}{2} & 11,0,0,0 & 6 & \frac{1+i}{2} \\
5,1,\frac{1}{2},\frac{1}{2} & 7,1,0,0 & 6,1,\frac{1}{2},\frac{1}{2} & 6,0,\frac{1}{2},0 & 6,0,0,\frac{1}{2} & 5,0,\frac{1}{2},0 & 12,0,\frac{1}{2},\frac{1}{2} & 11,0,0,0 & 6 & \frac{1+i}{2} \\
5,1,\frac{1}{2},\frac{1}{2} & 7,1,0,0 & 6,1,\frac{1}{2},\frac{1}{2} & 6,0,0,\frac{1}{2} & 6,0,\frac{1}{2},0 & 5,0,0,\frac{1}{2} & 12,0,\frac{1}{2},\frac{1}{2} & 11,0,0,0 & 6 & \frac{1+i}{2} \\
6,0,\frac{1}{2},0 & 5,0,\frac{1}{2},0 & 7,1,\frac{1}{2},0 & 6,1,\frac{1}{2},\frac{1}{2} & 5,1,\frac{1}{2},\frac{1}{2} & 7,0,\frac{1}{2},\frac{1}{2} & 11,0,0,0 & 13,1,0,0 & 6 & 0 \\
6,0,\frac{1}{2},0 & 6,0,0,\frac{1}{2} & 6,1,0,\frac{1}{2} & 6,1,\frac{1}{2},0 & 6,1,0,\frac{1}{2} & 6,0,0,\frac{1}{2} & 12,0,\frac{1}{2},\frac{1}{2} & 12,1,\frac{1}{2},\frac{1}{2} & 6 & 0 \\
6,0,\frac{1}{2},0 & 6,0,0,\frac{1}{2} & 6,1,0,\frac{1}{2} & 6,1,0,\frac{1}{2} & 6,1,\frac{1}{2},0 & 6,0,\frac{1}{2},0 & 12,0,\frac{1}{2},\frac{1}{2} & 12,1,\frac{1}{2},\frac{1}{2} & 6 & 0 \\
6,0,0,\frac{1}{2} & 5,0,0,\frac{1}{2} & 7,1,0,\frac{1}{2} & 6,1,\frac{1}{2},\frac{1}{2} & 5,1,\frac{1}{2},\frac{1}{2} & 7,0,\frac{1}{2},\frac{1}{2} & 11,0,0,0 & 13,1,0,0 & 6 & 0 \\
6,0,0,\frac{1}{2} & 6,0,\frac{1}{2},0 & 6,1,\frac{1}{2},0 & 6,1,\frac{1}{2},0 & 6,1,0,\frac{1}{2} & 6,0,0,\frac{1}{2} & 12,0,\frac{1}{2},\frac{1}{2} & 12,1,\frac{1}{2},\frac{1}{2} & 6 & 0 \\
6,0,0,\frac{1}{2} & 6,0,\frac{1}{2},0 & 6,1,\frac{1}{2},0 & 6,1,0,\frac{1}{2} & 6,1,\frac{1}{2},0 & 6,0,\frac{1}{2},0 & 12,0,\frac{1}{2},\frac{1}{2} & 12,1,\frac{1}{2},\frac{1}{2} & 6 & 0 \\
6,0,\frac{1}{2},\frac{1}{2} & 7,0,\frac{1}{2},\frac{1}{2} & 7,1,0,0 & 7,1,\frac{1}{2},0 & 6,1,\frac{1}{2},0 & 6,0,0,\frac{1}{2} & 13,0,0,0 & 13,1,\frac{1}{2},\frac{1}{2} & 7 & 0 \\
6,0,\frac{1}{2},\frac{1}{2} & 7,0,\frac{1}{2},\frac{1}{2} & 7,1,0,0 & 7,1,0,\frac{1}{2} & 6,1,0,\frac{1}{2} & 6,0,\frac{1}{2},0 & 13,0,0,0 & 13,1,\frac{1}{2},\frac{1}{2} & 7 & 0 \\
6,0,\frac{1}{2},\frac{1}{2} & 7,1,0,0 & 7,0,\frac{1}{2},\frac{1}{2} & 7,1,\frac{1}{2},0 & 6,0,0,\frac{1}{2} & 6,1,\frac{1}{2},0 & 13,1,\frac{1}{2},\frac{1}{2} & 13,0,0,0 & 7 & \frac{1+i}{2} \\
6,0,\frac{1}{2},\frac{1}{2} & 7,1,0,0 & 7,0,\frac{1}{2},\frac{1}{2} & 7,1,0,\frac{1}{2} & 6,0,\frac{1}{2},0 & 6,1,0,\frac{1}{2} & 13,1,\frac{1}{2},\frac{1}{2} & 13,0,0,0 & 7 & \frac{1+i}{2} \\
6,1,\frac{1}{2},0 & 5,0,0,\frac{1}{2} & 7,1,0,\frac{1}{2} & 6,1,\frac{1}{2},\frac{1}{2} & 5,0,0,0 & 7,1,0,0 & 11,1,\frac{1}{2},\frac{1}{2} & 13,0,\frac{1}{2},\frac{1}{2} & 6 & \frac{1+i}{2} \\
6,1,\frac{1}{2},0 & 6,1,0,\frac{1}{2} & 6,0,0,\frac{1}{2} & 6,0,\frac{1}{2},0 & 6,0,0,\frac{1}{2} & 6,1,0,\frac{1}{2} & 12,0,\frac{1}{2},\frac{1}{2} & 12,1,\frac{1}{2},\frac{1}{2} & 6 & \frac{1+i}{2} \\
6,1,\frac{1}{2},0 & 6,1,0,\frac{1}{2} & 6,0,0,\frac{1}{2} & 6,0,0,\frac{1}{2} & 6,0,\frac{1}{2},0 & 6,1,\frac{1}{2},0 & 12,0,\frac{1}{2},\frac{1}{2} & 12,1,\frac{1}{2},\frac{1}{2} & 6 & \frac{1+i}{2} \\
6,1,0,\frac{1}{2} & 5,0,\frac{1}{2},0 & 7,1,\frac{1}{2},0 & 6,1,\frac{1}{2},\frac{1}{2} & 5,0,0,0 & 7,1,0,0 & 11,1,\frac{1}{2},\frac{1}{2} & 13,0,\frac{1}{2},\frac{1}{2} & 6 & \frac{1+i}{2} \\
6,1,0,\frac{1}{2} & 6,1,\frac{1}{2},0 & 6,0,\frac{1}{2},0 & 6,0,\frac{1}{2},0 & 6,0,0,\frac{1}{2} & 6,1,0,\frac{1}{2} & 12,0,\frac{1}{2},\frac{1}{2} & 12,1,\frac{1}{2},\frac{1}{2} & 6 & \frac{1+i}{2} \\
6,1,0,\frac{1}{2} & 6,1,\frac{1}{2},0 & 6,0,\frac{1}{2},0 & 6,0,0,\frac{1}{2} & 6,0,\frac{1}{2},0 & 6,1,\frac{1}{2},0 & 12,0,\frac{1}{2},\frac{1}{2} & 12,1,\frac{1}{2},\frac{1}{2} & 6 & \frac{1+i}{2} \\ \hline
  \end{array} \notag
\end{align}
\caption{All quark and lepton flavor models satisfying $S$-symmetric vacuum conditions in Eqs.~(\ref{eq:conditions_for_SSmodsym}) and (\ref{eq:conditions_for_modsym}).
The first to eighth rows show the flux $M$ , $\mathbb{Z}_2$ parity $m$ (even, odd = 0, 1) and SS phases $(\alpha_1,\alpha_2)$ of the zero-modes of the fields.
$g_H$ denotes the number of Higgs fields.}
  \label{tab:classification_S_eig}
\end{table}

%-----------------------------------------------------
%-----------------------------------------------------
%-----------------------------------------------------

\section{Numerical example}
\label{sec:Numerical_example}

In this section, we study a flavor model shown in Table \ref{tab:ex-model} which can be realistic in the vicinity of $S$-eigenvector, and derive a realistic quark and lepton flavor structure.
In this model, quark doublets $Q$ have (Flux, $\mathbb{Z}_2$ parity, SS phases $\alpha_1,\alpha_2$) = ($6,0,0,\frac{1}{2}$); right-handed up-sector quarks $u_R$ have $(5,0,0,\frac{1}{2})$; right-handed down-sector quarks $d_R$ have $(6,0,\frac{1}{2},0)$; lepton doublets $L$ have $(6,0,\frac{1}{2},0)$; right-handed neutrinos $\nu_R$ have $(5,0,\frac{1}{2},0)$; right-handed charged leptons $e_R$ have $(6,0,0,\frac{1}{2})$; up type Higgs fields $H_u$ have $(11,0,0,0)$; down type Higgs fields $H_d$ have $(12,0,\frac{1}{2},\frac{1}{2})$.
The number of both up and down types Higgs fields are six.
\begin{table}[H]
\begin{align}
  \begin{array}{c|c|c|c|c|c|c|c|c} \hline
    B_Q & B_{u_R} & B_{d_R} & B_L & B_{\nu_R} & B_{e_R} & B_{H_u} & B_{H_d} & g_H \\ \hline\hline
    6,0,\frac{1}{2},0 & 6,0,0,\frac{1}{2} & 6,1,0,\frac{1}{2} & 6,1,0,\frac{1}{2} & 6,1,\frac{1}{2},0 & 6,0,\frac{1}{2},0 & 12,0,\frac{1}{2},\frac{1}{2} & 12,1,\frac{1}{2},\frac{1}{2} & 6 \\ \hline
  \end{array} \notag
\end{align}
\caption{Flux, $\mathbb{Z}_2$ parity (even, ood = 0, 1), SS phases $(\alpha_1,\alpha_2)$ of quarks, leptons and Higgs fields in the model.
$g_H$ denotes the number of Higgs fields.}
  \label{tab:ex-model}
\end{table}
Yukawa couplings $Y_u^{ijk}$, $Y_{d}^{ijk}$, $Y_\nu^{ijk}$ and $Y_e^{ijk}$ appearing in this model, are summarized in Appendix \ref{App:Yukawa-couplings}; the Majorana mass matrix of the right-handed neutrinos induced by D-brane instanton effect is shown in Appendix \ref{App:Majorana}.

In our numerical study, we fix the value of modulus by $\tau=i$ and use Higgs VEV directions as parameters.
Higgs VEV directions satisfying conditions I-IV, $h_{u,d}^k$, in this model are given by
\begin{align}
  &h_u^k = v_u(0.8464, 0.5014, 0.1759, 0.03657, 0.004504, 0.0003144), \\
  &h_d^k = v_d(0.4330, 0.7696, 0.4501, 0.1310, 0.02074, 0.001945),
\end{align}
where $h_u^k$ and $h_d^k$ are $S$-eigenbasis directions with eigenvalues $+1$ and $+i$, respectively.
Thus the modulus is $S$-symmetric vacuum, while  these Higgs VEV directions 
correspond to $S$-eigenstates.
Firstly, we try to realize flavor observations in exact $S$-eigenstate directions
in the Higgs VEV directions.
Six pairs of up (down) type Higgs fields include three $S$-eigenstates with eigenvalue $+1$ ($+i$) in total.
We use these three eigenstates as parameters for up and down type Higgs VEVs, respectively.
To obtain realistic flavors, let us choose the following Higgs VEV directions,
\begin{align}
  &\langle H_u^k \rangle = v_u(0.8466, 0.5009, 0.1762, 0.03715, 0.004794, 0.0003797), \label{eq:VEVHu_S} \\
  &\langle H_d^k \rangle = v_d(0.5006, 0.7890, 0.3521, 0.05382, -0.003787, -0.003709). \label{eq:VEVHd_S}
\end{align}
We note that again these directions are eigenbasis of $S$-transformation.
They lead to the following up quark, down quark and charged lepton mass ratios,
\begin{align}
  &(m_u,m_c,m_t)/m_t = (2.96\times 10^{-5},5.35\times 10^{-4},1), \\
  &(m_d,m_s,m_b)/m_b = (4.36\times 10^{-4},1.17\times 10^{-2},1), \\
  &(m_e,m_\mu,m_\tau)/m_\tau = (4.36\times 10^{-4},1.17\times 10^{-2},1),
\end{align}
and a ratio of the differences of the squares of the neutrino masses,
\begin{align}
  \sqrt{\frac{\Delta m_{\nu 12}^2}{\Delta m_{\nu 13}^2}}
  = \sqrt{\frac{|m_{\nu_1}^2-m_{\nu_2}^2|}{|m_{\nu_1}-m_{\nu_3}^2|}} = 0.179,
\end{align}
for normal ordering (NO), $m_{\nu_1}<m_{\nu_2}<m_{\nu_3}$.
Also the absolute values of the CKM matrix, $|V_{\rm CKM}|$, and the Pontecorvo-Maki-Nakagawa-Sakata (PMNS) matrix, $|V_{\rm PMNS}|$, are obtained as follows,
\begin{align}
  &|V_{\rm CKM}| = 
  \begin{pmatrix}
    0.972 & 0.235 & 0.00134 \\
    0.233 & 0.964 & 0.126 \\
    0.0309 & 0.122 & 0.992 \\
      \end{pmatrix}, \quad
  |V_{\rm PMNS}| = 
  \begin{pmatrix}
    0.990 & 0.137 & 0.0134 \\
    0.129 & 0.957 & 0.261 \\
    0.0487 & 0.257 & 0.965 \\
      \end{pmatrix}.
\end{align}
The mass ratios of quarks and leptons, and the absolute values of the CKM matrix are roughly realized, but the absolute values of the PMNS matrix are not realistic.
As a result, in this model it is difficult to realize both quark and lepton flavors in the exact $S$-eigenvectors of the Higgs VEV directions.

Next, we consider the vicinity of above $S$-eigenvector of Higgs VEV directions.
We use all six pairs of Higgs VEVs as parameters for both up and down types but fix the modulus at $\tau=i$ to simplify the analysis.
To obtain realistic flavors, in the vicinity of $h_{u,d}^k$, we have chosen the following Higgs VEV directions,
\begin{align}
  &\langle H_u^k \rangle = v_u(0.8509, 0.4970, 0.1679, 0.02805, -0.006762, -0.003731), \label{eq:VEVHu} \\
  &\langle H_d^k \rangle = v_d(0.4340, 0.7688, 0.4499, 0.1283, 0.02538, 0.03302). \label{eq:VEVHd}
\end{align}
The norm of $h_u^k$ in $\langle H_{u}^k \rangle$ is 0.9998 and one of $h_d^k$ in $\langle H_{d}^k \rangle$ is 0.9995.
In these directions, the mass matrices for quarks and leptons are given by
\begin{align}
&M_u/m_t =
\begin{pmatrix}
0.7202 & 0.5992 & 0.1214 \\
0.2492 & 0.2063 & 0.03922 \\
0.03057 & 0.02249 & -0.002550 \\
\end{pmatrix}, \quad
M_{d}/m_b =
\begin{pmatrix}
0.8675 & 0.3620 & 0.05514 \\
0.3053 & 0.1303 & 0.02287 \\
0.03861 & 0.03580 & 0.03967 \\
\end{pmatrix}, \\
&M_\nu/m_{\nu_3} =
\begin{pmatrix}
-0.3614 & -0.09456 & -0.3323 \\
-0.09456 & -0.1345 & -0.4077 \\
-0.3323 & -0.4077 & -0.5819 \\
\end{pmatrix}, \quad
M_e/m_\tau =
\begin{pmatrix}
0.8675 & 0.3053 & 0.03861 \\
0.3620 & 0.1303 & 0.03580 \\
0.05514 & 0.02287 & 0.03967 \\
\end{pmatrix}.
\end{align}
Then they lead to the following up quark, down quark and charged lepton mass ratios,
\begin{align}
  &(m_u,m_c,m_t)/m_t = (3.13\times 10^{-5},8.14\times 10^{-3},1), \\
  &(m_d,m_s,m_b)/m_b = (8.46\times 10^{-4},4.10\times 10^{-2},1), \\
  &(m_e,m_\mu,m_\tau)/m_\tau = (8.46\times 10^{-4},4.10\times 10^{-2},1),
\end{align}
and a ratio of the differences of the squares of the neutrino masses,
\begin{align}
  \sqrt{\frac{\Delta m_{\nu 12}^2}{\Delta m_{\nu 13}^2}}
  = \sqrt{\frac{|m_{\nu_1}^2-m_{\nu_2}^2|}{|m_{\nu_1}-m_{\nu_3}^2|}} = 0.162,
\end{align}
for NO.
For inverted ordering (IO), $m_{\nu_3}<m_{\nu_1}<m_{\nu_2}$, it is difficult to realize realistic 
the flavor structure.
Also the absolute values of the CKM matrix, $|V_{\rm CKM}|$, and the PMNS matrix, $|V_{\rm PMNS}|$, are obtained as follows,
\begin{align}
  &|V_{\rm CKM}| = 
  \begin{pmatrix}
    0.973 & 0.232 & 0.00234 \\
    0.232 & 0.973 & 0.0162 \\
    0.00603 & 0.0152 & 1.00 \\
      \end{pmatrix}, \quad
  |V_{\rm PMNS}| = 
  \begin{pmatrix}
    0.841 & 0.522 & 0.147 \\
    0.246 & 0.608 & 0.755 \\
    0.483 & 0.598 & 0.639 \\
      \end{pmatrix}.
\end{align}
The results are summarized in Table \ref{MassandCKMandPMNS}.
As a result, in this model we could realize quark and lepton flavor structure in the vicinity of $S$-eigenvector of Higgs VEV direction.
\begin{table}[H]
  \begin{center}
    \renewcommand{\arraystretch}{1.3}
    $\begin{array}{c|c|c} \hline
      & {\rm Obtained\ values} & {\rm Reference\ values} \\ \hline
      (m_u,m_c,m_t)/m_t & (3.13\times 10^{-5},8.14\times 10^{-3},1) & (5.58\times 10^{-6},2.69\times 10^{-3},1) \\ \hline
      (m_d,m_s,m_b)/m_b & (8.46\times 10^{-4},4.10\times 10^{-2},1) & (6.86\times 10^{-4},1.37\times 10^{-2},1) \\ \hline
      |V_{\rm CKM}| 
      &
      \begin{pmatrix}
        0.973 & 0.232 & 0.00234 \\
    0.232 & 0.973 & 0.0162 \\
    0.00603 & 0.0152 & 1.00 \\
      \end{pmatrix}
      & 
      \begin{pmatrix}
        0.974 & 0.227 & 0.00361 \\
        0.226 & 0.973 & 0.0405 \\
        0.00854 & 0.0398 & 0.999 
      \end{pmatrix}\\ \hline
      \sqrt{\Delta m_{\nu 12}^2/\Delta m_{\nu 13}^2} & 0.162~{\rm (NO)} & 0.173 \\ \hline
      (m_e,m_\mu,m_\tau)/m_\tau & (8.46\times 10^{-4},4.10\times 10^{-2},1) & (2.78\times 10^{-4},5.88\times 10^{-2},1) \\ \hline
      |V_{\rm PMNS}| 
      &
      \begin{pmatrix}
        0.841 & 0.522 & 0.147 \\
    0.246 & 0.608 & 0.755 \\
    0.483 & 0.598 & 0.639 \\
      \end{pmatrix}
      & 
      \begin{pmatrix}
        0.801{\text -}0.845 & 0.513{\text -}0.579 & 0.143{\text -}0.156 \\
        0.232{\text -}0.507 & 0.459{\text -}0.694 & 0.629{\text -}0.779 \\
        0.260{\text -}0.526 & 0.470{\text -}0.702 & 0.609{\text -}0.763 \\
      \end{pmatrix}\\ \hline
    \end{array}$
    \caption{The mass ratios of the quarks and leptons, and the absolute values of the CKM matrix and the PMNS matrix elements at $\tau=i$ under the vacuum alignments of Higgs fields in Eqs.~(\ref{eq:VEVHu}) and (\ref{eq:VEVHd}).
    Reference values of mass ratios are shown in Ref.~\cite{Bjorkeroth:2015ora}.
    Those of the CKM matrix and PMNS matrix elements are shown in Refs.~\cite{Zyla:2020zbs} and \cite{NuFIT:2021}.}
    \label{MassandCKMandPMNS}
  \end{center}
\end{table}

%-----------------------------------------------------
%-----------------------------------------------------
%-----------------------------------------------------

\section{Conclusion}
\label{sec:Conclusion}

In this paper, we have investigated the conditions to realize the quark and lepton flavor structure in magnetized orbifold models.
We have found four conditions I, II, III and IV.
The condition I demands the directions of up type Higgs VEVs $h_u^k$ leading to rank one mass matrix for up quark to realize its mass hierarchy.
The condition II demands the directions of down type Higgs VEVs $h_d^k$ leading to rank one mass matrices for both down-sector quarks and charged leptons to realize their mass hierarchies.
The condition III demands the equivalence between $u_L^u$ and $u_L^{d}$ which are unitary matrices diagonalizing rank one mass matrices to realize small quark mixing.
The condition IV demands that $h_u^k$ is also the direction leading to vanishing neutrino Dirac mass matrix to realize non small lepton mixing.
Note that the rank one mass matrices are favorable in the limit that we neglect masses of the first and second generations.
Through zero points analysis for zero-modes of each field, we could check whether the flavor models can satisfy these four conditions or not.
Consequently we have found the 408  flavor models which are consistent with the conditions I-IV.
In such models it is possible to realize the large hierarchy of up quark, down quark and charged lepton masses and realistic mixings of quark and lepton in the vicinity of $h_{u,d}^k$.

Also we have classified the flavor models which can be realistic in the vicinity of specific points under $S$-symmetry, where 
VEV of modulus lies on the fixed point of $S$-transformation, $\tau=i$, and Higgs VEVs are aligned in eigenbasis of $S$-transformation.
Indeed Higgs VEVs led by the leading $\mu$-term generated by D-brane instanton effects at $\tau=i$ are generally aligned in eigenbasis of $S$-transformation.
In this paper, we have classified the flavor models whose $h_{u,d}^k$ becomes eigenbasis of $S$-transformation.
As a result we have found 24 flavor models they have the possibilities to realize realistic flavor observations in the vicinity of $S$-eigenvector of Higgs VEV direction.

Here, we have given numerical studies on the model shown in Table \ref{tab:ex-model} in the exact and the vicinity of $S$-eigenvector of Higgs VEV direction.
In the exact $S$-eigenvector direction, we could roughly realize the values of the quark and lepton mass ratios and the CKM matrix but the PMNS matrix was not realistic.
In the vicinity of $S$-eigenvector of Higgs VEV direction, we could realize the values of quark and lepton mass ratios as well as the CKM and PMNS matrices.

Similar classifications through the zero point analysis can be applied for the flavor models in other orbifold models such as $T^2/\mathbb{Z}_3$, $T^2/\mathbb{Z}_4$ and $T^2/\mathbb{Z}_6$.
It would be possible for magnetized $T^4$ and its orbifold models.
Also we need to study Higgs $\mu$-term through D-brane instanton effects to check the direction of Higgs VEVs.
We would study them and examine the possibilities of realization of quark and lepton flavor 
structure elsewhere.

%-------- acknowledgement -------%
\vspace{1.5 cm}
\noindent
{\large\bf Acknowledgement}\\

H. U. was supported by Grant-in-Aid for JSPS Research Fellows No. 20J20388.
S. K. was supported by Grant-in-Aid for JSPS Research Fellows No. 22J10172.

%-----------------------------------------------------
%-----------------------------------------------------
%-----------------------------------------------------
\appendix
\section*{Appendix}

%-----------------------------------------------------
%-----------------------------------------------------
%-----------------------------------------------------

\section{Majorana neutrino masses by D-brane instanton effects}
\label{App:Neutrino_Majorana_mass_term}

Here we give a brief review of Majorana neutrino mass terms induced by D-brane instanton effects \cite{Blumenhagen:2006xt,Ibanez:2006da,Ibanez:2007rs,Antusch:2007jd,Kobayashi:2015siy}.

We consider two stacks of D-branes, $D_{N_1}$ and $D_{N_2}$, and assume D-brane instanton $D_{\rm inst}$ with magnetic fluxes.
Right-handed neutrinos $\nu_R$ appear as zero-modes of open strings  between $D_{N_1}$ and $D_{N_2}$; instanton zero-modes $\beta$ ($\gamma$) appear between $D_{N_1}$ ($D_{N_2}$) and $D_{\rm inst}$.
Then D-brane instanton effects give the following term,
\begin{align}
  \int d^2\beta d^2\gamma e^{-({\rm Im}\tau)^{-1/2} d_a^{ij}\beta^i\gamma^j \nu_R^a},
\end{align}
where $\beta$ and $\gamma$ are grasmannian, and $d_a^{ij}$ are the 3-point couplings among $\beta$, $\gamma$ and $\nu_R$ given by
\begin{align}
  d^{ij}_a = g({\rm Im}\tau)^{1/2} \int d^2z \psi^i_\beta(z) \cdot \psi^j_\gamma(z) \cdot (\psi^a_{\nu_R}(z))^*,
\end{align}
where $\psi$s are the zero-mode wavefunctions on $T^2/\mathbb{Z}_2$ corresponding to instanton zero-modes $\beta$, $\gamma$ and right-handed neutrinos $\nu_R$.
Mass terms can be generated only if each of $\beta$ and $\gamma$ has two zero-modes.
By grasmannian integral, we obtain Majorana mass term,
\begin{align}
  \Lambda e^{-S_{\rm inst}}\int d^2\beta d^2\gamma e^{-({\rm Im}\tau)^{-1/2}d_a^{ij}\beta^i\gamma^j \nu_R^a}
  &= \Lambda e^{-S_{\rm inst}} ({\rm Im}\tau)^{-1} \varepsilon_{ij}\varepsilon_{k\ell} d^{ik}_a d^{j\ell}_b \nu_R^a\nu_R^b 
  = M_{RR}^{ab} \nu_R^a\nu_R^b, \label{eq:m_RR}
\end{align}
where $S_{\rm inst}$ denotes the instanton action and $\Lambda$ denotes a
typical sale as the compactification scale.
The possible instanton zero-modes configurations are given by the following non-vanishing conditions for 3-point couplings,
\begin{align}
  &M_\beta \pm M_\gamma = \pm M_{\nu_R}, \quad m_\beta + m_\gamma = m_{\nu_R}, \quad
  (\alpha_1,\alpha_2)_\beta + (\alpha_1,\alpha_2)_\gamma = (\alpha_1,\alpha_2)_{\nu_R},
\end{align}
where $M_f$, $m_f$ and $(\alpha_1,\alpha_2)_f$, $f=\beta,\gamma,\nu_R$ denote the magnetic fluxes, $\mathbb{Z}_2$ twist parities and SS phases for zero-modes of $\beta$, $\gamma$ and $\nu_R$.

%-----------------------------------------------------
%-----------------------------------------------------
%-----------------------------------------------------

\section{Higgs $\mu$-term by D-brane instanton effects}
\label{App:Higgs_mu_term}

Here we give a brief review of Higgs $\mu$-terms induced by D-brane instanton effects 
\cite{Blumenhagen:2006xt,Ibanez:2006da,Ibanez:2007rs,Antusch:2007jd,Kobayashi:2015siy}.

We consider three stacks of D-branes, $D_a$, $D_b$ and $D_c$ with magnetic fluxes.
The D-brane $D_b$ is parallel to $D_c$.
Up (down) type Higgs fields, $H_u$ ($H_d$), appear as zero-modes of open strings  between $D_a$ and $D_b$ ($D_c$).
To generate $\mu$-terms, we also assume the D-brane instanton $D_{\rm inst}$ with magnetic flux which has a single zero-mode with each of  other branes .
The instanton zero-modes $\alpha$, $\beta$ and $\gamma$ appear as zero-modes of open strings between $D_a$ and $D_{\rm inst}$, $D_b$ and $D_{\rm inst}$, and $D_c$ and $D_{\rm inst}$.
Then D-brane instanton effects give the following term,
\begin{align}
  \int d^2\alpha d\beta d\gamma e^{({\rm Im}\tau)^{-1/2} (Y_u^i\alpha\cdot H_u^i\beta+ Y_d^j\alpha\cdot H_d^j\gamma)},
\end{align}
where $\alpha,\beta$ and $\gamma$ are grasmannian and $Y_u^i$ ($Y_d^j$) are the 3-point couplings among $\alpha$, $\beta$ ($\gamma$) and $H_u^i$ ($H_d^j$) given by
\begin{align}
  Y_u^{i} = g({\rm Im}\tau)^{1/2} \int d^2z \psi_\alpha(z) \cdot \psi_\beta(z) \cdot (\psi_{H_u}^i(z))^*, \quad
  Y_d^{j} = g({\rm Im}\tau)^{1/2} \int d^2z \psi_\alpha(z) \cdot \psi_\gamma(z) \cdot (\psi_{H_d}^j(z))^*,
\end{align}
where $\psi$s are the zero-mode wavefunctions on $T^2/\mathbb{Z}_2$ corresponding to instanton zero-modes $\alpha$, $\beta$ $(\gamma)$ and Higgs fields $H_u$ ($H_d$).
Mass terms can be generated only if each of $\alpha$, $\beta$ and $\gamma$ has a single zero-mode.
By grasmannian integral, we obtain the Higgs $\mu$ term,
\begin{align}
 \Lambda e^{-S_{\rm inst}} \int d^2\alpha d\beta d\gamma e^{({\rm Im}\tau)^{-1/2}(Y_u^i\alpha\cdot H_u^i+Y_d^j\alpha\cdot H_d^j\gamma)}
  =  \Lambda e^{-S_{\rm inst}} ({\rm Im}\tau)^{-1}(Y_u^iY_d^j)\varepsilon_{nm}H_{um}^iH_{dn}^j 
  = \mu^{ij}\varepsilon_{nm}H_{um}^iH_{dn}^j,
\end{align}
where $m,n\in\{1,2\}$ denote components of $SU(2)_L$ doublet.
The possible instanton zero-modes configurations are given by the following non-vanishing conditions for 3-point couplings,
\begin{align}
  &M_\alpha \pm M_\beta = \pm M_{H_u}, \quad m_\alpha + m_\beta = m_{H_u}, \quad
  (\alpha_1,\alpha_2)_\alpha + (\alpha_1,\alpha_2)_\beta = (\alpha_1,\alpha_2)_{H_u}, \\
  &M_\alpha \pm M_\gamma = \pm M_{H_d}, \quad m_\alpha + m_\gamma = m_{H_d}, \quad
  (\alpha_1,\alpha_2)_\alpha + (\alpha_1,\alpha_2)_\gamma = (\alpha_1,\alpha_2)_{H_d},
\end{align}
where $M_f$, $m_f$ and $(\alpha_1,\alpha_2)_f$, $f=\alpha,\beta,\gamma,H_u,H_d$, denote the magnetic fluxes, $\mathbb{Z}_2$ twist parities and SS phases for zero-modes of $\alpha$, $\beta$, $\gamma$, $H_u$ and $H_d$.

%-----------------------------------------------------
%-----------------------------------------------------
%-----------------------------------------------------

\section{The model in the numerical example}
\label{App:The_model_in_the_numerical_example}

%-----------------------------------------------------
%-----------------------------------------------------
\subsection{Yukawa couplings}
\label{App:Yukawa-couplings}

Here we summarize Yukawa couplings of up-sector quarks, down-sector quarks, neutrinos and charged leptons, $Y_u^{ijk}$, $Y_{d}^{ijk}$, $Y_\nu^{ijk}$ and $Y_e^{ijk}$ in our model.

%-----------------------------------------------------
\subsubsection{Up quark: $B_Q{\text -}B_{u_R}{\text -}B_{H_u}=(6,0,\frac{1}{2},0){\text -}(6,0,0,\frac{1}{2}){\text -}(12,0,\frac{1}{2},\frac{1}{2})$}

The following table shows the zero-mode assignments for quark doublets $Q^i$, right-handed up-sector quarks $u_R^j$ and up type Higgs fields $H_u^k$.
\begin{table}[H]
\begin{center}
\renewcommand{\arraystretch}{1.2}
\begin{tabular}{c|c|c|c}
& $Q^i$ & $u_R^j$ & $H^k_u$ \\ \hline
0 & $\frac{1}{\sqrt{2}}(\psi_{T^2}^{(1/2,0),6}+\psi_{T^2}^{(11/2,0),6})$ & $\psi_{T^2}^{(0,1/2),6}$ & $\frac{1}{\sqrt{2}}(\psi_{T^2}^{(1/2,1/2),12}-\psi_{T^2}^{(23/2,1/2),12})$ \\
1 & $\frac{1}{\sqrt{2}}(\psi_{T^2}^{(3/2,0),6}+\psi_{T^2}^{(9/2,0),6})$ & $\frac{1}{\sqrt{2}}(\psi_{T^2}^{(1,1/2),6}-\psi_{T^2}^{(5,1/2),6})$ & $\frac{1}{\sqrt{2}}(\psi_{T^2}^{(3/2,1/2),12}-\psi_{T^2}^{(21/2,1/2),12})$ \\
2 & $\frac{1}{\sqrt{2}}(\psi_{T^2}^{(5/2,0),6}+\psi_{T^2}^{(7/2,0),6})$ & $\frac{1}{\sqrt{2}}(\psi_{T^2}^{(2,1/2),6}-\psi_{T^2}^{(4,1/2),6})$ & $\frac{1}{\sqrt{2}}(\psi_{T^2}^{(5/2,1/2),12}-\psi_{T^2}^{(19/2,1/2),12})$ \\
3 & & & $\frac{1}{\sqrt{2}}(\psi_{T^2}^{(7/2,1/2),12}-\psi_{T^2}^{(17/2,1/2),12})$ \\
4 & & & $\frac{1}{\sqrt{2}}(\psi_{T^2}^{(9/2,1/2),12}-\psi_{T^2}^{(15/2,1/2),12})$ \\
5 & & & $\frac{1}{\sqrt{2}}(\psi_{T^2}^{(11/2,1/2),12}-\psi_{T^2}^{(13/2,1/2),12})$ \\
\end{tabular}
\end{center}
%\caption{Zero-mode wave functions in ``(6,e;1/2,0)-(6,e;0,1/2)-(12,e;1/2,1/2)'' model.}
%\label{tab:three-model-ChargedL}
\end{table}
Yukawa couplings are given by
\begin{align}
  Y^{ijk}_{u}H_u^k = Y^{ij0}_{u}H_u^0+Y^{ij1}_{u}H_u^1+Y^{ij2}_{u}H_u^2+Y^{ij3}_{u}H_u^3+Y^{ij4}_{u}H_u^4+Y^{ij5}_{u}H_u^5,
\end{align}
with
\begin{align}
&Y_u^{ij0} = c_{(6{\text -}6{\text -}12)}
\begin{pmatrix}
  X_0 & \frac{1}{\sqrt{2}}X_1 & 0 \\
  0 & \frac{1}{\sqrt{2}}X_2 & \frac{1}{\sqrt{2}}X_3 \\
  0 & 0 & \frac{1}{\sqrt{2}}X_4 \\
\end{pmatrix}, \quad
Y_u^{ij1} = c_{(6{\text -}6{\text -}12)}
\begin{pmatrix}
  0 & \frac{1}{\sqrt{2}}X_0 & \frac{1}{\sqrt{2}}X_2 \\
  X_1 & 0 & 0 \\
  0 & \frac{1}{\sqrt{2}}X_3 & \frac{1}{\sqrt{2}}X_5 \\
\end{pmatrix},
\notag \\
&Y_u^{ij2} = c_{(6{\text -}6{\text -}12)}
\begin{pmatrix}
  0 & 0 & \frac{1}{\sqrt{2}}X_1 \\
  0 & \frac{1}{\sqrt{2}}X_0 & -\frac{1}{\sqrt{2}}X_5 \\
  X_2 & \frac{1}{\sqrt{2}}X_4 & 0 \\
\end{pmatrix},
\quad
Y_u^{ij3} = c_{(6{\text -}6{\text -}12)}
\begin{pmatrix}
  0 & 0 & -\frac{1}{\sqrt{2}}X_4 \\
  0 & \frac{1}{\sqrt{2}}X_5 & \frac{1}{\sqrt{2}}X_0 \\
  X_3 & \frac{1}{\sqrt{2}}X_1 & 0 \\
\end{pmatrix},
\notag \\
&Y_u^{ij4} = c_{(6{\text -}6{\text -}12)}
\begin{pmatrix}
  0 & -\frac{1}{\sqrt{2}}X_5 & -\frac{1}{\sqrt{2}}X_3 \\
  X_4 & 0 & 0 \\
  0 & \frac{1}{\sqrt{2}}X_2 & \frac{1}{\sqrt{2}}X_0 \\
\end{pmatrix},
\quad
Y_u^{ij5} = c_{(6{\text -}6{\text -}12)}
\begin{pmatrix}
  X_5 & -\frac{1}{\sqrt{2}}X_4 & 0 \\
  0 & \frac{1}{\sqrt{2}}X_3 & -\frac{1}{\sqrt{2}}X_2 \\
  0 & 0 & \frac{1}{\sqrt{2}}X_1 \\
\end{pmatrix},
\notag
\end{align}
where
\begin{align}
  X_N \equiv \sum_{n=0}^5 (-1)^n \eta_{6(N+1/2)+72n}, \quad
  \eta_N \equiv \vartheta
  \begin{bmatrix}
    \frac{N}{432} \\ 0 \\
  \end{bmatrix}
  (0,432\tau).
\end{align}

%-----------------------------------------------------

\subsubsection{Down quark: $B_Q{\text -}B_{d_R}{\text -}B_{H_d}=(6,0,\frac{1}{2},0){\text -}(6,1,0,\frac{1}{2}){\text -}(12,1,\frac{1}{2},\frac{1}{2})$}

The following table shows the zero-mode assignments for quark doublets $Q^i$, right-handed down-sector quarks $d_R^j$ and down type Higgs fields $H_d^k$.
\begin{table}[H]
\begin{center}
\renewcommand{\arraystretch}{1.2}
\begin{tabular}{c|c|c|c}
& $Q^i$ & $d_R^j$ & $H^k_d$ \\ \hline
0 & $\frac{1}{\sqrt{2}}(\psi_{T^2}^{(1/2,0),6}+\psi_{T^2}^{(11/2,0),6})$ & $\frac{1}{\sqrt{2}}(\psi_{T^2}^{(1,1/2),6}+\psi_{T^2}^{(5,1/2),6})$ & $\frac{1}{\sqrt{2}}(\psi_{T^2}^{(1/2,1/2),12}+\psi_{T^2}^{(23/2,1/2),12})$ \\
1 & $\frac{1}{\sqrt{2}}(\psi_{T^2}^{(3/2,0),6}+\psi_{T^2}^{(9/2,0),6})$ & $\frac{1}{\sqrt{2}}(\psi_{T^2}^{(2,1/2),6}+\psi_{T^2}^{(4,1/2),6})$ & $\frac{1}{\sqrt{2}}(\psi_{T^2}^{(3/2,1/2),12}+\psi_{T^2}^{(21/2,1/2),12})$ \\
2 & $\frac{1}{\sqrt{2}}(\psi_{T^2}^{(5/2,0),6}+\psi_{T^2}^{(7/2,0),6})$ & $\psi_{T^2}^{(3,1/2),6}$ & $\frac{1}{\sqrt{2}}(\psi_{T^2}^{(5/2,1/2),12}+\psi_{T^2}^{(19/2,1/2),12})$ \\
3 & & & $\frac{1}{\sqrt{2}}(\psi_{T^2}^{(7/2,1/2),12}+\psi_{T^2}^{(17/2,1/2),12})$ \\
4 & & & $\frac{1}{\sqrt{2}}(\psi_{T^2}^{(9/2,1/2),12}+\psi_{T^2}^{(15/2,1/2),12})$ \\
5 & & & $\frac{1}{\sqrt{2}}(\psi_{T^2}^{(11/2,1/2),12}+\psi_{T^2}^{(13/2,1/2),12})$ \\
\end{tabular}
\end{center}
%\caption{Zero-mode wave functions in ``(6,e;0,1/2)-(6,e;1/2,0)-(12,e;1/2,1/2)'' model.}
%\label{tab:three-model-DownQ}
\end{table}
Yukawa couplings are given by
\begin{align}
  Y^{ijk}_{d}H_d^k = Y^{ij0}_{d}H_d^0+Y^{ij1}_{d}H_d^1+Y^{ij2}_{d}H_d^2+Y^{ij3}_{d}H_d^3+Y^{ij4}_{d}H_d^4+Y^{ij5}_{d}H_d^5,
\end{align}
with
\begin{align}
&Y_{d}^{ij0} = c_{(6{\text -}6{\text -}12)}
\begin{pmatrix}
\frac{1}{\sqrt{2}}X_1 & 0 & 0 \\
-\frac{1}{\sqrt{2}}X_2 & \frac{1}{\sqrt{2}}X_3 & 0 \\
0 & -\frac{1}{\sqrt{2}}X_4 & X_5 \\
\end{pmatrix},
&Y_{d}^{ij1} = c_{(6{\text -}6{\text -}12)}
\begin{pmatrix}
\frac{1}{\sqrt{2}}X_0 & \frac{1}{\sqrt{2}}X_2 & 0 \\
0 & 0 & X_4 \\
-\frac{1}{\sqrt{2}}X_3 & -\frac{1}{\sqrt{2}}X_5 & 0 \\
\end{pmatrix},
\notag \\
&Y_{d}^{ij2} = c_{(6{\text -}6{\text -}12)}
\begin{pmatrix}
0 & \frac{1}{\sqrt{2}}X_1 & X_3 \\
\frac{1}{\sqrt{2}}X_0 & \frac{1}{\sqrt{2}}X_5 & 0 \\
-\frac{1}{\sqrt{2}}X_4 & 0 & 0 \\
\end{pmatrix},
&Y_{d}^{ij3} = c_{(6{\text -}6{\text -}12)}
\begin{pmatrix}
0 & \frac{1}{\sqrt{2}}X_4 & X_2 \\
-\frac{1}{\sqrt{2}}X_5 & \frac{1}{\sqrt{2}}X_0 & 0 \\
\frac{1}{\sqrt{2}}X_1 & 0 & 0 \\
\end{pmatrix},
\notag \\
&Y_{d}^{ij4} = c_{(6{\text -}6{\text -}12)}
\begin{pmatrix}
\frac{1}{\sqrt{2}}X_5 & \frac{1}{\sqrt{2}}X_3 & 0 \\
0 & 0 & X_1 \\
\frac{1}{\sqrt{2}}X_2 & \frac{1}{\sqrt{2}}X_0 & 0 \\
\end{pmatrix},
&Y_{d}^{ij5} = c_{(6{\text -}6{\text -}12)}
\begin{pmatrix}
\frac{1}{\sqrt{2}}X_4 & 0 & 0 \\
\frac{1}{\sqrt{2}}X_3 & \frac{1}{\sqrt{2}}X_2 & 0 \\
0 & \frac{1}{\sqrt{2}}X_1 & X_0 \\
\end{pmatrix},
\notag
\end{align}
where
\begin{align}
  X_N \equiv \sum_{n=0}^5 (-1)^n \eta_{6(N+1/2)+72n}, \quad
  \eta_N \equiv \vartheta
  \begin{bmatrix}
    \frac{N}{432} \\ 0 \\
  \end{bmatrix}
  (0,432\tau).
\end{align}

%-----------------------------------------------------

\subsubsection{Neutrino: $B_L{\text -}B_{\nu_R}{\text -}B_{H_u}=(6,1,0,\frac{1}{2}){\text -}(6,1,\frac{1}{2},0){\text -}(12,0,\frac{1}{2},\frac{1}{2})$}

The following table shows the zero-mode assignments for lepton doublets $L^i$, right-handed neutrinos $\nu_R^j$ and up type Higgs fields $H_u^k$.
\begin{table}[H]
\begin{center}
\renewcommand{\arraystretch}{1.2}
\begin{tabular}{c|c|c|c}
& $L^i$ & $\nu_R^j$ & $H^k_u$ \\ \hline
0 & $\frac{1}{\sqrt{2}}(\psi_{T^2}^{(1,1/2),6}+\psi_{T^2}^{(5,1/2),6})$ & $\frac{1}{\sqrt{2}}(\psi_{T^2}^{(1/2,0),6}-\psi_{T^2}^{(11/2,0),6})$ & $\frac{1}{\sqrt{2}}(\psi_{T^2}^{(1/2,1/2),12}-\psi_{T^2}^{(23/2,1/2),12})$ \\
1 & $\frac{1}{\sqrt{2}}(\psi_{T^2}^{(2,1/2),6}+\psi_{T^2}^{(4,1/2),6})$ & $\frac{1}{\sqrt{2}}(\psi_{T^2}^{(3/2,0),6}-\psi_{T^2}^{(9/2,0),6})$ & $\frac{1}{\sqrt{2}}(\psi_{T^2}^{(3/2,1/2),12}-\psi_{T^2}^{(21/2,1/2),12})$ \\
2 & $\psi_{T^2}^{(3,1/2),6}$ & $\frac{1}{\sqrt{2}}(\psi_{T^2}^{(5/2,0),6}-\psi_{T^2}^{(7/2,0),6})$ & $\frac{1}{\sqrt{2}}(\psi_{T^2}^{(5/2,1/2),12}-\psi_{T^2}^{(19/2,1/2),12})$ \\
3 & & & $\frac{1}{\sqrt{2}}(\psi_{T^2}^{(7/2,1/2),12}-\psi_{T^2}^{(17/2,1/2),12})$ \\
4 & & & $\frac{1}{\sqrt{2}}(\psi_{T^2}^{(9/2,1/2),12}-\psi_{T^2}^{(15/2,1/2),12})$ \\
5 & & & $\frac{1}{\sqrt{2}}(\psi_{T^2}^{(11/2,1/2),12}-\psi_{T^2}^{(13/2,1/2),12})$ \\
\end{tabular}
\end{center}
%\caption{Zero-mode wave functions in ``(6,e;1/2,0)-(5,e;1/2,0)-(11,e;0,0)'' model.}
%\label{tab:three-model-Neutrino}
\end{table}
Yukawa couplings are given by
\begin{align}
  Y^{ijk}_\nu H_u^k = Y^{ij0}_\nu H_u^0+Y^{ij1}_\nu H_u^1+Y^{ij2}_\nu H_u^2+Y^{ij3}_\nu H_u^3+Y^{ij4}_\nu H_u^4+Y^{ij5}_\nu H_u^5,
\end{align}
with
\begin{align}
&Y_\nu^{ij0} = c_{(6{\text -}6{\text -}12)}
\begin{pmatrix}
-\frac{1}{\sqrt{2}}X_1 & -\frac{1}{\sqrt{2}}X_2 & 0 \\
0 & -\frac{1}{\sqrt{2}}X_3 & -\frac{1}{\sqrt{2}}X_4 \\
0 & 0 & -X_5 \\
\end{pmatrix},
&Y_\nu^{ij1} = c_{(6{\text -}6{\text -}12)}
\begin{pmatrix}
\frac{1}{\sqrt{2}}X_0 & 0 & -\frac{1}{\sqrt{2}}X_3 \\
-\frac{1}{\sqrt{2}}X_2 & 0 & \frac{1}{\sqrt{2}}X_5 \\
0 & -X_4 & 0 \\
\end{pmatrix},
\notag \\
&Y_\nu^{ij2} = c_{(6{\text -}6{\text -}12)}
\begin{pmatrix}
0 & \frac{1}{\sqrt{2}}X_0 & \frac{1}{\sqrt{2}}X_4 \\
\frac{1}{\sqrt{2}}X_1 & -\frac{1}{\sqrt{2}}X_5 & 0 \\
-X_3 & 0 & 0 \\
\end{pmatrix},
&Y_\nu^{ij3} = c_{(6{\text -}6{\text -}12)}
\begin{pmatrix}
0 & \frac{1}{\sqrt{2}}X_5 & \frac{1}{\sqrt{2}}X_1 \\
-\frac{1}{\sqrt{2}}X_4 & \frac{1}{\sqrt{2}}X_0 & 0 \\
X_2 & 0 & 0 \\
\end{pmatrix},
\notag \\
&Y_\nu^{ij4} = c_{(6{\text -}6{\text -}12)}
\begin{pmatrix}
-\frac{1}{\sqrt{2}}X_5 & 0 & -\frac{1}{\sqrt{2}}X_2 \\
\frac{1}{\sqrt{2}}X_3 & 0 & \frac{1}{\sqrt{2}}X_0 \\
0 & X_1 & 0 \\
\end{pmatrix},
&Y_\nu^{ij5} = c_{(6{\text -}6{\text -}12)}
\begin{pmatrix}
\frac{1}{\sqrt{2}}X_4 & -\frac{1}{\sqrt{2}}X_3 & 0 \\
0 & \frac{1}{\sqrt{2}}X_2 & -\frac{1}{\sqrt{2}}X_1 \\
0 & 0 & X_0 \\
\end{pmatrix},
\notag
\end{align}
where
\begin{align}
  X_N \equiv \sum_{n=0}^5 (-1)^n \eta_{6(N+1/2)+72n}, \quad
  \eta_N \equiv \vartheta
  \begin{bmatrix}
    \frac{N}{432} \\ 0 \\
  \end{bmatrix}
  (0,432\tau).
\end{align}

%-----------------------------------------------------

\subsubsection{Charged lepton: $B_L{\text -}B_{e_R}{\text -}B_{H_d}=(6,1,0,\frac{1}{2}){\text -}(6,0,\frac{1}{2},0){\text -}(12,1,\frac{1}{2},\frac{1}{2})$}

The following table shows the zero-mode assignments for lepton doublets $L^i$, right-handed charged leptons $e_R^j$ and down type Higgs fields $H_d^k$.
\begin{table}[H]
\begin{center}
\renewcommand{\arraystretch}{1.2}
\begin{tabular}{c|c|c|c}
& $L^i$ & $e_R^j$ & $H^k_d$ \\ \hline
0 & $\frac{1}{\sqrt{2}}(\psi_{T^2}^{(1,1/2),6}+\psi_{T^2}^{(5,1/2),6})$ & $\frac{1}{\sqrt{2}}(\psi_{T^2}^{(1/2,0),6}+\psi_{T^2}^{(11/2,0),6})$ & $\frac{1}{\sqrt{2}}(\psi_{T^2}^{(1/2,1/2),12}+\psi_{T^2}^{(23/2,1/2),12})$ \\
1 & $\frac{1}{\sqrt{2}}(\psi_{T^2}^{(2,1/2),6}+\psi_{T^2}^{(4,1/2),6})$ & $\frac{1}{\sqrt{2}}(\psi_{T^2}^{(3/2,0),6}+\psi_{T^2}^{(9/2,0),6})$ & $\frac{1}{\sqrt{2}}(\psi_{T^2}^{(3/2,1/2),12}+\psi_{T^2}^{(21/2,1/2),12})$ \\
2 & $\psi_{T^2}^{(3,1/2),6}$ & $\frac{1}{\sqrt{2}}(\psi_{T^2}^{(5/2,0),6}+\psi_{T^2}^{(7/2,0),6})$ & $\frac{1}{\sqrt{2}}(\psi_{T^2}^{(5/2,1/2),12}+\psi_{T^2}^{(19/2,1/2),12})$ \\
3 & & & $\frac{1}{\sqrt{2}}(\psi_{T^2}^{(7/2,1/2),12}+\psi_{T^2}^{(17/2,1/2),12})$ \\
4 & & & $\frac{1}{\sqrt{2}}(\psi_{T^2}^{(9/2,1/2),12}+\psi_{T^2}^{(15/2,1/2),12})$ \\
5 & & & $\frac{1}{\sqrt{2}}(\psi_{T^2}^{(11/2,1/2),12}+\psi_{T^2}^{(13/2,1/2),12})$ \\
\end{tabular}
\end{center}
%\caption{Zero-mode wave functions in ``(6,e;1/2,0)-(6,e;0,1/2)-(12,e;1/2,1/2)'' model.}
%\label{tab:three-model-ChargedL}
\end{table}
Yukawa couplings are given by
\begin{align}
  Y^{ijk}_{e} = Y^{jik}_{d}.
\end{align}

%-----------------------------------------------------
%-----------------------------------------------------

\subsection{Majorana mass of right-handed neutrino}
\label{App:Majorana}

Majorana mass of right-handed neutrinos can be induced by D-brane instanton effects as shown in Appendix \ref{App:Neutrino_Majorana_mass_term}.
For the right-handed neutrinos in our model, there are two possible instanton zero-mode configurations, $\beta_1,\gamma_1$ and $\beta_2,\gamma_2$,
\begin{align}
  B_{\beta_1} {\text -} B_{\gamma_1} {\text -} B_{\nu_R} = (3,0,0,\textstyle\frac{1}{2}) {\text -} (3,1,\textstyle\frac{1}{2},\textstyle\frac{1}{2}) {\text -} (6,1,\textstyle\frac{1}{2},0), \quad 
  B_{\beta_2} {\text -} B_{\gamma_2} {\text -} B_{\nu_R} = (2,0,0,0) {\text -} (4,1,\textstyle\frac{1}{2},0) {\text -} (6,1,\textstyle\frac{1}{2},0).
\end{align}
However, these two configurations give same Majorana mass matrix up to overall factor; therefore we concentrate on the former configuration, $\beta_1,\gamma_1$.
The following table shows the zero-mode assignments for two zero-modes of them.
\begin{table}[H]
\begin{center}
\renewcommand{\arraystretch}{1.2}
\begin{tabular}{c|c|c|c}
& $\beta_1^i$ & $\gamma_1^j$ & $\nu_R^a$ \\ \hline
0 & $\psi_{T^2}^{(0,1/2),3}$ & $\frac{1}{\sqrt{2}}(\psi_{T^2}^{(1/2,1/2),3}+\psi_{T^2}^{(5/2,1/2),3})$ & $\frac{1}{\sqrt{2}}(\psi_{T^2}^{(1/2,0),6}-\psi_{T^2}^{(11/2,0),6})$ \\
1 & $\frac{1}{\sqrt{2}}(\psi_{T^2}^{(1,1/2),3}-\psi_{T^2}^{(2,1/2),3})$ & $\psi_{T^2}^{(3/2,1/2),3}$ & $\frac{1}{\sqrt{2}}(\psi_{T^2}^{(3/2,0),6}-\psi_{T^2}^{(9/2,0),6})$ \\
2 & & & $\frac{1}{\sqrt{2}}(\psi_{T^2}^{(5/2,0),6}-\psi_{T^2}^{(7/2,0),6})$ \\
\end{tabular}
\end{center}
\end{table}
The 3-point couplings $d_a^{ij}$ are given by
\begin{align}
&d_0^{ij} = c_{(3{\text -}3{\text -}6)}
\begin{pmatrix}
  \eta_{1.5}+\eta_{16.5}+\eta_{19.5} & 0 \\
  -\frac{1}{\sqrt{2}}(\eta_{4.5}+\eta_{13.5}+\eta_{22.5}) & \eta_{7.5}+\eta_{10.5}+\eta_{25.5} \\
\end{pmatrix}, \\
&d_1^{ij} = c_{(3{\text -}3{\text -}6)}
\begin{pmatrix}
  0 & \sqrt{2}(\eta_{4.5}+\eta_{13.5}+\eta_{22.5}) \\
  \frac{1}{\sqrt{2}}(\eta_{1.5}+\eta_{16.5}+\eta_{19.5}+\eta_{7.5}+\eta_{10.5}+\eta_{25.5}) & 0 \\
\end{pmatrix}, \\
&d_2^{ij} = c_{(3{\text -}3{\text -}6)}
\begin{pmatrix}
  \eta_{7.5}+\eta_{10.5}+\eta_{25.5} & 0 \\
  -\frac{1}{\sqrt{2}}(\eta_{4.5}+\eta_{13.5}+\eta_{22.5}) & \eta_{1.5}+\eta_{16.5}+\eta_{19.5} \\
\end{pmatrix},
\end{align}
where
\begin{align}
  \eta_N \equiv \vartheta
  \begin{bmatrix}
    \frac{N}{54} \\ 0 \\
  \end{bmatrix}
  (0,54\tau).
\end{align}
Using above $d_a^{ij}$, Majorana masses of right-handed neutrinos can be calculated by Eq.~(\ref{eq:m_RR}).

%-----------------------------------------------------
%-----------------------------------------------------
%-----------------------------------------------------

\clearpage
\section{Quark and lepton flavor models satisfying the conditions I, II, III and IV}

\label{apx:CLASSMODELS}

\begin{table}[H]
  \caption{Quark and lepton flavor models satisfying the conditions for quark and lepton flavors.
  The first to eighth rows show the flux, $\mathbb{Z}_2$ parity (even, odd = 0, 1) and SS phases $(\alpha_1,\alpha_2)$ of the zero-modes of the fields.
Those are denoted in the same was as Table \ref{tab:classification_S_eig}.
  $g_H$ denotes number of Higgs fields and $p$ denotes the fixed points satisfying Eq.~(\ref{eq:consistent_p}).}
  \label{tab:classification}
\begin{align}
\begin{array}{c|c|c|c|c|c|c|c|c|c} \hline
    B_Q & B_{u_R} & B_{d_R} & B_L & B_{\nu_R} & B_{e_R} & B_{H_u} & B_{H_d} & g_H & p \\ \hline \hline
    5,0,0,0 & 6,0,\frac{1}{2},0 & 6,0,0,\frac{1}{2} & 6,0,\frac{1}{2},\frac{1}{2} & 5,0,0,\frac{1}{2} & 5,0,\frac{1}{2},0 & 11,0,\frac{1}{2},0 & 11,0,0,\frac{1}{2} & 6 & \frac{\tau}{2} \\
5,0,0,0 & 6,0,\frac{1}{2},0 & 6,1,\frac{1}{2},\frac{1}{2} & 6,1,0,\frac{1}{2} & 5,1,\frac{1}{2},\frac{1}{2} & 5,0,\frac{1}{2},0 & 11,0,\frac{1}{2},0 & 11,1,\frac{1}{2},\frac{1}{2} & 6 & 0 \\
5,0,0,0 & 6,0,\frac{1}{2},0 & 7,1,\frac{1}{2},0 & 6,0,\frac{1}{2},\frac{1}{2} & 5,0,0,\frac{1}{2} & 6,1,0,\frac{1}{2} & 11,0,\frac{1}{2},0 & 12,1,\frac{1}{2},0 & 6 & \frac{\tau}{2} \\
5,0,0,0 & 6,0,\frac{1}{2},0 & 7,1,\frac{1}{2},0 & 6,1,0,\frac{1}{2} & 5,1,\frac{1}{2},\frac{1}{2} & 6,0,\frac{1}{2},\frac{1}{2} & 11,0,\frac{1}{2},0 & 12,1,\frac{1}{2},0 & 6 & 0 \\
5,0,0,0 & 6,0,0,\frac{1}{2} & 6,0,\frac{1}{2},0 & 6,0,\frac{1}{2},\frac{1}{2} & 5,0,\frac{1}{2},0 & 5,0,0,\frac{1}{2} & 11,0,0,\frac{1}{2} & 11,0,\frac{1}{2},0 & 6 & \frac{1}{2} \\
5,0,0,0 & 6,0,0,\frac{1}{2} & 6,1,\frac{1}{2},\frac{1}{2} & 6,1,\frac{1}{2},0 & 5,1,\frac{1}{2},\frac{1}{2} & 5,0,0,\frac{1}{2} & 11,0,0,\frac{1}{2} & 11,1,\frac{1}{2},\frac{1}{2} & 6 & 0 \\
5,0,0,0 & 6,0,0,\frac{1}{2} & 7,1,0,\frac{1}{2} & 6,0,\frac{1}{2},\frac{1}{2} & 5,0,\frac{1}{2},0 & 6,1,\frac{1}{2},0 & 11,0,0,\frac{1}{2} & 12,1,0,\frac{1}{2} & 6 & \frac{1}{2} \\
\hline
  \end{array} \notag
\end{align}
\end{table}
\begin{table}[h]
\begin{align}
  \begin{array}{c|c|c|c|c|c|c|c|c|c} \hline
    B_Q & B_{u_R} & B_{d_R} & B_L & B_{\nu_R} & B_{e_R} & B_{H_u} & B_{H_d} & g_H & p \\ \hline \hline
5,0,0,0 & 7,1,0,\frac{1}{2} & 7,1,\frac{1}{2},0 & 6,0,\frac{1}{2},\frac{1}{2} & 6,1,\frac{1}{2},0 & 6,1,0,\frac{1}{2} & 12,1,0,\frac{1}{2} & 12,1,\frac{1}{2},0 & 6 & \frac{\tau}{2} \\
5,0,0,0 & 7,1,0,\frac{1}{2} & 7,1,\frac{1}{2},0 & 7,1,0,0 & 5,0,0,\frac{1}{2} & 5,0,\frac{1}{2},0 & 12,1,0,\frac{1}{2} & 12,1,\frac{1}{2},0 & 6 & \frac{\tau}{2} \\
5,0,\frac{1}{2},0 & 6,0,\frac{1}{2},0 & 6,0,\frac{1}{2},\frac{1}{2} & 6,0,0,\frac{1}{2} & 5,0,0,\frac{1}{2} & 5,0,0,0 & 11,0,0,0 & 11,0,0,\frac{1}{2} & 6 & \frac{\tau}{2} \\
5,0,\frac{1}{2},0 & 6,0,\frac{1}{2},0 & 6,1,0,\frac{1}{2} & 6,1,\frac{1}{2},\frac{1}{2} & 5,1,\frac{1}{2},\frac{1}{2} & 5,0,0,0 & 11,0,0,0 & 11,1,\frac{1}{2},\frac{1}{2} & 6 & 0 \\
5,0,\frac{1}{2},0 & 6,0,\frac{1}{2},0 & 7,1,0,0 & 6,0,0,\frac{1}{2} & 5,0,0,\frac{1}{2} & 6,1,\frac{1}{2},\frac{1}{2} & 11,0,0,0 & 12,1,\frac{1}{2},0 & 6 & \frac{\tau}{2} \\
5,0,\frac{1}{2},0 & 6,0,\frac{1}{2},0 & 7,1,0,0 & 6,1,\frac{1}{2},\frac{1}{2} & 5,1,\frac{1}{2},\frac{1}{2} & 6,0,0,\frac{1}{2} & 11,0,0,0 & 12,1,\frac{1}{2},0 & 6 & 0 \\
5,0,\frac{1}{2},0 & 6,0,\frac{1}{2},\frac{1}{2} & 6,0,\frac{1}{2},0 & 6,0,0,\frac{1}{2} & 5,0,0,0 & 5,0,0,\frac{1}{2} & 11,0,0,\frac{1}{2} & 11,0,0,0 & 6 & \frac{1+\tau}{2} \\
\hline
  \end{array} \notag
\end{align}
\end{table}
\begin{table}[h]
\begin{align}
  \begin{array}{c|c|c|c|c|c|c|c|c|c} \hline
    B_Q & B_{u_R} & B_{d_R} & B_L & B_{\nu_R} & B_{e_R} & B_{H_u} & B_{H_d} & g_H & p \\ \hline \hline
5,0,0,\frac{1}{2} & 6,0,\frac{1}{2},\frac{1}{2} & 7,1,\frac{1}{2},0 & 6,0,\frac{1}{2},0 & 5,0,0,0 & 6,1,0,\frac{1}{2} & 11,0,\frac{1}{2},0 & 12,1,\frac{1}{2},\frac{1}{2} & 6 & \frac{1+\tau}{2} \\
5,0,0,\frac{1}{2} & 6,0,\frac{1}{2},\frac{1}{2} & 7,1,\frac{1}{2},0 & 6,1,0,\frac{1}{2} & 5,1,\frac{1}{2},\frac{1}{2} & 6,0,\frac{1}{2},0 & 11,0,\frac{1}{2},0 & 12,1,\frac{1}{2},\frac{1}{2} & 6 & 0 \\
5,0,0,\frac{1}{2} & 6,1,\frac{1}{2},0 & 6,0,0,\frac{1}{2} & 6,1,\frac{1}{2},\frac{1}{2} & 5,0,0,0 & 5,1,\frac{1}{2},\frac{1}{2} & 11,1,\frac{1}{2},\frac{1}{2} & 11,0,0,0 & 6 & \frac{1+\tau}{2} \\
5,0,0,\frac{1}{2} & 6,1,\frac{1}{2},0 & 6,0,\frac{1}{2},\frac{1}{2} & 6,1,0,\frac{1}{2} & 5,0,\frac{1}{2},0 & 5,1,\frac{1}{2},\frac{1}{2} & 11,1,\frac{1}{2},\frac{1}{2} & 11,0,\frac{1}{2},0 & 6 & \frac{1}{2} \\
5,0,0,\frac{1}{2} & 6,1,\frac{1}{2},0 & 7,0,\frac{1}{2},\frac{1}{2} & 6,1,0,\frac{1}{2} & 5,0,\frac{1}{2},0 & 6,1,\frac{1}{2},\frac{1}{2} & 11,1,\frac{1}{2},\frac{1}{2} & 12,0,\frac{1}{2},0 & 6 & \frac{1}{2} \\
5,0,0,\frac{1}{2} & 6,1,\frac{1}{2},0 & 7,0,\frac{1}{2},\frac{1}{2} & 6,1,\frac{1}{2},\frac{1}{2} & 5,0,0,0 & 6,1,0,\frac{1}{2} & 11,1,\frac{1}{2},\frac{1}{2} & 12,0,\frac{1}{2},0 & 6 & \frac{1+\tau}{2} \\
5,0,0,\frac{1}{2} & 7,0,\frac{1}{2},\frac{1}{2} & 6,1,\frac{1}{2},0 & 6,1,0,\frac{1}{2} & 6,1,\frac{1}{2},\frac{1}{2} & 5,0,\frac{1}{2},0 & 12,0,\frac{1}{2},0 & 11,1,\frac{1}{2},\frac{1}{2} & 6 & 0 \\
\hline
  \end{array} \notag
\end{align}
\end{table}
\begin{table}[h]
\begin{align}
  \begin{array}{c|c|c|c|c|c|c|c|c|c} \hline
    B_Q & B_{u_R} & B_{d_R} & B_L & B_{\nu_R} & B_{e_R} & B_{H_u} & B_{H_d} & g_H & p \\ \hline \hline
5,1,\frac{1}{2},\frac{1}{2} & 7,1,\frac{1}{2},0 & 7,1,0,\frac{1}{2} & 6,0,\frac{1}{2},\frac{1}{2} & 6,0,\frac{1}{2},0 & 6,0,0,\frac{1}{2} & 12,0,0,\frac{1}{2} & 12,0,\frac{1}{2},0 & 6 & \frac{1}{2} \\
5,1,\frac{1}{2},\frac{1}{2} & 7,1,\frac{1}{2},0 & 7,1,0,\frac{1}{2} & 7,0,\frac{1}{2},\frac{1}{2} & 5,0,\frac{1}{2},0 & 5,0,0,\frac{1}{2} & 12,0,0,\frac{1}{2} & 12,0,\frac{1}{2},0 & 6 & \frac{1}{2} \\
5,1,\frac{1}{2},\frac{1}{2} & 7,1,\frac{1}{2},0 & 7,1,0,0 & 6,0,\frac{1}{2},0 & 6,0,\frac{1}{2},\frac{1}{2} & 6,0,0,\frac{1}{2} & 12,0,0,\frac{1}{2} & 12,0,\frac{1}{2},\frac{1}{2} & 6 & \frac{1}{2} \\
5,1,\frac{1}{2},\frac{1}{2} & 7,1,\frac{1}{2},0 & 7,1,0,0 & 7,0,\frac{1}{2},\frac{1}{2} & 5,0,\frac{1}{2},0 & 5,0,0,0 & 12,0,0,\frac{1}{2} & 12,0,\frac{1}{2},\frac{1}{2} & 6 & \frac{1}{2} \\
5,1,\frac{1}{2},\frac{1}{2} & 7,1,0,\frac{1}{2} & 6,1,\frac{1}{2},0 & 6,0,0,\frac{1}{2} & 6,0,\frac{1}{2},\frac{1}{2} & 5,0,0,0 & 12,0,\frac{1}{2},0 & 11,0,0,\frac{1}{2} & 6 & \frac{\tau}{2} \\
5,1,\frac{1}{2},\frac{1}{2} & 7,1,0,\frac{1}{2} & 6,1,\frac{1}{2},0 & 6,0,\frac{1}{2},\frac{1}{2} & 6,0,0,\frac{1}{2} & 5,0,\frac{1}{2},0 & 12,0,\frac{1}{2},0 & 11,0,0,\frac{1}{2} & 6 & \frac{\tau}{2} \\
5,1,\frac{1}{2},\frac{1}{2} & 7,1,0,\frac{1}{2} & 7,1,\frac{1}{2},0 & 6,0,\frac{1}{2},\frac{1}{2} & 6,0,0,\frac{1}{2} & 6,0,\frac{1}{2},0 & 12,0,\frac{1}{2},0 & 12,0,0,\frac{1}{2} & 6 & \frac{\tau}{2} \\
\hline
  \end{array} \notag
\end{align}
\end{table}
\begin{table}[h]
\begin{align}
  \begin{array}{c|c|c|c|c|c|c|c|c|c} \hline
    B_Q & B_{u_R} & B_{d_R} & B_L & B_{\nu_R} & B_{e_R} & B_{H_u} & B_{H_d} & g_H & p \\ \hline \hline
6,0,\frac{1}{2},0 & 6,1,0,\frac{1}{2} & 6,0,\frac{1}{2},\frac{1}{2} & 6,1,\frac{1}{2},0 & 6,0,0,\frac{1}{2} & 6,1,\frac{1}{2},\frac{1}{2} & 12,1,\frac{1}{2},\frac{1}{2} & 12,0,0,\frac{1}{2} & 6 & \frac{\tau}{2} \\
6,0,\frac{1}{2},0 & 6,1,0,\frac{1}{2} & 6,0,\frac{1}{2},\frac{1}{2} & 7,1,\frac{1}{2},0 & 5,0,0,\frac{1}{2} & 5,1,\frac{1}{2},\frac{1}{2} & 12,1,\frac{1}{2},\frac{1}{2} & 12,0,0,\frac{1}{2} & 6 & \frac{\tau}{2} \\
6,0,\frac{1}{2},0 & 6,1,0,\frac{1}{2} & 7,1,\frac{1}{2},0 & 6,0,0,\frac{1}{2} & 6,1,\frac{1}{2},0 & 7,1,0,\frac{1}{2} & 12,1,\frac{1}{2},\frac{1}{2} & 13,1,0,0 & 6 & \frac{\tau}{2} \\
6,0,\frac{1}{2},0 & 6,1,0,\frac{1}{2} & 7,1,\frac{1}{2},0 & 7,1,\frac{1}{2},0 & 5,0,0,\frac{1}{2} & 6,0,\frac{1}{2},0 & 12,1,\frac{1}{2},\frac{1}{2} & 13,1,0,0 & 6 & \frac{\tau}{2} \\
6,0,\frac{1}{2},0 & 6,1,\frac{1}{2},\frac{1}{2} & 6,0,0,\frac{1}{2} & 6,1,\frac{1}{2},0 & 6,0,\frac{1}{2},\frac{1}{2} & 6,1,0,\frac{1}{2} & 12,1,0,\frac{1}{2} & 12,0,\frac{1}{2},\frac{1}{2} & 6 & \frac{\tau}{2} \\
6,0,\frac{1}{2},0 & 6,1,\frac{1}{2},\frac{1}{2} & 6,0,0,\frac{1}{2} & 7,1,0,0 & 5,0,0,\frac{1}{2} & 5,1,\frac{1}{2},\frac{1}{2} & 12,1,0,\frac{1}{2} & 12,0,\frac{1}{2},\frac{1}{2} & 6 & \frac{\tau}{2} \\
6,0,\frac{1}{2},0 & 6,1,\frac{1}{2},\frac{1}{2} & 6,0,\frac{1}{2},\frac{1}{2} & 6,0,\frac{1}{2},\frac{1}{2} & 6,1,\frac{1}{2},0 & 6,0,\frac{1}{2},0 & 12,1,0,\frac{1}{2} & 12,0,0,\frac{1}{2} & 6 & \frac{\tau}{2} \\
\hline
  \end{array} \notag
\end{align}
\end{table}
\begin{table}[h]
\begin{align}
  \begin{array}{c|c|c|c|c|c|c|c|c|c} \hline
    B_Q & B_{u_R} & B_{d_R} & B_L & B_{\nu_R} & B_{e_R} & B_{H_u} & B_{H_d} & g_H & p \\ \hline \hline
6,0,0,\frac{1}{2} & 6,1,\frac{1}{2},0 & 7,1,0,\frac{1}{2} & 6,0,\frac{1}{2},0 & 6,1,0,\frac{1}{2} & 7,1,\frac{1}{2},0 & 12,1,\frac{1}{2},\frac{1}{2} & 13,1,0,0 & 6 & \frac{1}{2} \\
6,0,0,\frac{1}{2} & 6,1,\frac{1}{2},0 & 7,1,0,\frac{1}{2} & 7,1,0,\frac{1}{2} & 5,0,\frac{1}{2},0 & 6,0,0,\frac{1}{2} & 12,1,\frac{1}{2},\frac{1}{2} & 13,1,0,0 & 6 & \frac{1}{2} \\
6,0,0,\frac{1}{2} & 6,1,\frac{1}{2},\frac{1}{2} & 6,0,\frac{1}{2},0 & 6,1,0,\frac{1}{2} & 6,0,\frac{1}{2},\frac{1}{2} & 6,1,\frac{1}{2},0 & 12,1,\frac{1}{2},0 & 12,0,\frac{1}{2},\frac{1}{2} & 6 & \frac{1}{2} \\
6,0,0,\frac{1}{2} & 6,1,\frac{1}{2},\frac{1}{2} & 6,0,\frac{1}{2},0 & 7,1,0,0 & 5,0,\frac{1}{2},0 & 5,1,\frac{1}{2},\frac{1}{2} & 12,1,\frac{1}{2},0 & 12,0,\frac{1}{2},\frac{1}{2} & 6 & \frac{1}{2} \\
6,0,0,\frac{1}{2} & 6,1,\frac{1}{2},\frac{1}{2} & 6,0,\frac{1}{2},\frac{1}{2} & 6,0,\frac{1}{2},\frac{1}{2} & 6,1,0,\frac{1}{2} & 6,0,0,\frac{1}{2} & 12,1,\frac{1}{2},0 & 12,0,\frac{1}{2},0 & 6 & \frac{1}{2} \\
6,0,0,\frac{1}{2} & 6,1,\frac{1}{2},\frac{1}{2} & 6,0,\frac{1}{2},\frac{1}{2} & 6,1,0,\frac{1}{2} & 6,0,\frac{1}{2},\frac{1}{2} & 6,1,\frac{1}{2},\frac{1}{2} & 12,1,\frac{1}{2},0 & 12,0,\frac{1}{2},0 & 6 & \frac{1}{2} \\
6,0,0,\frac{1}{2} & 6,1,\frac{1}{2},\frac{1}{2} & 7,1,0,0 & 6,0,\frac{1}{2},\frac{1}{2} & 6,1,0,\frac{1}{2} & 7,1,\frac{1}{2},0 & 12,1,\frac{1}{2},0 & 13,1,0,\frac{1}{2} & 6 & \frac{1}{2} \\
\hline
  \end{array} \notag
\end{align}
\end{table}
\begin{table}[h]
\begin{align}
  \begin{array}{c|c|c|c|c|c|c|c|c|c} \hline
    B_Q & B_{u_R} & B_{d_R} & B_L & B_{\nu_R} & B_{e_R} & B_{H_u} & B_{H_d} & g_H & p \\ \hline \hline
6,0,\frac{1}{2},\frac{1}{2} & 6,1,0,\frac{1}{2} & 6,0,\frac{1}{2},0 & 6,1,\frac{1}{2},\frac{1}{2} & 6,0,0,\frac{1}{2} & 6,1,\frac{1}{2},0 & 12,1,\frac{1}{2},0 & 12,0,0,\frac{1}{2} & 6 & \frac{1+\tau}{2} \\
6,0,\frac{1}{2},\frac{1}{2} & 6,1,0,\frac{1}{2} & 6,0,\frac{1}{2},0 & 7,1,\frac{1}{2},0 & 5,0,0,0 & 5,1,\frac{1}{2},\frac{1}{2} & 12,1,\frac{1}{2},0 & 12,0,0,\frac{1}{2} & 6 & \frac{1+\tau}{2} \\
6,0,\frac{1}{2},\frac{1}{2} & 6,1,0,\frac{1}{2} & 6,0,0,\frac{1}{2} & 6,0,0,\frac{1}{2} & 6,1,\frac{1}{2},\frac{1}{2} & 6,0,\frac{1}{2},\frac{1}{2} & 12,1,\frac{1}{2},0 & 12,0,\frac{1}{2},0 & 6 & \frac{1+\tau}{2} \\
6,0,\frac{1}{2},\frac{1}{2} & 6,1,0,\frac{1}{2} & 6,0,0,\frac{1}{2} & 6,1,\frac{1}{2},\frac{1}{2} & 6,0,0,\frac{1}{2} & 6,1,0,\frac{1}{2} & 12,1,\frac{1}{2},0 & 12,0,\frac{1}{2},0 & 6 & \frac{1+\tau}{2} \\
6,0,\frac{1}{2},\frac{1}{2} & 6,1,0,\frac{1}{2} & 7,1,\frac{1}{2},0 & 6,0,0,\frac{1}{2} & 6,1,\frac{1}{2},\frac{1}{2} & 7,1,0,0 & 12,1,\frac{1}{2},0 & 13,1,0,\frac{1}{2} & 6 & \frac{1+\tau}{2} \\
6,0,\frac{1}{2},\frac{1}{2} & 6,1,0,\frac{1}{2} & 7,1,\frac{1}{2},0 & 7,1,\frac{1}{2},0 & 5,0,0,0 & 6,0,\frac{1}{2},\frac{1}{2} & 12,1,\frac{1}{2},0 & 13,1,0,\frac{1}{2} & 6 & \frac{1+\tau}{2} \\
6,0,\frac{1}{2},\frac{1}{2} & 7,0,\frac{1}{2},\frac{1}{2} & 7,1,0,0 & 7,1,\frac{1}{2},0 & 6,1,\frac{1}{2},0 & 6,0,0,\frac{1}{2} & 13,0,0,0 & 13,1,\frac{1}{2},\frac{1}{2} & 7 & 0 \\
\hline
  \end{array} \notag
\end{align}
\end{table}
\begin{table}[h]
\begin{align}
  \begin{array}{c|c|c|c|c|c|c|c|c|c} \hline
    B_Q & B_{u_R} & B_{d_R} & B_L & B_{\nu_R} & B_{e_R} & B_{H_u} & B_{H_d} & g_H & p \\ \hline \hline
6,1,\frac{1}{2},0 & 6,1,\frac{1}{2},\frac{1}{2} & 6,1,0,\frac{1}{2} & 6,0,\frac{1}{2},0 & 6,0,\frac{1}{2},\frac{1}{2} & 6,0,0,\frac{1}{2} & 12,0,0,\frac{1}{2} & 12,0,\frac{1}{2},\frac{1}{2} & 6 & \frac{1}{2} \\
6,1,\frac{1}{2},0 & 6,1,\frac{1}{2},\frac{1}{2} & 6,1,0,\frac{1}{2} & 7,0,\frac{1}{2},\frac{1}{2} & 5,0,\frac{1}{2},0 & 5,0,0,0 & 12,0,0,\frac{1}{2} & 12,0,\frac{1}{2},\frac{1}{2} & 6 & \frac{1}{2} \\
6,1,\frac{1}{2},0 & 6,1,\frac{1}{2},\frac{1}{2} & 7,0,\frac{1}{2},\frac{1}{2} & 6,0,\frac{1}{2},\frac{1}{2} & 6,0,\frac{1}{2},0 & 7,1,\frac{1}{2},0 & 12,0,0,\frac{1}{2} & 13,1,0,\frac{1}{2} & 6 & \frac{1}{2} \\
6,1,\frac{1}{2},0 & 6,1,\frac{1}{2},\frac{1}{2} & 7,0,\frac{1}{2},\frac{1}{2} & 7,0,\frac{1}{2},\frac{1}{2} & 5,0,\frac{1}{2},0 & 6,1,\frac{1}{2},0 & 12,0,0,\frac{1}{2} & 13,1,0,\frac{1}{2} & 6 & \frac{1}{2} \\
6,1,\frac{1}{2},0 & 7,1,\frac{1}{2},0 & 7,1,0,0 & 7,0,\frac{1}{2},\frac{1}{2} & 6,0,\frac{1}{2},\frac{1}{2} & 6,0,0,\frac{1}{2} & 13,0,0,0 & 13,0,\frac{1}{2},0 & 7 & \frac{1}{2} \\
6,1,\frac{1}{2},0 & 7,1,\frac{1}{2},0 & 7,1,0,0 & 7,1,0,\frac{1}{2} & 6,1,0,\frac{1}{2} & 6,1,\frac{1}{2},\frac{1}{2} & 13,0,0,0 & 13,0,\frac{1}{2},0 & 7 & \frac{1}{2} \\
6,1,\frac{1}{2},0 & 7,1,0,0 & 7,1,\frac{1}{2},0 & 7,0,\frac{1}{2},\frac{1}{2} & 6,0,0,\frac{1}{2} & 6,0,\frac{1}{2},\frac{1}{2} & 13,0,\frac{1}{2},0 & 13,0,0,0 & 7 & \frac{1+\tau}{2} \\
\hline
  \end{array} \notag
\end{align}
\end{table}
\begin{table}[h]
\begin{align}
  \begin{array}{c|c|c|c|c|c|c|c|c|c} \hline
    B_Q & B_{u_R} & B_{d_R} & B_L & B_{\nu_R} & B_{e_R} & B_{H_u} & B_{H_d} & g_H & p \\ \hline \hline
6,1,0,\frac{1}{2} & 6,1,\frac{1}{2},\frac{1}{2} & 7,0,\frac{1}{2},\frac{1}{2} & 6,0,\frac{1}{2},\frac{1}{2} & 6,0,0,\frac{1}{2} & 7,1,0,\frac{1}{2} & 12,0,\frac{1}{2},0 & 13,1,\frac{1}{2},0 & 6 & \frac{\tau}{2} \\
6,1,0,\frac{1}{2} & 6,1,\frac{1}{2},\frac{1}{2} & 7,0,\frac{1}{2},\frac{1}{2} & 7,0,\frac{1}{2},\frac{1}{2} & 5,0,0,\frac{1}{2} & 6,1,0,\frac{1}{2} & 12,0,\frac{1}{2},0 & 13,1,\frac{1}{2},0 & 6 & \frac{\tau}{2} \\
6,1,0,\frac{1}{2} & 7,1,0,\frac{1}{2} & 7,1,0,0 & 7,0,\frac{1}{2},\frac{1}{2} & 6,0,\frac{1}{2},\frac{1}{2} & 6,0,\frac{1}{2},0 & 13,0,0,0 & 13,0,0,\frac{1}{2} & 7 & \frac{\tau}{2} \\
6,1,0,\frac{1}{2} & 7,1,0,\frac{1}{2} & 7,1,0,0 & 7,1,\frac{1}{2},0 & 6,1,\frac{1}{2},0 & 6,1,\frac{1}{2},\frac{1}{2} & 13,0,0,0 & 13,0,0,\frac{1}{2} & 7 & \frac{\tau}{2} \\
6,1,0,\frac{1}{2} & 7,1,0,0 & 7,1,0,\frac{1}{2} & 7,0,\frac{1}{2},\frac{1}{2} & 6,0,\frac{1}{2},0 & 6,0,\frac{1}{2},\frac{1}{2} & 13,0,0,\frac{1}{2} & 13,0,0,0 & 7 & \frac{1+\tau}{2} \\
6,1,0,\frac{1}{2} & 7,1,0,0 & 7,1,0,\frac{1}{2} & 7,1,\frac{1}{2},0 & 6,1,\frac{1}{2},\frac{1}{2} & 6,1,\frac{1}{2},0 & 13,0,0,\frac{1}{2} & 13,0,0,0 & 7 & \frac{1+\tau}{2} \\
6,1,\frac{1}{2},\frac{1}{2} & 5,0,0,0 & 6,1,\frac{1}{2},0 & 6,1,\frac{1}{2},0 & 5,0,0,\frac{1}{2} & 6,1,\frac{1}{2},\frac{1}{2} & 11,1,\frac{1}{2},\frac{1}{2} & 12,0,0,\frac{1}{2} & 6 & \frac{\tau}{2} \\
\hline
  \end{array} \notag
\end{align}
\end{table}
\begin{table}[h]
\begin{align}
  \begin{array}{c|c|c|c|c|c|c|c|c|c} \hline
    B_Q & B_{u_R} & B_{d_R} & B_L & B_{\nu_R} & B_{e_R} & B_{H_u} & B_{H_d} & g_H & p \\ \hline \hline
6,1,\frac{1}{2},\frac{1}{2} & 7,1,\frac{1}{2},0 & 7,1,0,\frac{1}{2} & 7,0,\frac{1}{2},\frac{1}{2} & 6,0,\frac{1}{2},0 & 6,0,0,\frac{1}{2} & 13,0,0,\frac{1}{2} & 13,0,\frac{1}{2},0 & 7 & \frac{1}{2} \\
6,1,\frac{1}{2},\frac{1}{2} & 7,1,\frac{1}{2},0 & 7,1,0,\frac{1}{2} & 7,1,0,0 & 6,1,0,\frac{1}{2} & 6,1,\frac{1}{2},0 & 13,0,0,\frac{1}{2} & 13,0,\frac{1}{2},0 & 7 & \frac{1}{2} \\
6,1,\frac{1}{2},\frac{1}{2} & 7,1,0,\frac{1}{2} & 7,1,\frac{1}{2},0 & 7,0,\frac{1}{2},\frac{1}{2} & 6,0,0,\frac{1}{2} & 6,0,\frac{1}{2},0 & 13,0,\frac{1}{2},0 & 13,0,0,\frac{1}{2} & 7 & \frac{\tau}{2} \\
6,1,\frac{1}{2},\frac{1}{2} & 7,1,0,\frac{1}{2} & 7,1,\frac{1}{2},0 & 7,1,0,0 & 6,1,\frac{1}{2},0 & 6,1,0,\frac{1}{2} & 13,0,\frac{1}{2},0 & 13,0,0,\frac{1}{2} & 7 & \frac{\tau}{2} \\
7,0,\frac{1}{2},\frac{1}{2} & 5,0,0,0 & 6,1,\frac{1}{2},0 & 6,1,\frac{1}{2},0 & 6,1,0,\frac{1}{2} & 7,0,\frac{1}{2},\frac{1}{2} & 12,0,\frac{1}{2},\frac{1}{2} & 13,1,0,\frac{1}{2} & 6 & 0 \\
7,0,\frac{1}{2},\frac{1}{2} & 5,0,0,0 & 6,1,\frac{1}{2},0 & 7,1,0,0 & 5,1,\frac{1}{2},\frac{1}{2} & 6,0,0,\frac{1}{2} & 12,0,\frac{1}{2},\frac{1}{2} & 13,1,0,\frac{1}{2} & 6 & 0 \\
7,0,\frac{1}{2},\frac{1}{2} & 5,0,0,0 & 6,1,0,\frac{1}{2} & 6,1,0,\frac{1}{2} & 6,1,\frac{1}{2},0 & 7,0,\frac{1}{2},\frac{1}{2} & 12,0,\frac{1}{2},\frac{1}{2} & 13,1,\frac{1}{2},0 & 6 & 0 \\
\hline
  \end{array} \notag
\end{align}
\end{table}
\begin{table}[h]
\begin{align}
  \begin{array}{c|c|c|c|c|c|c|c|c|c} \hline
    B_Q & B_{u_R} & B_{d_R} & B_L & B_{\nu_R} & B_{e_R} & B_{H_u} & B_{H_d} & g_H & p \\ \hline \hline
7,1,\frac{1}{2},0 & 6,1,\frac{1}{2},\frac{1}{2} & 7,0,\frac{1}{2},\frac{1}{2} & 7,0,\frac{1}{2},\frac{1}{2} & 6,0,\frac{1}{2},0 & 7,1,\frac{1}{2},0 & 13,0,0,\frac{1}{2} & 14,1,0,\frac{1}{2} & 7 & \frac{1}{2} \\
7,1,\frac{1}{2},0 & 6,1,\frac{1}{2},\frac{1}{2} & 7,1,0,0 & 7,1,0,0 & 6,1,0,\frac{1}{2} & 7,1,\frac{1}{2},0 & 13,0,0,\frac{1}{2} & 14,0,\frac{1}{2},0 & 7 & \frac{1}{2} \\
7,1,0,\frac{1}{2} & 5,0,0,0 & 6,0,\frac{1}{2},\frac{1}{2} & 6,0,\frac{1}{2},\frac{1}{2} & 6,1,\frac{1}{2},0 & 7,1,0,\frac{1}{2} & 12,1,0,\frac{1}{2} & 13,1,\frac{1}{2},0 & 6 & \frac{\tau}{2} \\
7,1,0,\frac{1}{2} & 5,0,0,0 & 6,0,\frac{1}{2},\frac{1}{2} & 7,1,0,0 & 5,0,0,\frac{1}{2} & 6,0,\frac{1}{2},0 & 12,1,0,\frac{1}{2} & 13,1,\frac{1}{2},0 & 6 & \frac{\tau}{2} \\
7,1,0,\frac{1}{2} & 5,0,0,0 & 6,1,\frac{1}{2},0 & 6,1,\frac{1}{2},0 & 6,0,\frac{1}{2},\frac{1}{2} & 7,1,0,\frac{1}{2} & 12,1,0,\frac{1}{2} & 13,0,\frac{1}{2},\frac{1}{2} & 6 & \frac{\tau}{2} \\
7,1,0,\frac{1}{2} & 5,0,0,0 & 6,1,\frac{1}{2},0 & 7,1,0,0 & 5,0,0,\frac{1}{2} & 6,1,\frac{1}{2},\frac{1}{2} & 12,1,0,\frac{1}{2} & 13,0,\frac{1}{2},\frac{1}{2} & 6 & \frac{\tau}{2} \\
7,1,0,\frac{1}{2} & 5,0,\frac{1}{2},0 & 6,0,0,\frac{1}{2} & 6,0,0,\frac{1}{2} & 6,1,\frac{1}{2},0 & 7,1,0,\frac{1}{2} & 12,1,\frac{1}{2},\frac{1}{2} & 13,1,0,0 & 6 & \frac{\tau}{2} \\
\hline
  \end{array} \notag
\end{align}
\end{table}

\clearpage
%-----------------------------------------------------
%-----------------------------------------------------
%-----------------------------------------------------


\begin{thebibliography}{99}



%%%%%%% magnetized models %%%%%%%%%%%%%%%%%%

%\cite{Cremades:2004wa}
\bibitem{Cremades:2004wa}
D.~Cremades, L.~E.~Ibanez and F.~Marchesano,
%``Computing Yukawa couplings from magnetized extra dimensions,''
JHEP \textbf{05} (2004), 079
%doi:10.1088/1126-6708/2004/05/079
[arXiv:hep-th/0404229 [hep-th]].


%\cite{Abe:2008fi}
\bibitem{Abe:2008fi}
H.~Abe, T.~Kobayashi and H.~Ohki,
%``Magnetized orbifold models,''
JHEP \textbf{09} (2008), 043
%doi:10.1088/1126-6708/2008/09/043
[arXiv:0806.4748 [hep-th]].


%\cite{Abe:2013bca}
\bibitem{Abe:2013bca} 
  T.~H.~Abe, Y.~Fujimoto, T.~Kobayashi, T.~Miura, K.~Nishiwaki and M.~Sakamoto,
  %``$Z_N$ twisted orbifold models with magnetic flux,''
  JHEP {\bf 1401}, 065 (2014)
%  doi:10.1007/JHEP01(2014)065
  [arXiv:1309.4925 [hep-th]].
  %%CITATION = doi:10.1007/JHEP01(2014)065;%%
 
  
%\cite{Abe:2014noa}
\bibitem{Abe:2014noa} 
  T.~h.~Abe, Y.~Fujimoto, T.~Kobayashi, T.~Miura, K.~Nishiwaki and M.~Sakamoto,
  %``Operator analysis of physical states on magnetized $T^{2}/Z_{N}$ orbifolds,''
  Nucl.\ Phys.\ B {\bf 890}, 442 (2014)
%  doi:10.1016/j.nuclphysb.2014.11.022
  [arXiv:1409.5421 [hep-th]].
  %%CITATION = doi:10.1016/j.nuclphysb.2014.11.022;%%


%%%%%%%%% quark mass in magnetized models %%%%%%%%%%%%




%\cite{Abe:2012fj}
\bibitem{Abe:2012fj}
H.~Abe, T.~Kobayashi, H.~Ohki, A.~Oikawa and K.~Sumita,
%``Phenomenological aspects of 10D SYM theory with magnetized extra dimensions,''
Nucl. Phys. B \textbf{870}, 30-54 (2013)
%doi:10.1016/j.nuclphysb.2013.01.014
[arXiv:1211.4317 [hep-ph]].  

%\cite{Abe:2014vza}
\bibitem{Abe:2014vza}
H.~Abe, T.~Kobayashi, K.~Sumita and Y.~Tatsuta,
%``Gaussian Froggatt-Nielsen mechanism on magnetized orbifolds,''
Phys. Rev. D \textbf{90}, no.10, 105006 (2014)
%doi:10.1103/PhysRevD.90.105006
[arXiv:1405.5012 [hep-ph]].





%\cite{Fujimoto:2016zjs}
\bibitem{Fujimoto:2016zjs}
Y.~Fujimoto, T.~Kobayashi, K.~Nishiwaki, M.~Sakamoto and Y.~Tatsuta,
%``Comprehensive analysis of Yukawa hierarchies on $T^2/Z_N$ with magnetic fluxes,''
Phys. Rev. D \textbf{94}, no.3, 035031 (2016)
%doi:10.1103/PhysRevD.94.035031
[arXiv:1605.00140 [hep-ph]].




%\cite{Kobayashi:2016qag}
\bibitem{Kobayashi:2016qag}
T.~Kobayashi, K.~Nishiwaki and Y.~Tatsuta,
%``CP-violating phase on magnetized toroidal orbifolds,''
JHEP \textbf{04}, 080 (2017)
%doi:10.1007/JHEP04(2017)080
[arXiv:1609.08608 [hep-th]].

%\cite{Kikuchi:2021yog}
\bibitem{Kikuchi:2021yog}
S.~Kikuchi, T.~Kobayashi, Y.~Ogawa and H.~Uchida,
%``Yukawa textures in modular symmetric vacuum of magnetized orbifold models,''
PTEP \textbf{2022}, no.3, 033B10 (2022)
%doi:10.1093/ptep/ptac035
[arXiv:2112.01680 [hep-ph]].

%\cite{Kikuchi:2022geu}
\bibitem{Kikuchi:2022geu}
S.~Kikuchi, T.~Kobayashi, M.~Tanimoto and H.~Uchida,
%``Mass matrices with CP phase in modular flavor symmetry,''
[arXiv:2206.08538 [hep-ph]].





%%%%%%%%%%  finite modular symmetry %%%%%%%%%%%%

%\cite{deAdelhartToorop:2011re}
\bibitem{deAdelhartToorop:2011re}
R.~de Adelhart Toorop, F.~Feruglio and C.~Hagedorn,
%``Finite Modular Groups and Lepton Mixing,''
Nucl. Phys. B \textbf{858} (2012), 437-467
%doi:10.1016/j.nuclphysb.2012.01.017
[arXiv:1112.1340 [hep-ph]].
%222 citations counted in INSPIRE as of 21 Nov 2021

%%%%%%%%% modular flavor model %%%%%%%%%%%%%%%%%%%%

%\cite{Feruglio:2017spp}
\bibitem{Feruglio:2017spp} 
  F.~Feruglio,
  %``Are neutrino masses modular forms?,''
%  doi:10.1142/9789813238053_0012
  arXiv:1706.08749 [hep-ph].
  %%CITATION = doi:10.1142/9789813238053_0012;%%

%\cite{Kobayashi:2018vbk}
\bibitem{Kobayashi:2018vbk} 
T.~Kobayashi, K.~Tanaka and T.~H.~Tatsuishi,
%``Neutrino mixing from finite modular groups,''
Phys.\ Rev.\ D {\bf 98}, no. 1, 016004 (2018)
%  doi:10.1103/PhysRevD.98.016004
[arXiv:1803.10391 [hep-ph]].
%%CITATION = doi:10.1103/PhysRevD.98.016004;%%

%\cite{Penedo:2018nmg}
\bibitem{Penedo:2018nmg} 
J.~T.~Penedo and S.~T.~Petcov,
%``Lepton Masses and Mixing from Modular $S_4$ Symmetry,''
Nucl.\ Phys.\ B {\bf 939}, 292 (2019)
%  doi:10.1016/j.nuclphysb.2018.12.016
[arXiv:1806.11040 [hep-ph]].
%%CITATION = doi:10.1016/j.nuclphysb.2018.12.016;%%


%\cite{Novichkov:2018nkm}
\bibitem{Novichkov:2018nkm} 
P.~P.~Novichkov, J.~T.~Penedo, S.~T.~Petcov and A.~V.~Titov,
%``Modular A$_{5}$ symmetry for flavour model building,''
JHEP {\bf 1904}, 174 (2019)
%  doi:10.1007/JHEP04(2019)174
[arXiv:1812.02158 [hep-ph]].
%%CITATION = doi:10.1007/JHEP04(2019)174;%%

%\cite{Criado:2018thu}
\bibitem{Criado:2018thu}
J.~C.~Criado and F.~Feruglio,
%``Modular Invariance Faces Precision Neutrino Data,''
SciPost Phys.\  {\bf 5} (2018) no.5,  042
%doi:10.21468/SciPostPhys.5.5.042
[arXiv:1807.01125 [hep-ph]].



%\cite{Kobayashi:2018scp}
\bibitem{Kobayashi:2018scp}
T.~Kobayashi, N.~Omoto, Y.~Shimizu, K.~Takagi, M.~Tanimoto and T.~H.~Tatsuishi,
%``Modular A$_{4}$ invariance and neutrino mixing,''
JHEP \textbf{11} (2018), 196
%doi:10.1007/JHEP11(2018)196
[arXiv:1808.03012 [hep-ph]].


%\cite{Ding:2019zxk}
\bibitem{Ding:2019zxk}
G.~J.~Ding, S.~F.~King and X.~G.~Liu,
%``Modular A$_{4}$ symmetry models of neutrinos and charged leptons,''
JHEP {\bf 1909} (2019) 074
%doi:10.1007/JHEP09(2019)074
[arXiv:1907.11714 [hep-ph]].


%\cite{Novichkov:2018ovf}
\bibitem{Novichkov:2018ovf}
P.~P.~Novichkov, J.~T.~Penedo, S.~T.~Petcov and A.~V.~Titov,
%``Modular S$_{4}$ models of lepton masses and mixing,''
JHEP {\bf 1904} (2019) 005
%doi:10.1007/JHEP04(2019)005
[arXiv:1811.04933 [hep-ph]].

%%%%%%%%%%%%%% A_4 originated from broken S_4 %%%%%%%%%%%%%%%%%%%%%%%
%\cite{Kobayashi:2019mna}
\bibitem{Kobayashi:2019mna}
T.~Kobayashi, Y.~Shimizu, K.~Takagi, M.~Tanimoto and T.~H.~Tatsuishi,
%``New $A_4$ lepton flavor model from $S_4$ modular symmetry,''
JHEP \textbf{02} (2020), 097
%doi:10.1007/JHEP02(2020)097
[arXiv:1907.09141 [hep-ph]].


%%%%%%%%%%%%%%%%%%%%% S4 modular  %%%%%%%%%%%%%%%%%%%%%%%%%%%%%%%%%%%%
%\cite{Wang:2019ovr}
\bibitem{Wang:2019ovr}
X.~Wang and S.~Zhou,
%``The minimal seesaw model with a modular S$_{4}$ symmetry,''
JHEP \textbf{05} (2020), 017
%doi:10.1007/JHEP05(2020)017
[arXiv:1910.09473 [hep-ph]].
%%%%%%%%%%%%%%%%%%%%%%%%%%%%%%%%%%%%%%%%%%%%%%%%%%%%%%%%%%%%%%%%%%%%%%

%\cite{Ding:2019xna}
\bibitem{Ding:2019xna}
G.~J.~Ding, S.~F.~King and X.~G.~Liu,
%``Neutrino mass and mixing with $A_5$ modular symmetry,''
Phys.\ Rev.\ D {\bf 100} (2019) no.11,  115005
%doi:10.1103/PhysRevD.100.115005
[arXiv:1903.12588 [hep-ph]].
%%CITATION = doi:10.1103/PhysRevD.100.115005;%%


%%%%%%%%%%%%%% T' (double covering of A_4) S4'  %%%%%%%%%%%%%%%%%%%%%

%\cite{Liu:2019khw}
\bibitem{Liu:2019khw}
X.~G.~Liu and G.~J.~Ding,
%``Neutrino Masses and Mixing from Double Covering of Finite Modular Groups,''
JHEP {\bf 1908} (2019) 134
%doi:10.1007/JHEP08(2019)134
[arXiv:1907.01488 [hep-ph]].
%%CITATION = doi:10.1007/JHEP08(2019)134;%%
%20 citations counted in INSPIRE as of 19 Dec 2019


%\cite{Chen:2020udk}
\bibitem{Chen:2020udk}
P.~Chen, G.~J.~Ding, J.~N.~Lu and J.~W.~F.~Valle,
%``Predictions from warped flavor dynamics based on the $T??��?�� family group,''
Phys. Rev. D \textbf{102} (2020) no.9, 095014
%doi:10.1103/PhysRevD.102.095014
[arXiv:2003.02734 [hep-ph]].



%\cite{Novichkov:2020eep}
\bibitem{Novichkov:2020eep}
P.~P.~Novichkov, J.~T.~Penedo and S.~T.~Petcov,
%``Double cover of modular $S_4$ for flavour model building,''
Nucl. Phys. B \textbf{963} (2021), 115301
%doi:10.1016/j.nuclphysb.2020.115301
[arXiv:2006.03058 [hep-ph]].


%\cite{Liu:2020akv}
\bibitem{Liu:2020akv}
X.~G.~Liu, C.~Y.~Yao and G.~J.~Ding,
%``Modular invariant quark and lepton models in double covering of $S_4$ modular group,''
Phys. Rev. D \textbf{103} (2021) no.5, 056013
%doi:10.1103/PhysRevD.103.056013
[arXiv:2006.10722 [hep-ph]].


%\cite{deMedeirosVarzielas:2019cyj}
\bibitem{deMedeirosVarzielas:2019cyj}
I.~de Medeiros Varzielas, S.~F.~King and Y.~L.~Zhou,
%``Multiple modular symmetries as the origin of flavor,''
Phys.\ Rev.\ D {\bf 101} (2020) no.5,  055033
%doi:10.1103/PhysRevD.101.055033
[arXiv:1906.02208 [hep-ph]].

%\cite{Asaka:2019vev}
\bibitem{Asaka:2019vev}
T.~Asaka, Y.~Heo, T.~H.~Tatsuishi and T.~Yoshida,
%``Modular $A_4$ invariance and leptogenesis,''
JHEP {\bf 2001} (2020) 144
%doi:10.1007/JHEP01(2020)144
[arXiv:1909.06520 [hep-ph]].

%\cite{Ding:2020msi}
\bibitem{Ding:2020msi}
G.~J.~Ding, S.~F.~King, C.~C.~Li and Y.~L.~Zhou,
%``Modular Invariant Models of Leptons at Level 7,''
JHEP \textbf{08} (2020), 164
%doi:10.1007/JHEP08(2020)164
[arXiv:2004.12662 [hep-ph]].
%%%%%%%%%%%%%%%%%%%%%%%%%%%%%%%%%%%%

%\cite{Asaka:2020tmo}
\bibitem{Asaka:2020tmo}
T.~Asaka, Y.~Heo and T.~Yoshida,
%``Lepton flavor model with modular $A_4$ symmetry in large volume limit,''
Phys. Lett. B \textbf{811} (2020), 135956
%doi:10.1016/j.physletb.2020.135956
[arXiv:2009.12120 [hep-ph]].




%%%%%%%%%  GUT %%%%%%%%%%%%%%%%%%

%\cite{deAnda:2018ecu}
\bibitem{deAnda:2018ecu}
F.~J.~de Anda, S.~F.~King and E.~Perdomo,
%``$SU(5)$ grand unified theory with $A_4$ modular symmetry,''
Phys. Rev. D \textbf{101} (2020) no.1, 015028
%doi:10.1103/PhysRevD.101.015028
[arXiv:1812.05620 [hep-ph]].


%\cite{Kobayashi:2019rzp}
\bibitem{Kobayashi:2019rzp}
T.~Kobayashi, Y.~Shimizu, K.~Takagi, M.~Tanimoto and T.~H.~Tatsuishi,
%``Modular $S_3$-invariant flavor model in SU(5) grand unified theory,''
PTEP \textbf{2020}, no.5, 053B05 (2020)
%doi:10.1093/ptep/ptaa055
[arXiv:1906.10341 [hep-ph]].
%70 citations counted in INSPIRE as of 28 Nov 2021

%
%\cite{Novichkov:2018yse}
\bibitem{Novichkov:2018yse}
P.~P.~Novichkov, S.~T.~Petcov and M.~Tanimoto,
%``Trimaximal Neutrino Mixing from Modular A4 Invariance with Residual Symmetries,''
Phys.\ Lett.\ B {\bf 793} (2019) 247
%doi:10.1016/j.physletb.2019.04.043
[arXiv:1812.11289 [hep-ph]].

%\cite{Kobayashi:2018wkl}
\bibitem{Kobayashi:2018wkl}
T.~Kobayashi, Y.~Shimizu, K.~Takagi, M.~Tanimoto, T.~H.~Tatsuishi and H.~Uchida,
%``Finite modular subgroups for fermion mass matrices and baryon/lepton number violation,''
Phys.\ Lett.\ B {\bf 794} (2019) 114
%  doi:10.1016/j.physletb.2019.05.034
[arXiv:1812.11072 [hep-ph]].
%%CITATION = doi:10.1016/j.physletb.2019.05.034;%%

%%%%%%%%%%%%%%%%%%%%  Quark  %%%%%%%%%%%%%%%%%%%%
%\cite{Okada:2018yrn}
\bibitem{Okada:2018yrn}
H.~Okada and M.~Tanimoto,
%``CP violation of quarks in $A_4$ modular invariance,''
Phys.\ Lett.\ B {\bf 791} (2019) 54
%doi:10.1016/j.physletb.2019.02.028
[arXiv:1812.09677 [hep-ph]].

%\cite{Okada:2019uoy}
\bibitem{Okada:2019uoy}
H.~Okada and M.~Tanimoto,
%``Towards unification of quark and lepton flavors in $A_4$ modular invariance,''
Eur. Phys. J. C \textbf{81} (2021) no.1, 52
%doi:10.1140/epjc/s10052-021-08845-y
[arXiv:1905.13421 [hep-ph]].

%\cite{Nomura:2019jxj}
\bibitem{Nomura:2019jxj}
T.~Nomura and H.~Okada,
%``A modular $A_4$ symmetric model of dark matter and neutrino,''
Phys. Lett. B \textbf{797}, 134799 (2019)
%doi:10.1016/j.physletb.2019.134799
[arXiv:1904.03937 [hep-ph]].

%\cite{Okada:2019xqk}
\bibitem{Okada:2019xqk}
H.~Okada and Y.~Orikasa,
%``Modular $S_3$ symmetric radiative seesaw model,''
Phys. Rev. D \textbf{100}, no.11, 115037 (2019)
%doi:10.1103/PhysRevD.100.115037
[arXiv:1907.04716 [hep-ph]].




%\cite{Nomura:2019yft}
\bibitem{Nomura:2019yft}
T.~Nomura and H.~Okada,
%``A two loop induced neutrino mass model with modular $A_4$ symmetry,''
Nucl. Phys. B \textbf{966} (2021), 115372
%doi:10.1016/j.nuclphysb.2021.115372
[arXiv:1906.03927 [hep-ph]].


%\cite{Nomura:2019lnr}
\bibitem{Nomura:2019lnr}
T.~Nomura, H.~Okada and O.~Popov,
%``A modular $A_4$ symmetric scotogenic model,''
Phys.\ Lett.\ B {\bf 803} (2020) 135294
%doi:10.1016/j.physletb.2020.135294
[arXiv:1908.07457 [hep-ph]].

%\cite{Criado:2019tzk}
\bibitem{Criado:2019tzk}
J.~C.~Criado, F.~Feruglio and S.~J.~D.~King,
%``Modular Invariant Models of Lepton Masses at Levels 4 and 5,''
JHEP {\bf 2002} (2020) 001
%doi:10.1007/JHEP02(2020)001
[arXiv:1908.11867 [hep-ph]].

%\cite{King:2019vhv}
\bibitem{King:2019vhv}
S.~F.~King and Y.~L.~Zhou,
%``Trimaximal TM$_1$ mixing with two modular $S_4$ groups,''
Phys. Rev. D \textbf{101} (2020) no.1, 015001
%doi:10.1103/PhysRevD.101.015001
[arXiv:1908.02770 [hep-ph]].

%cite{Gui-JunDing:2019wap}
\bibitem{Gui-JunDing:2019wap}
G.~J.~Ding, S.~F.~King, X.~G.~Liu and J.~N.~Lu,
%``Modular S$_{4}$ and A$_{4}$ symmetries and their fixed points: new predictive examples of lepton mixing,''
JHEP {\bf 1912} (2019) 030
%doi:10.1007/JHEP12(2019)030
[arXiv:1910.03460 [hep-ph]].

%\cite{deMedeirosVarzielas:2020kji}
\bibitem{deMedeirosVarzielas:2020kji}
I.~de Medeiros Varzielas, M.~Levy and Y.~L.~Zhou,
%``Symmetries and stabilisers in modular invariant flavour models,''
JHEP \textbf{11} (2020), 085
%doi:10.1007/JHEP11(2020)085
[arXiv:2008.05329 [hep-ph]].

%\cite{Zhang:2019ngf}
\bibitem{Zhang:2019ngf}
D.~Zhang,
%``A modular $A_4$ symmetry realization of two-zero textures of the Majorana neutrino mass matrix,''
Nucl.\ Phys.\ B {\bf 952} (2020) 114935
%doi:10.1016/j.nuclphysb.2020.114935
[arXiv:1910.07869 [hep-ph]].

%\cite{Nomura:2019xsb}
\bibitem{Nomura:2019xsb}
T.~Nomura, H.~Okada and S.~Patra,
%``An inverse seesaw model with $A_4$ -modular symmetry,''
Nucl. Phys. B \textbf{967} (2021), 115395
%doi:10.1016/j.nuclphysb.2021.115395
[arXiv:1912.00379 [hep-ph]].

%\cite{Kobayashi:2019gtp}
\bibitem{Kobayashi:2019gtp}
T.~Kobayashi, T.~Nomura and T.~Shimomura,
%``Type II seesaw models with modular $A_4$ symmetry,''
Phys. Rev. D \textbf{102} (2020) no.3, 035019
%doi:10.1103/PhysRevD.102.035019
[arXiv:1912.00637 [hep-ph]].


%\cite{Lu:2019vgm}
\bibitem{Lu:2019vgm}
J.~N.~Lu, X.~G.~Liu and G.~J.~Ding,
%``Modular symmetry origin of texture zeros and quark lepton unification,''
Phys. Rev. D \textbf{101} (2020) no.11, 115020
%	doi:10.1103/PhysRevD.101.115020
[arXiv:1912.07573 [hep-ph]].

%\cite{Wang:2019xbo}
\bibitem{Wang:2019xbo}
X.~Wang,
%``Lepton flavor mixing and CP violation in the minimal type-(I+II) seesaw model with a modular $A_4$ symmetry,''
Nucl. Phys. B \textbf{957} (2020), 115105
%doi:10.1016/j.nuclphysb.2020.115105
[arXiv:1912.13284 [hep-ph]].


\bibitem{King:2020qaj}
S.~J.~D.~King and S.~F.~King,
%``Fermion mass hierarchies from modular symmetry,''
JHEP \textbf{09} (2020), 043
%doi:10.1007/JHEP09(2020)043
[arXiv:2002.00969 [hep-ph]].

%\cite{Abbas:2020qzc}
\bibitem{Abbas:2020qzc}
M.~Abbas,
%``Fermion masses and mixing in modular A4 Symmetry,''
Phys. Rev. D \textbf{103} (2021) no.5, 056016
%doi:10.1103/PhysRevD.103.056016
[arXiv:2002.01929 [hep-ph]].


%\cite{Okada:2020oxh}
\bibitem{Okada:2020oxh}
H.~Okada and Y.~Shoji,
%``Dirac dark matter in a radiative neutrino model,''
Phys. Dark Univ. \textbf{31} (2021), 100742
%doi:10.1016/j.dark.2020.100742
[arXiv:2003.11396 [hep-ph]].



%\cite{Okada:2020dmb}
\bibitem{Okada:2020dmb}
H.~Okada and Y.~Shoji,
%``A radiative seesaw model with three Higgs doublets in modular $A_4$ symmetry,''
Nucl. Phys. B \textbf{961} (2020), 115216
%	doi:10.1016/j.nuclphysb.2020.115216
[arXiv:2003.13219 [hep-ph]].

%\cite{Ding:2020yen}
\bibitem{Ding:2020yen}
G.~J.~Ding and F.~Feruglio,
%``Testing Moduli and Flavon Dynamics with Neutrino Oscillations,''
JHEP \textbf{06} (2020), 134
%doi:10.1007/JHEP06(2020)134
[arXiv:2003.13448 [hep-ph]].

%%%%%%%%%%%%%%%%%%%%%%%%%%%%%%%%%%%%%%%%%%%%%%%

%\cite{Okada:2020rjb}
\bibitem{Okada:2020rjb}
H.~Okada and M.~Tanimoto,
%``Quark and lepton flavors with common modulus $\tau$ in $A_4$ modular symmetry,''
[arXiv:2005.00775 [hep-ph]].



%\cite{Okada:2020ukr}
\bibitem{Okada:2020ukr}
H.~Okada and M.~Tanimoto,
%``Modular invariant flavor model of $A_4$ and hierarchical structures at nearby fixed points,''
Phys. Rev. D \textbf{103} (2021) no.1, 015005
% doi:10.1103/PhysRevD.103.015005
[arXiv:2009.14242 [hep-ph]].

%\cite{Nagao:2020azf}
\bibitem{Nagao:2020azf}
K.~I.~Nagao and H.~Okada,
%``Neutrino and dark matter in a gauged $U(1)_R$ symmetry,''
JCAP \textbf{05} (2021), 063
%doi:10.1088/1475-7516/2021/05/063
[arXiv:2008.13686 [hep-ph]].







%\cite{Wang:2020lxk}
\bibitem{Wang:2020lxk}
X.~Wang, B.~Yu and S.~Zhou,
%``Double covering of the modular $A_5$ group and lepton flavor mixing in the minimal seesaw model,''
Phys. Rev. D \textbf{103} (2021) no.7, 076005
%doi:10.1103/PhysRevD.103.076005
[arXiv:2010.10159 [hep-ph]].



%\cite{Okada:2020brs}
\bibitem{Okada:2020brs}
H.~Okada and M.~Tanimoto,
%``Spontaneous CP violation by modulus $\tau$ in $A_4$ model of lepton flavors,''
JHEP \textbf{03} (2021), 010
%doi:10.1007/JHEP03(2021)010
[arXiv:2012.01688 [hep-ph]].
%%%%%%%%%%%%%%%%%%%%%%


%\cite{Yao:2020qyy}
\bibitem{Yao:2020qyy}
C.~Y.~Yao, J.~N.~Lu and G.~J.~Ding,
%``Modular Invariant $A_{4}$ Models for Quarks and Leptons with Generalized CP Symmetry,''
JHEP \textbf{05} (2021), 102
%doi:10.1007/JHEP05(2021)102
[arXiv:2012.13390 [hep-ph]].

%%%%%%%%%%%% end of modular flavor model



%\cite{Kikuchi:2022svo}
\bibitem{Kikuchi:2022svo}
S.~Kikuchi, T.~Kobayashi, M.~Tanimoto and H.~Uchida,
%``Texture zeros of quark mass matrices at fixed point $\tau=\omega$ in modular flavor symmetry,''
[arXiv:2207.04609 [hep-ph]].




%%%%  modular symmetry %%%%%%%%%%%%%%%%%%%%

%\cite{Kobayashi:2018rad}
\bibitem{Kobayashi:2018rad} 
 T.~Kobayashi, S.~Nagamoto, S.~Takada, S.~Tamba and T.~H.~Tatsuishi,
 %``Modular symmetry and non-Abelian discrete flavor symmetries in string compactification,''
 Phys.\ Rev.\ D {\bf 97}, no. 11, 116002 (2018)
% doi:10.1103/PhysRevD.97.116002
 [arXiv:1804.06644 [hep-th]].
 %%CITATION = doi:10.1103/PhysRevD.97.116002;%% 

%\cite{Kobayashi:2018bff}
\bibitem{Kobayashi:2018bff}
T.~Kobayashi and S.~Tamba,
%``Modular forms of finite modular subgroups from magnetized D-brane models,''
Phys.\ Rev.\ D {\bf 99} (2019) no.4, 046001
%doi:10.1103/PhysRevD.99.046001
[arXiv:1811.11384 [hep-th]].

%\cite{Kariyazono:2019ehj}
\bibitem{Kariyazono:2019ehj}
Y.~Kariyazono, T.~Kobayashi, S.~Takada, S.~Tamba and H.~Uchida,
%``Modular symmetry anomaly in magnetic flux compactification,''
Phys. Rev. D \textbf{100}, no.4, 045014 (2019)
%doi:10.1103/PhysRevD.100.045014
[arXiv:1904.07546 [hep-th]].




%\cite{Ohki:2020bpo}
\bibitem{Ohki:2020bpo}
H.~Ohki, S.~Uemura and R.~Watanabe,
%``Modular flavor symmetry on a magnetized torus,''
Phys. Rev. D \textbf{102}, no.8, 085008 (2020)
%doi:10.1103/PhysRevD.102.085008
[arXiv:2003.04174 [hep-th]].



%\cite{Kikuchi:2020frp}
\bibitem{Kikuchi:2020frp}
S.~Kikuchi, T.~Kobayashi, S.~Takada, T.~H.~Tatsuishi and H.~Uchida,
%``Revisiting modular symmetry in magnetized torus and orbifold compactifications,''
Phys. Rev. D \textbf{102}, no.10, 105010 (2020)
%doi:10.1103/PhysRevD.102.105010
[arXiv:2005.12642 [hep-th]].




%\cite{Kikuchi:2020nxn}
\bibitem{Kikuchi:2020nxn}
S.~Kikuchi, T.~Kobayashi, H.~Otsuka, S.~Takada and H.~Uchida,
%``Modular symmetry by orbifolding magnetized $T^2\times T^2$: realization of double cover of $\Gamma_N$,''
JHEP \textbf{11} (2020), 101
%doi:10.1007/JHEP11(2020)101
[arXiv:2007.06188 [hep-th]].


%\cite{Kikuchi:2021ogn}
\bibitem{Kikuchi:2021ogn}
S.~Kikuchi, T.~Kobayashi and H.~Uchida,
%``Modular flavor symmetries of three-generation modes on magnetized toroidal orbifolds,''
Phys. Rev. D \textbf{104}, no.6, 065008 (2021)
%doi:10.1103/PhysRevD.104.065008
[arXiv:2101.00826 [hep-th]].

%\cite{Almumin:2021fbk}
\bibitem{Almumin:2021fbk}
Y.~Almumin, M.~C.~Chen, V.~Knapp-Perez, S.~Ramos-Sanchez, M.~Ratz and S.~Shukla,
%``Metaplectic Flavor Symmetries from Magnetized Tori,'' 
JHEP \textbf{05} (2021), 078
%doi:10.1007/JHEP05(2021)078
[arXiv:2102.11286 [hep-th]].




%%%%%%%%%% D-brane instanton %%%%%%%%%%%%%%%%%%%%

%\cite{Blumenhagen:2006xt}
\bibitem{Blumenhagen:2006xt}
R.~Blumenhagen, M.~Cvetic and T.~Weigand,
%``Spacetime instanton corrections in 4D string vacua: The Seesaw mechanism for D-Brane models,''
Nucl. Phys. B \textbf{771}, 113-142 (2007)
%doi:10.1016/j.nuclphysb.2007.02.016
[arXiv:hep-th/0609191 [hep-th]].


%\cite{Ibanez:2006da}
\bibitem{Ibanez:2006da}
L.~E.~Ibanez and A.~M.~Uranga,
%``Neutrino Majorana Masses from String Theory Instanton Effects,''
JHEP \textbf{03}, 052 (2007)
%doi:10.1088/1126-6708/2007/03/052
[arXiv:hep-th/0609213 [hep-th]].

%\cite{Ibanez:2007rs}
\bibitem{Ibanez:2007rs}
L.~E.~Ibanez, A.~N.~Schellekens and A.~M.~Uranga,
%``Instanton Induced Neutrino Majorana Masses in CFT Orientifolds with MSSM-like spectra,''
JHEP \textbf{06}, 011 (2007)
%doi:10.1088/1126-6708/2007/06/011
[arXiv:0704.1079 [hep-th]].

%\cite{Antusch:2007jd}
\bibitem{Antusch:2007jd}
S.~Antusch, L.~E.~Ibanez and T.~Macri,
%``Neutrino masses and mixings from string theory instantons,''
JHEP \textbf{09}, 087 (2007)
%doi:10.1088/1126-6708/2007/09/087
[arXiv:0706.2132 [hep-ph]].

%\cite{Kobayashi:2015siy}
\bibitem{Kobayashi:2015siy}
T.~Kobayashi, Y.~Tatsuta and S.~Uemura,
%``Majorana neutrino mass structure induced by rigid instantons on toroidal orbifold,''
Phys. Rev. D \textbf{93}, no.6, 065029 (2016)
%doi:10.1103/PhysRevD.93.065029
[arXiv:1511.09256 [hep-ph]].






%%%%%%%%%%%%%%% Majorana mass by D-brane instanton effects %%%%%%%%%%%%%%%%%%%%%%%%%%%


%\cite{Hoshiya:2021nux}
\bibitem{Hoshiya:2021nux}
K.~Hoshiya, S.~Kikuchi, T.~Kobayashi, K.~Nasu, H.~Uchida and S.~Uemura,
%``Majorana neutrino masses by D-brane instanton effects in magnetized orbifold models,''
PTEP \textbf{2022}, no.1, 013B04 (2022)
%doi:10.1093/ptep/ptab152
[arXiv:2103.07147 [hep-th]].





%\cite{Abe:2017gye}
\bibitem{Abe:2017gye}
H.~Abe, T.~Kobayashi, K.~Sumita and S.~Uemura,
%``K\"ahler moduli stabilization in semirealistic magnetized orbifold models,''
Phys. Rev. D \textbf{96}, no.2, 026019 (2017)
%doi:10.1103/PhysRevD.96.026019
[arXiv:1703.03402 [hep-th]].




%\cite{Bjorkeroth:2015ora}
\bibitem{Bjorkeroth:2015ora}
F.~Bj\"orkeroth, F.~J.~de Anda, I.~de Medeiros Varzielas and S.~F.~King,
%``Towards a complete A$_{4} \times$ SU(5) SUSY GUT,''
JHEP \textbf{06} (2015), 141
doi:10.1007/JHEP06(2015)141
[arXiv:1503.03306 [hep-ph]].
%81 citations counted in INSPIRE as of 20 Apr 2021


%%%%%%%%%%%%%%% Particle Data Group %%%%%%%%%%%%%%%%%%%%%%%%%%%

%\cite{Zyla:2020zbs}
\bibitem{Zyla:2020zbs}
P.~A.~Zyla \textit{et al.} [Particle Data Group],
%``Review of Particle Physics,''
PTEP \textbf{2020} (2020) no.8, 083C01
%doi:10.1093/ptep/ptaa104

%%%%%%%%%%%%%%% NuFIT 2021 %%%%%%%%%%%%%%%%%%%%%%%%%%%

\bibitem{NuFIT:2021}
NuFIT 5.1, http://www.nu-fit.org/?q=node/238, 2021

\end{thebibliography}
\end{document}